\documentclass{article}
\usepackage{arxiv}
\usepackage[T1]{fontenc}    
\usepackage{hyperref}       
\usepackage{url}            
\usepackage{booktabs}       
\usepackage{amsfonts}       
\usepackage{nicefrac}       
\usepackage{microtype}      
\usepackage{lipsum}
\usepackage{algpseudocode}
\usepackage{amsmath}
\usepackage{float}
\usepackage{amsmath,amssymb,amsfonts}
\usepackage{multirow}
\usepackage{algorithm}
\usepackage{graphicx}
\usepackage{subcaption}
\usepackage{xcolor}
\usepackage{setspace}
\usepackage[algo2e,linesnumbered,lined,boxed,commentsnumbered, ruled]{algorithm2e}
\usepackage{cite}
\graphicspath{ {./images/} }

\title{Probabilistic Assessment of West Nile Virus Spillover Risk Using a
Compartmental Mechanistic Model}
\author{
  Saman Hosseini* \\
  Department of Electrical and Computer Engineering \\
  Kansas State University \\
  Manhattan, KS, USA \\
  \texttt{*corresponding author:} \\
  shosseini@ksu.edu \\
  \And
  Lee W. Cohnstaedt \\
  Foreign Arthropod-Borne Animal Diseases Research Unit \\
  National Bio- and Agro-defense Facility, USDA ARS \\
  Manhattan, KS, USA \\
  \And
  Matin Marjani \\
  Department of Electrical and Computer Engineering \\
  Kansas State University \\
  Manhattan, KS, USA \\
  \And
  Caterina Scoglio \\
  Department of Electrical and Computer Engineering \\
  Kansas State University \\
  Manhattan, KS, USA \\
}
\date{}

\begin{document}
\maketitle

\begin{abstract}
This paper presents a novel probabilistic approach for assessing the risk of West Nile Disease (WND) spillover to the human population. The assessment has been conducted under two different scenarios: (1) assessment of the onset of spillover, and (2) assessment of the severity of the epidemic after the onset of the disease.
A compartmental model of differential equations is developed to describe the disease transmission mechanism, and a probability density function for pathogen spillover to humans is derived based on the model for the assessment of the risk of the spillover onset and the severity of the epidemic.  The prediction strategy involves making a long-term forecast and then updating it with a short-term (lead time of two weeks or daily). The methodology is demonstrated using detailed outbreak data from high-case counties in California, including Orange County, Los Angeles County, and Kern County. The predicted results are compared with actual infection dates reported by the California Department of Public Health for 2022–2024 to assess prediction accuracy.
The performance accuracy is evaluated using a logarithmic scoring system and compared with one of the most renowned predictive models to assess its effectiveness. In all prediction scenarios, the model demonstrated strong performance. Lastly, the method is applied to explore the impact of global warming on spillover risk, revealing an increasing trend in the number of high-risk days and a shift toward a greater proportion of these days over time for the onset of the disease.
\end{abstract}

\flushbottom
\maketitle
%
%

\section{Introduction}
The first recorded case of WND in the United States was observed in 1999 in New York \cite{sejvar2003west}. 
Over time, disease became a major public health issue in many states in the United States, causing both health risks and economic losses. The increasing prevalence of the disease led to increased investment in mosquito control programs and the implementation of widespread public health campaigns to mitigate its impact. Consequently, since 1999, authorities in the United States have mainly been concerned with assessing the various aspects of the risk of WND to the human population. The West Nile virus is transmitted primarily by mosquitoes, the main vector species belonging to the \textit{Culex} family.
The disease's mechanism of spread relies on interactions between mosquitoes and birds, with occasional transmission to humans as spillover cases. Mosquitoes and birds can mutually infect each other, while mosquitoes serve as the only agents that transmit the infection to mammals. In particular, mammals cannot infect mosquitoes and act as dead ends of the disease \cite{campbell2002west}.\\

Significant work has been done on the risk assessment of West Nile disease, focusing on various aspects and parameters. For example, \cite{kilpatrick2005west} focused on various types of mosquitoes to determine which species significantly affect the transmission of WND to the human population. This work proposed a measure based on the abundance, prevalence of infection, vector competence, and bite behavior of vectors. \cite{harrigan2014continental} assessed the effect of climate and climate change on the risk of spreading of WND using a comprehensive set of species distribution models.
For this work, CDC information on the West Nile virus (WNV) about vectors, birds, and humans, along with climate data, have been used to predict the future of the disease in different climate scenarios. \cite{theophilides2003identifying} used a localized Knox test to identify areas of high risk for WND in New York City. \cite{gorris2023assessing} used a mean random forest (RF) model to assess the influence of climate parameters on the geographical pattern of (WNV) in the USA. In addition, \cite{garcia2021predicting} used two risk models, one environmental and the other spatial-environmental with fuzzy logic, to assess the risk and concluded that the spatial-environmental model is the most useful for predicting the location of the outbreak.\\
For network base works, Das et al. \cite{das2024sir} proposed a model to simulate and predict spillover dynamics across interconnected networks, emphasizing the level of interconnection as a critical factor. Using the FastGEMF framework, developed by Samaei et al. \cite{samaei2025fastgemf}, for efficient simulation of mechanistic models over multilayer networks, they explored the spillover as a function of the connections between human and animal networks. Their findings demonstrated how these interconnections influence transmission processes, providing valuable insight into the dynamics of networked systems. However, these insights are most beneficial when detailed information about reservoir and host networks is readily available. For diseases like WND, whose reservoir networks are highly complex and difficult to model, this approach may be less applicable.\\

To address this gap, Alizadeh et al. proposed a data-driven hybrid CNN-LSTM model 
to forecast short-term disease dynamics \cite{alizadeh2025epidemic}. Their method demonstrated promising performance using Covid-19 data, particularly with short horizons and clean data. However, its performance is not guaranteed across datasets with varying forecasting windows and risks assessment can not be evaluated.

One of the most important aspects of risk assessment (RA) is the evaluation of the timing of risk, that is, the determination when there is a potential for infection in human (or even animal) populations. An accurate understanding of the timing of spillover to humans helps related organizations prepare for potential epidemics. In addition to the accuracy of the risk assessment, another important factor is the lead time during which it remains reliable. In other words, not only must the risk assessment be accurate, but its validity should also last long enough, which is absent in most risk models, to allow effective decision-making about the situation. Third, an important consideration in risk assessment is the potential to overestimate infection risk by using mechanistic parameters such as the basic reproduction number ($R_0$). For example, there have been surprising cases involving infectious arboviruses such as dengue, chikungunya, and Zika, where the value of $R_0$ exceeded one and infections were reported, yet the diseases did not escalate into epidemics. This indicates that parameters such as $R_0$ should be used more cautiously and perhaps interpreted differently \cite{fox2019downgrading}. In addition, risk assessments must account for the severity of possible epidemics. Binary evaluations, which simply indicate whether or not a risk exists, may not offer the nuanced insight needed. \\

In recent years, a shocking surge of WND infections has been observed in some states of the United States, and even other arbovirus diseases. In some cases, not only has the number of cases been surprisingly high, but also the onset of the disease has shifted. An important question arising from the recent changes in disease behavior is whether there is any evidence linking global warming to these surges in infections and changes in timing. In addition, this new behavior of diseases and epidemics places additional pressure on the healthcare system. To reduce the pressure and stress of facing a sudden high number of cases, new methods can be developed to assess risk in a way that makes it possible to predict the prospects of what will happen during the hot seasons and months at the beginning of the year.\\
This paper proposes a novel approach to predict the risk of infection over time in human populations. Specifically, the innovation of this work lies in the development of a probabilistic method that establishes a joint probability density function for the assessment of spillover risk. 
To perform this assessment, two scenarios are considered. The first scenario occurs at the beginning of the year or well before the outbreak season, when a prediction is needed to define general policies for mitigation strategies, requiring a long-term forecast. The second scenario takes place closer to the outbreak season, where more accurate predictions are necessary to adjust general policies, necessitating a short-term forecast.
It helps us avoid overestimating the risk of spillover by providing a probabilistic perspective based on the function of $R_0$, the abundance of mosquitoes, and the weather parameter. In other words, in the proposed risk assessment framework, $R_0$ (also the abundance of mosquitoes and the weather parameter) plays the role of a random variable rather than a mechanistic parameter. 
Finally, using this probabilistic approach allows us to identify a trend that shows the effect of global warming on recent WND epidemics, in terms of the number of high-risk days and the time of infection.\\
To demonstrate the method, we have chosen Orange County, California, because of its high number of cases. For the creation and calibration of the model, we used data from 2006 to 2021. 

For validation purposes, predictions were generated and compared with actual data for Orange County, Los Angeles County, and Kern County in California for the years 2022, 2023, and 2024. A similar validation was conducted for Dallas County and Harris County in Texas, but only for the year 2024, due to the shorter historical dataset available compared to California. In both states, the results indicate high precision of the method. All data were collected from the California Department of Public Health (CDPH) and the Texas Department of State Health Services (DSHS).

Sections 3 and 4 represent the Bayesian assessment methodology of the risk of WND spillover onset and WND spillover severity and their validation.
The last section focuses on global warming and its effect on WND, which shows that there is evidence that WND has been affected by global warming in terms of earlier onset and higher level of risk.

\section{The Model}
One of the most well-known approaches to studying an epidemic and the dynamics of disease, while simultaneously considering the profiles of agents involved in the model, such as vectors, birds, and zoonotic agents, is to develop a mechanistic model based on nonlinear differential equations.
In this section, our objective is to explain the compartmental differential equation model used to describe the mechanism of the disease and its behavior between different populations. \\
Mechanistic models for WND are constructed based on interactions between three populations of mosquitoes, birds, and humans (or mammals). We have established our models using a single type of mosquito (\textit{Culex} mosquitoes, \textit{pipiens} family), one species of bird (American crows), and the human population.

\begin{figure}[htbp]
    \centering
    \begin{subfigure}[b]{0.45\linewidth}
         \centering
         \includegraphics[width=\linewidth]{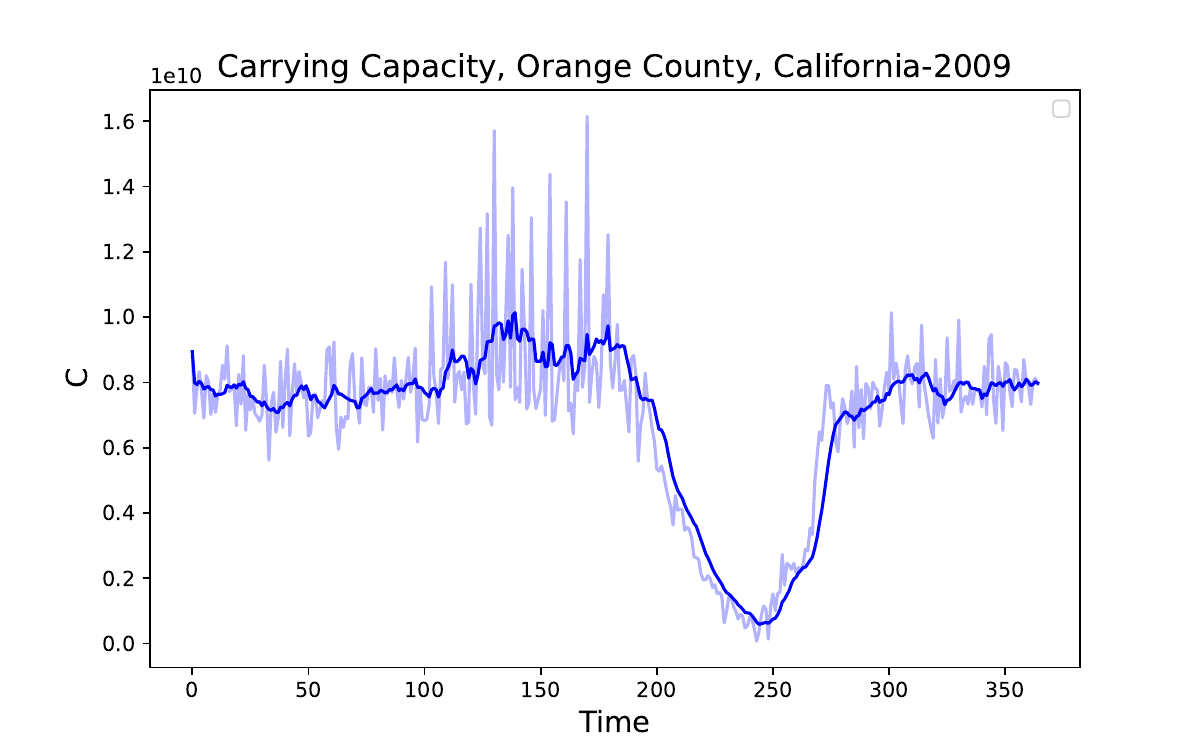}
         \caption{Carrying Capacity}
         \label{fig:carrying_capacity}
    \end{subfigure}
    \hfill
    \begin{subfigure}[b]{0.45\linewidth}
         \centering
         \includegraphics[width=\linewidth]{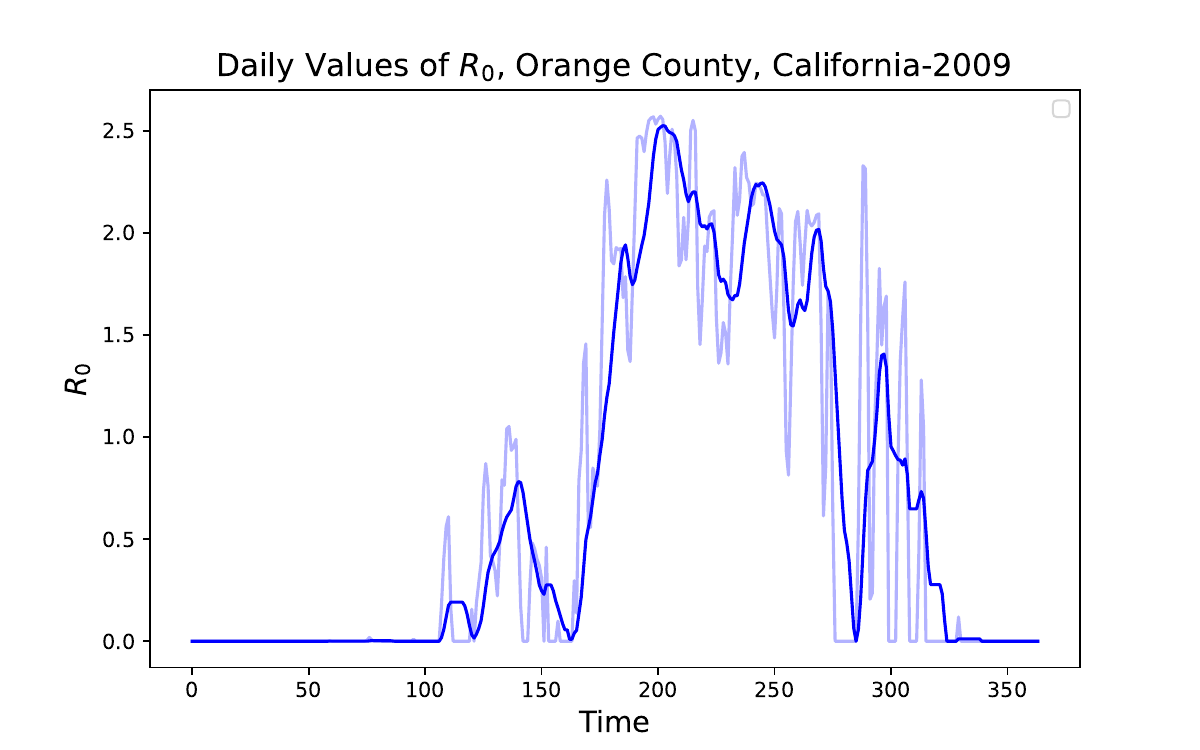}
         \caption{Daily $R_0$}
         \label{fig:R0_moving_average}
    \end{subfigure}
    \hfill
    \begin{subfigure}[b]{0.45\linewidth}
         \centering
         \includegraphics[width=\linewidth]{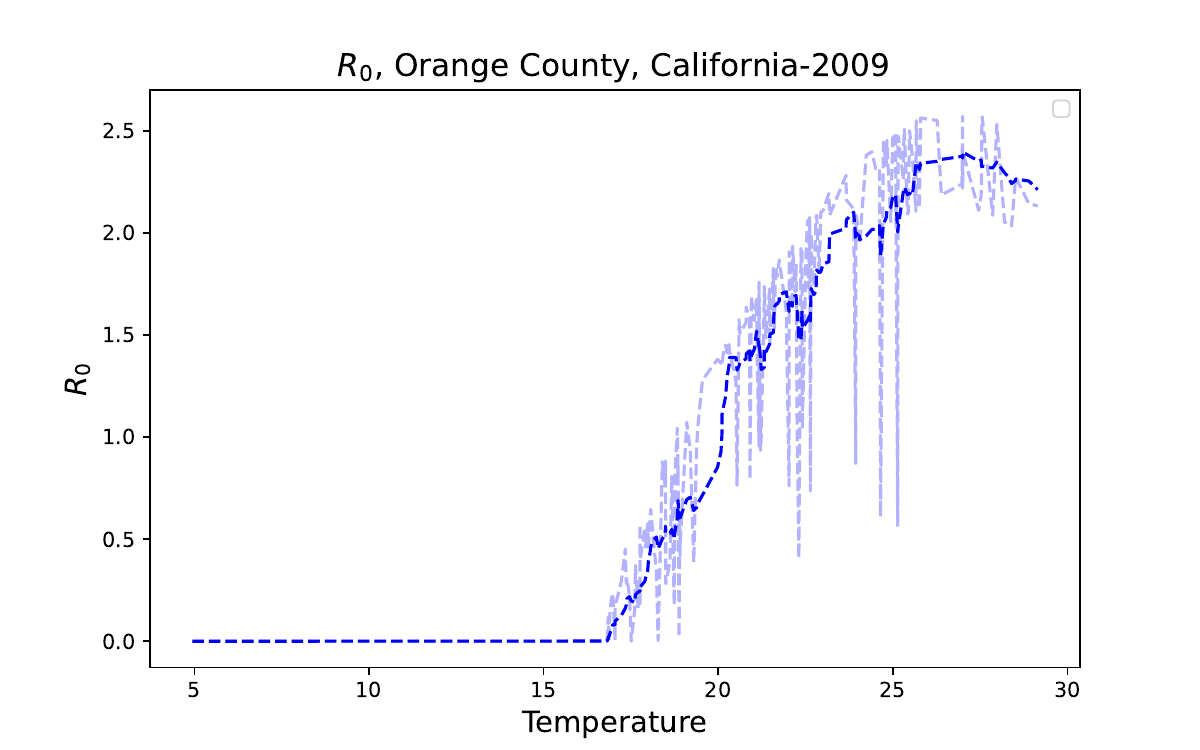}
         \caption{$R_0$ vs Temperature}
         \label{fig:R0_vs_Temperature}
    \end{subfigure}
    \caption{(a) Estimates of the carrying capacity; (b) and (c) Estimates of $R_0$ over time and by temperature, respectively, for Orange County, California, 2009.}
    \label{fig:R0 and Carrying capacity}
\end{figure}

Knowing these populations, the model consists of the following compartments.\\
\begin{enumerate}
    \item The human population consists of four compartments of susceptible, exposed, infected, and recovered. 
    \item Mosquito population that involves the life cycle and disease compartments of the eggs, aquatic phase, mosquitoes (susceptible, exposed, and infectious).
    \item Birds consisting of eggs, fledglings, and adult birds (susceptible, exposed, infectious, and recovered).
\end{enumerate}
It should be noted that there is no recovery compartment in the mosquito population. This is because the lifespan of mosquitoes is shorter than the duration required to recover from the disease.\\\\
Taking into account the above compartments, the goal is to model the disease with a realistic representation of its mechanism to obtain as accurate as possible results. To minimize the impact of noise (error) related to the coefficients used for the mosquito life cycle and disease transmission in non-linear equations, temperature-dependent mechanistic coefficients obtained from laboratory data have been incorporated \cite{shocket2020transmission}. In addition, incorporating temperature-dependent coefficients into the model allows its use as a temporal model. This means that it is possible to obtain accurate inferences for each geographical region by considering factors such as temperature (generally the weather parameter). The general form of the mechanistic differential equation model has been described in the Appendix and is not the main topic of this work. \\\\
We need to obtain $R_0$ for the model, but we have to note that in WND the human population cannot be infectious and serves as a dead end host for the disease (Their viremia levels are not significant enough to make them infectious.), and the meaning of $R_{0}$ changes from a directly transmitted disease. $R_0$ represents the average number of secondary infections in mosquitoes and birds from one initially infected host (mosquito or bird) in a completely susceptible population.
Statistically speaking, if the variable $X$ represents the number of secondary infections by an infected host, $R_{0}$ is the mathematical expectation (mean) of that random variable. For the model presented in this work, $R_{0}$ has been obtained in a closed form using the next-generation matrix concept as follows \cite{diekmann1990definition}. More details about the method and the process of obtaining the closed form of $R_{0}$ are provided in the appendix.\\
The general form of the $R_{0}$ of the model is like:
\begin{equation}
\label{eq:R0}
\begin{aligned}
  {R}_{0} &= \sqrt{{{R}_{0}}^{B}{{R}_{0}}^{M}} 
\end{aligned}
\end{equation}
in which:
\begin{equation}\label{eq:RB0}
{{R}_{0}}^{B}=\frac{{{\beta }_{B\to M}}{{M}_{S}}{{\delta }_{B}}}{({{\delta }_{B}}+{{\mu }_{B}})({{\lambda }_{B}}+{{\mu }_{WND-B}}+{{\mu }_{B}})},
\end{equation}
and
\begin{equation*}
{{R}_{0}}^{M}=\frac{{{\beta }_{M\to B}}{{B}_{S}}PDR}{{{\mu }_{M}}(PDR+{{\mu }_{M}})}
\end{equation*}
Based on what we explained, it is evident that (\ref{eq:R0}) does not include any parameters related to the human population or its compartments.\\
It is possible to explain the general formula of ${R_{0}}$ as follows. First, in the bird-based fraction of the basic reproduction number ($R_{0}^{B}$), the term $\frac{1}{\lambda_{B}+\mu_{B}+\mu_{WND-B}}$ signifies the average duration of the birds in the infectious phase. This is based on the fact that the time spent in this phase follows an exponential distribution with the parameter ($\lambda_{B}+\mu_{B}+\mu_{WND-B}$). Therefore, the mean of this exponential variable is equal to $\frac{1}{\lambda_{B}+\mu_{B}+\mu_{WND-B}}$. Second, $R_{0}^{B}$ is directly proportional to the probability of survival from the exposed phase ($\frac{\delta_{B}}{\delta_{B}+\mu_{B}}$). To demonstrate this, consider random variables $T_{\delta_{B}}$ and $T_{\mu_{B}}$ that represent the time to transition from exposed phase to infectious and death, respectively. The probability of transitioning from the exposed to the infectious phase is then derived as follows.\\
$P(E\to I)=P({{T}_{{{\delta }_{B}}}}<{{T}_{{{\mu }_{B}}}})$\\
It is known that both random variables of $T_{\delta_{B}}$ and $T_{\mu_{B}}$ are distributed exponentially (${{T}_{{{\delta }_{B}}}}\sim \exp ({{\delta }_{B}}),{{T}_{{{\mu }_{B}}}}\sim \exp ({{\mu }_{B}})$).\\
Based on these facts:
\begin{align*}
  P(E \to I) &= P(T_{\delta_B} < T_{\mu_B}) \\
  &= \int_{0}^{+\infty} P(T_{\delta_B} < T_{\mu_B} \mid T_{\mu_B} = t) f_{T_{\mu_B}}(t) \, dt \\
  &= \int_{0}^{+\infty} P(T_{\delta_B} < t \mid T_{\mu_B} = t) f_{T_{\mu_B}}(t) \, dt \\
  &= \int_{0}^{+\infty} P(T_{\delta_B} < t) f_{T_{\mu_B}}(t) \, dt \\
  &= \int_{0}^{+\infty} (1 - e^{-\delta_B t}) \mu_B e^{-\mu_B t} \, dt \\
  &= \frac{\delta_B}{\delta_B + \mu_B}.
\end{align*}
Using the same reasoning, the following results can be inferred from the mosquito-based fraction of the basic reproduction number ($R_{0}^{M}$):\\
\begin{itemize}
    \item Probability of an Infected Mosquito Surviving the Exposed Phase: $\frac{PDR}{PDR+\mu_{M}}$.\\
\end{itemize}
\begin{itemize}
    \item The average duration a mosquito spends in the infectious phase: $\frac{1}{\mu_{M}}$.\\
\end{itemize}
\begin{itemize}
    \item Furthermore, it can be observed that the value of $R_{0}$ is a function of the number of susceptible mosquitoes $M_{S}$. Consequently, it is a function of the carrying capacity defined in the aquatic phase of the mosquito development population.\\
\end{itemize}
In the following, the basic reproduction number is plotted for a given year in Figure \ref{fig:R0 and Carrying capacity}.
If we review the structure of $R_{0}$, all the components in the formula make logical sense. For the infection to spread initially, a bird must survive the exposed phase, and the proportion of birds that survive is given by $\frac{\delta_{B}}{\delta_{B}+\mu_{B}}$. Following survival, the duration of time that a bird remains infectious is crucial to the transmission of the pathogen. During this period, the bird is infectious and can spread the pathogen, with a mean duration equal to $\frac{1}{\mu_{M}}$ (importantly, the lifetime of the mosquitoes is shorter than the recovery time of the disease). An infectious bird during the period of time spent in the infectious phase infects the mosquitoes at the rate of $\beta_{B\to M}$ and the number of all infected mosquitoes is obtained by multiplying $M_{S}$.\\ 
Applying the same logic to the mosquito fraction of $R_{0}$, to have a spillover from mosquitoes to birds, mosquitoes must survive the exposed phase, which its probability is $\frac{PDR}{PDR+\mu_{M}}$.\\
After survival, the average time it takes a mosquito to infect birds is $\frac{1}{\mu_{M}}$.
During this infectious period, mosquitoes can infect the bird population at a rate of $\beta_{M\to B}$, and the total number of infected birds is obtained by multiplying this rate by the susceptible number of the bird population $B_{S}$.

Although the mechanistic model is a useful tool for analyzing disease dynamics over time, it is evident that many factors influence the disease mechanism. Therefore, for risk assessment, it is reasonable to use a model that accounts for stochasticity. Based on this idea, we will develop a statistical model using our mechanistic framework.

\section{Bayesian Assessment of WND Spillover Onset Risk}
Preparing for potential threats to humans and their lives is crucial, as it allows planned responses rather than reactive measures to outbreaks. Epidemics and diseases are no exceptions to this necessity. Therefore, a key component of this preparation involves developing methods to anticipate disease outbreaks, especially within human communities.\\
Consequently, this section is dedicated to a description of approaches to assess the risk of spreading WND to the human population.\\

We will address this in two main subsections: first, by assessing the risk of the first-time spillover, and second, by assessing the severity of the subsequent spillover, after the onset of the disease, in the human population.
The focus on spillover in these two subsections arises from the fact that global warming has altered the dynamics of the disease. By analyzing the timing of disease emergence, we can estimate the temporal boundaries of the outbreak, which may have changed due to climate-induced changes in recent years, helping to prevent unexpected outbreaks and their consequences while developing appropriate mitigation strategies. In the end of the paper, we will further discuss the impact of global warming and the specific changes it has caused in greater detail.
To develop a stochastic model, we used the mosquito population predicted by the model and the basic reproduction number ($R_{0}$) using historical data (in fact, we used a function of them).\\
To explain the generality of what we are going to do, it is reminded that the West Nile virus mainly circulates between mosquitoes and birds. However, in specific circumstances, such as when avian hosts are scarce, mosquitoes can seek blood meals from alternative sources, including mammals and, consequently, humans \cite{colpitts2012west}. 
This situation is referred to as the spillover of the virus into the human population. Logically, higher values of the basic reproduction number (or even its variables) suggest that, in the presence of the pathogen, the probability of spreading to the human population increases. This fundamental insight has guided us in the establishment of a risk assessment method that involves calculating the likelihood of spillover.\\
\subsection{Spillover Onset} \label{sec: initial}
\begin{figure}
    \centering
    \begin{subfigure}{0.45\textwidth}
        \centering
        \includegraphics[width=\linewidth]{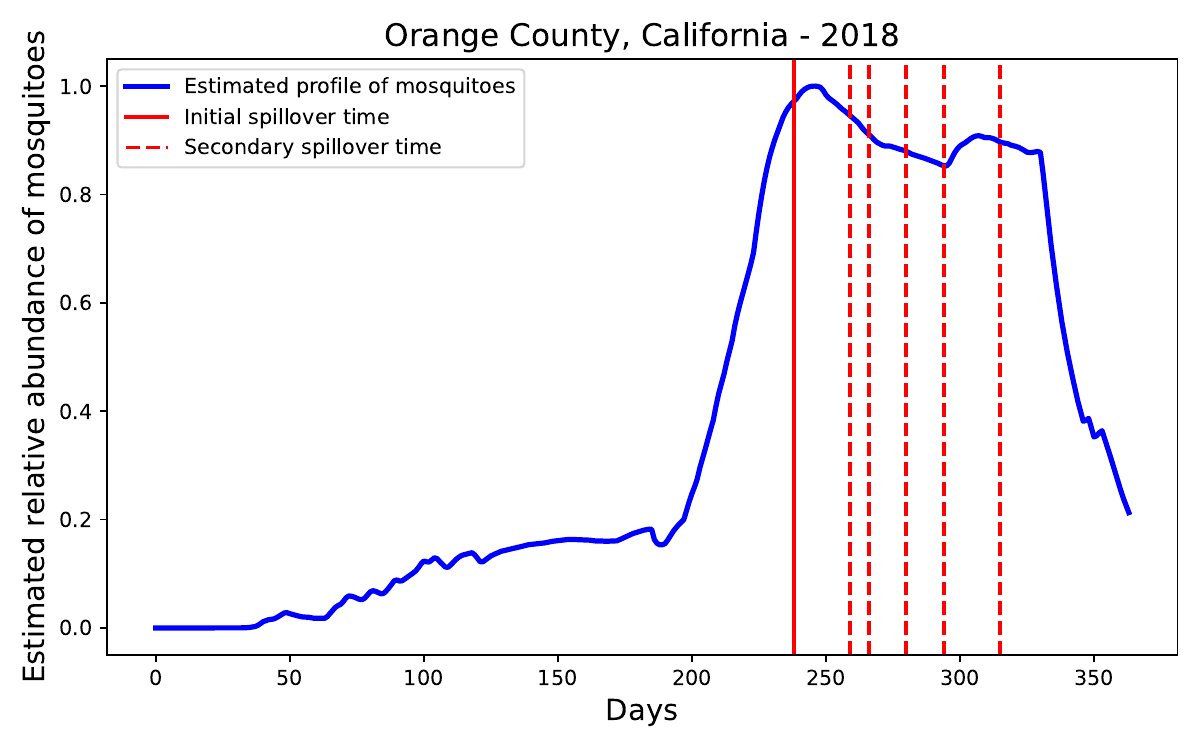}
        \caption{}
        \label{fig:mosquito_profile}
    \end{subfigure}
    \hfill
    \begin{subfigure}{0.45\textwidth}
        \centering
        \includegraphics[width=\linewidth]{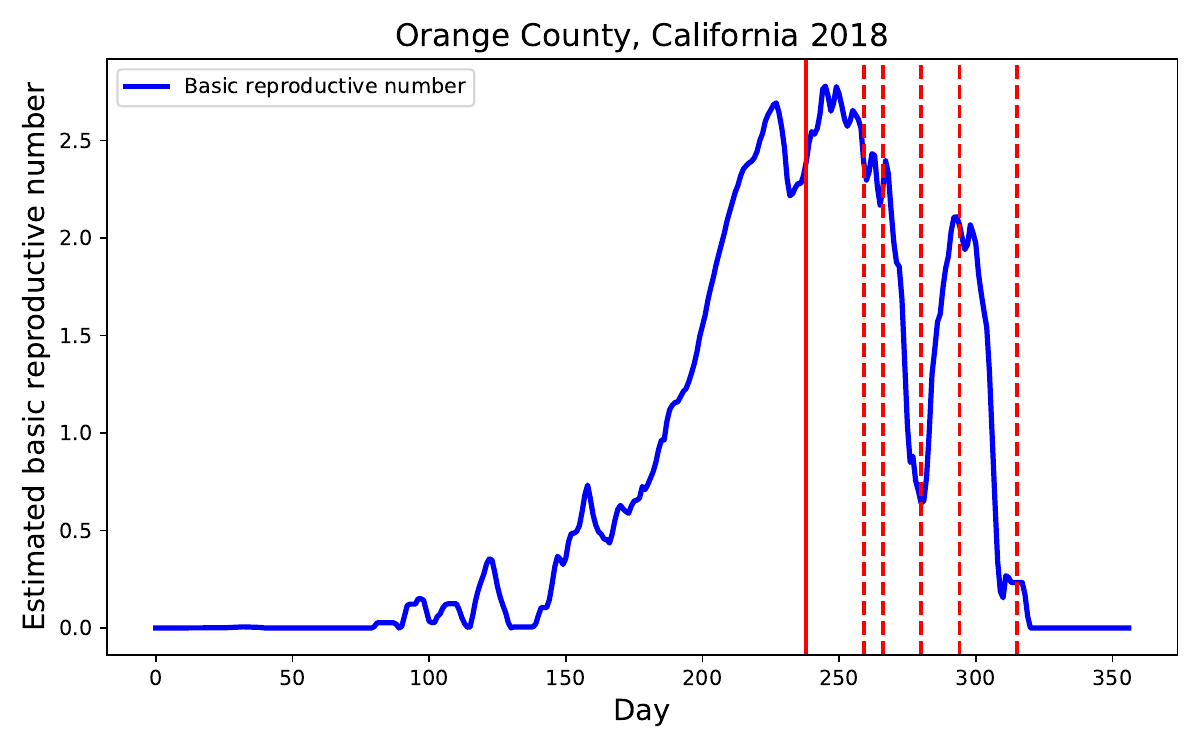}
        \caption{}
        \label{fig:R0_profile}
    \end{subfigure}
    
    \caption{Mosquito profile, basic reproduction number, and reported times of spillover to human population (first-time spillover and after the first time), Orange County, California, 2018.}
    \label{fig: Mosquito profile and effective numbers}
\end{figure}

By running the compartmental model using historical climate parameters, we will collect the values of the effective mosquito population ($M$), the effective $R_{0}$ (the value of the mosquito profile and $R_{0}$ in which human cases have been reported), and the number of cases reported at the time when human infection was reported for the first time (from 2006 to 2020).\\
Figure \ref{fig: Mosquito profile and effective numbers} presents the time of the starting risk of spillover (first-time spillover), the estimated effective mosquito population profile, and the effective basic reproduction numbers for Orange County, California, for 2006, along with the other reported spillover times. 

\begin{figure*}[!ht]
    \centering
    \begin{subfigure}[t]{0.32\textwidth}
        \centering
        \includegraphics[width=\textwidth]{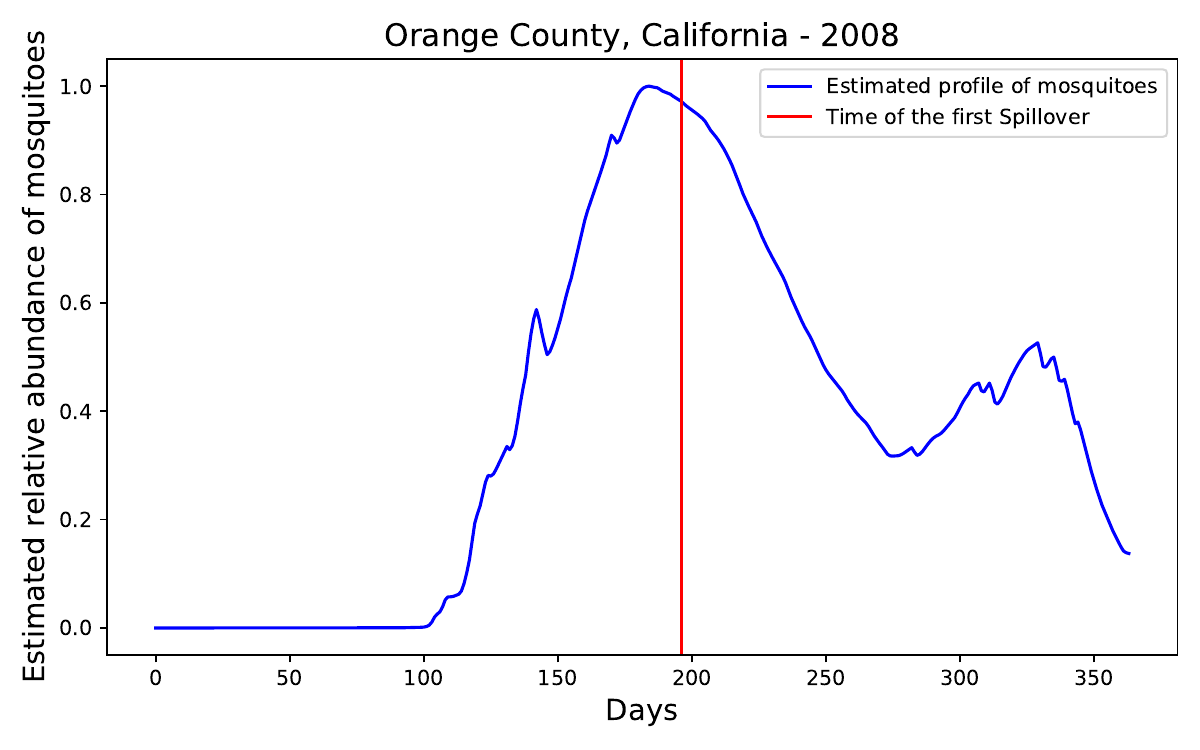}
        \caption{}
        \label{fig:january}
    \end{subfigure}
    \hfill
    \begin{subfigure}[t]{0.32\textwidth}
        \centering
        \includegraphics[width=\textwidth]{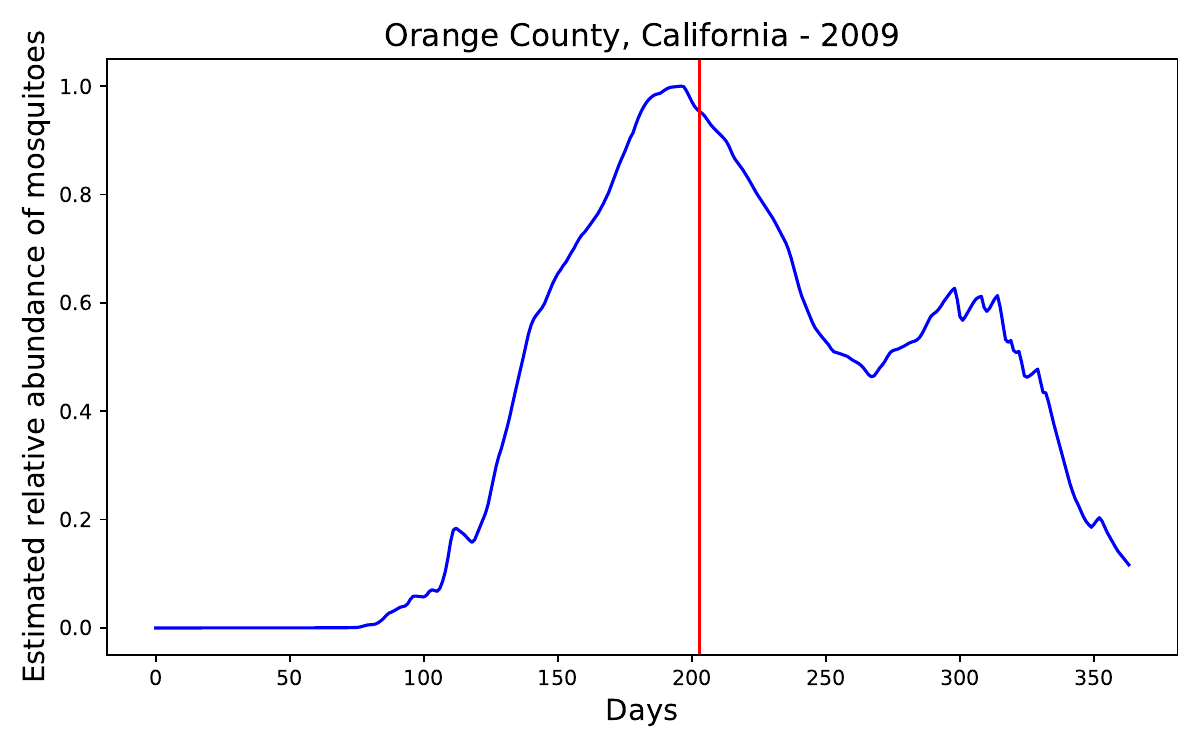}
        \caption{}
        \label{fig:february}
    \end{subfigure}
    \hfill
    \begin{subfigure}[t]{0.32\textwidth}
        \centering
        \includegraphics[width=\textwidth]{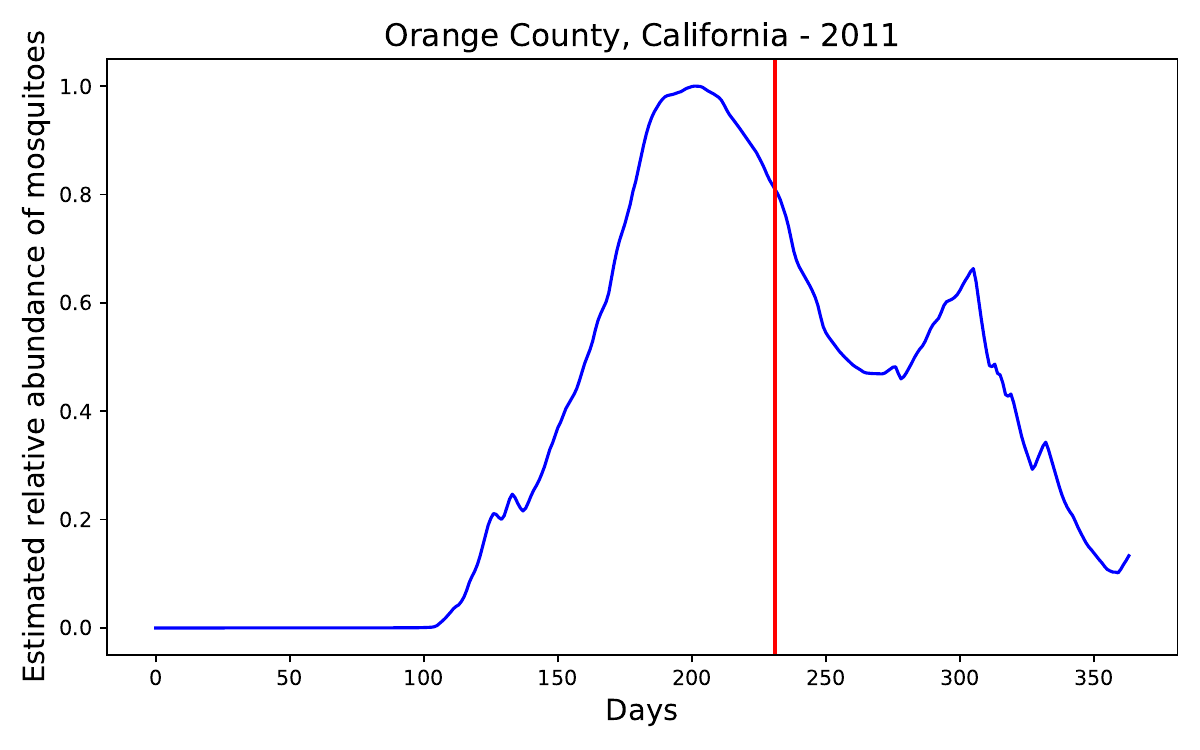}
        \caption{}
        \label{fig:april}
    \end{subfigure}
    \hfill
    \begin{subfigure}[t]{0.32\textwidth}
        \centering
        \includegraphics[width=\textwidth]{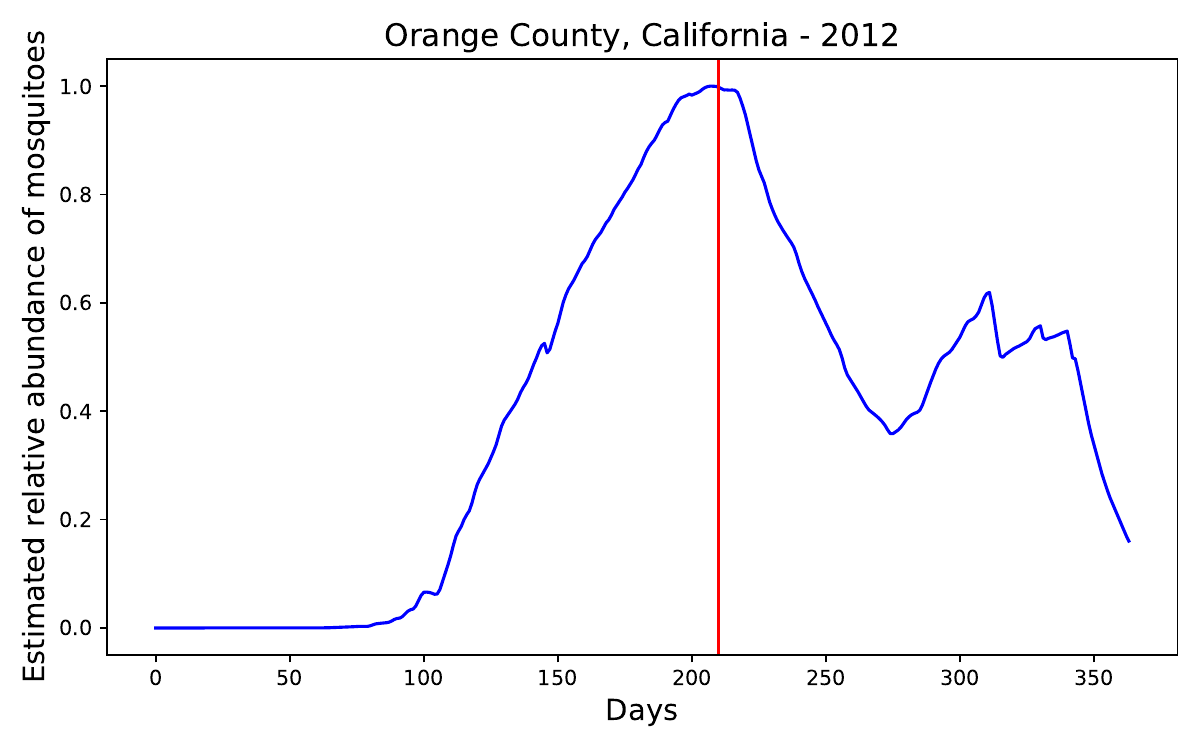}
        \caption{}
        \label{fig:may}
    \end{subfigure}
    \hfill
    \begin{subfigure}[t]{0.32\textwidth}
        \centering
        \includegraphics[width=\textwidth]{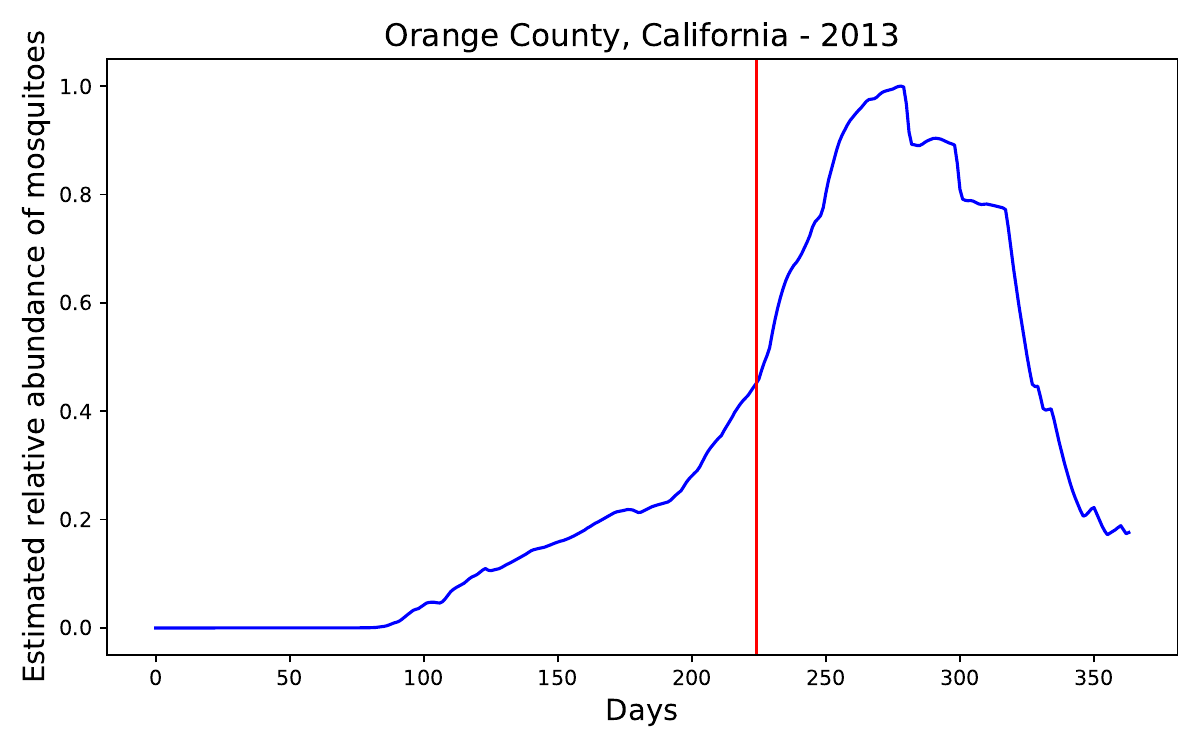}
        \caption{}
        \label{fig:june}
    \end{subfigure}
        \hfill
    \begin{subfigure}[t]{0.32\textwidth}
        \centering
        \includegraphics[width=\textwidth]{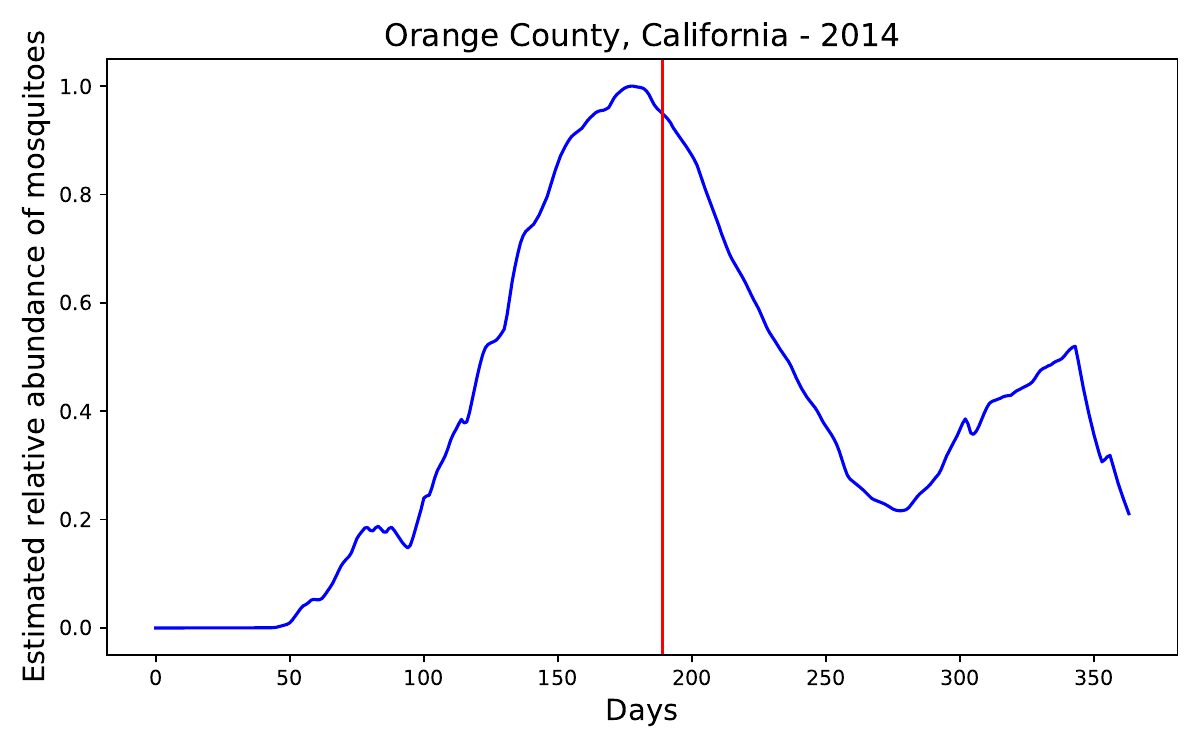}
        \caption{}
        \label{fig:june}
    \end{subfigure}
        \hfill
    \begin{subfigure}[t]{0.32\textwidth}
        \centering
        \includegraphics[width=\textwidth]{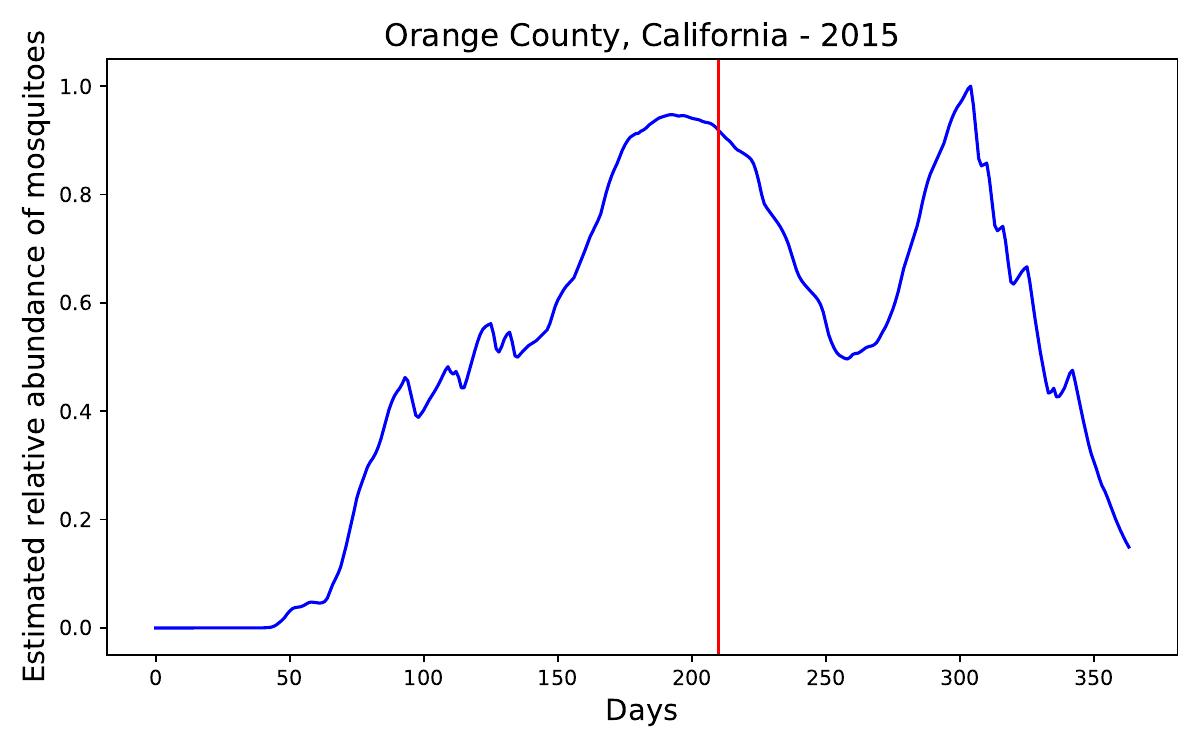}
        \caption{}
        \label{fig:june}
    \end{subfigure}
        \hfill
    \begin{subfigure}[t]{0.32\textwidth}
        \centering
        \includegraphics[width=\textwidth]{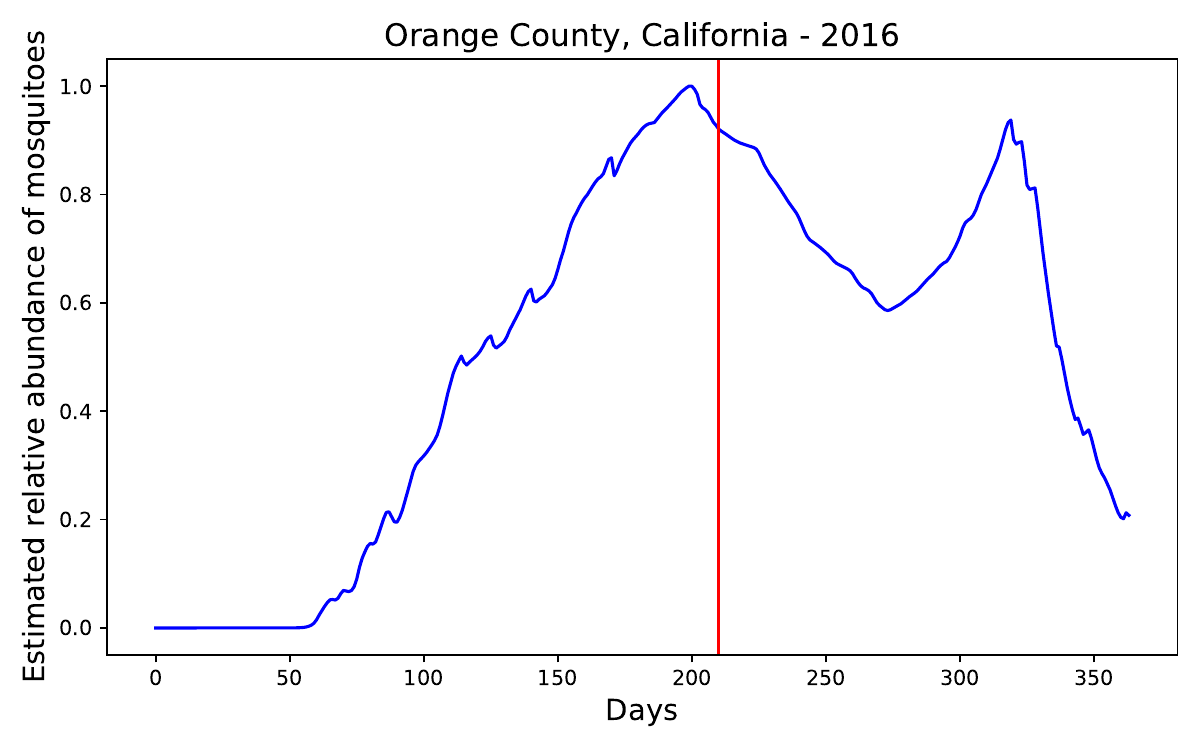}
        \caption{}
        \label{fig:june}
    \end{subfigure}
        \hfill
    \begin{subfigure}[t]{0.32\textwidth}
        \centering
        \includegraphics[width=\textwidth]{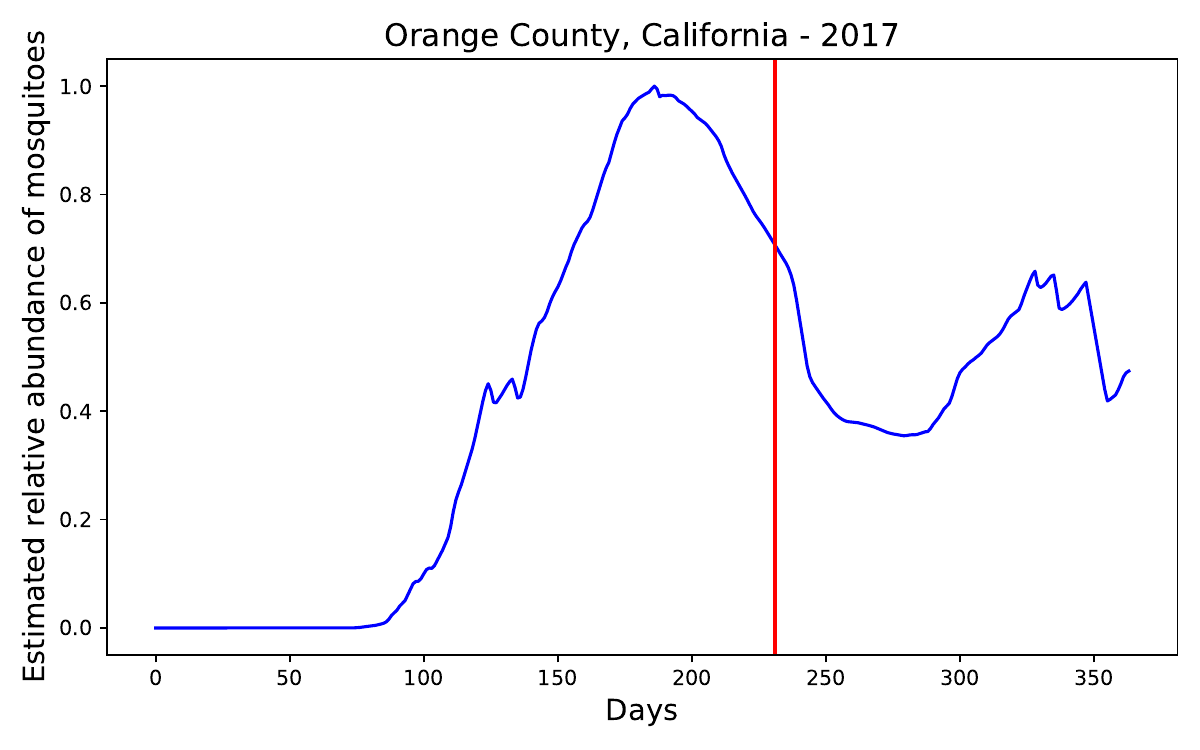}
        \caption{}
        \label{fig:june}
    \end{subfigure}
        \hfill
    \begin{subfigure}[t]{0.32\textwidth}
        \centering
        \includegraphics[width=\textwidth]{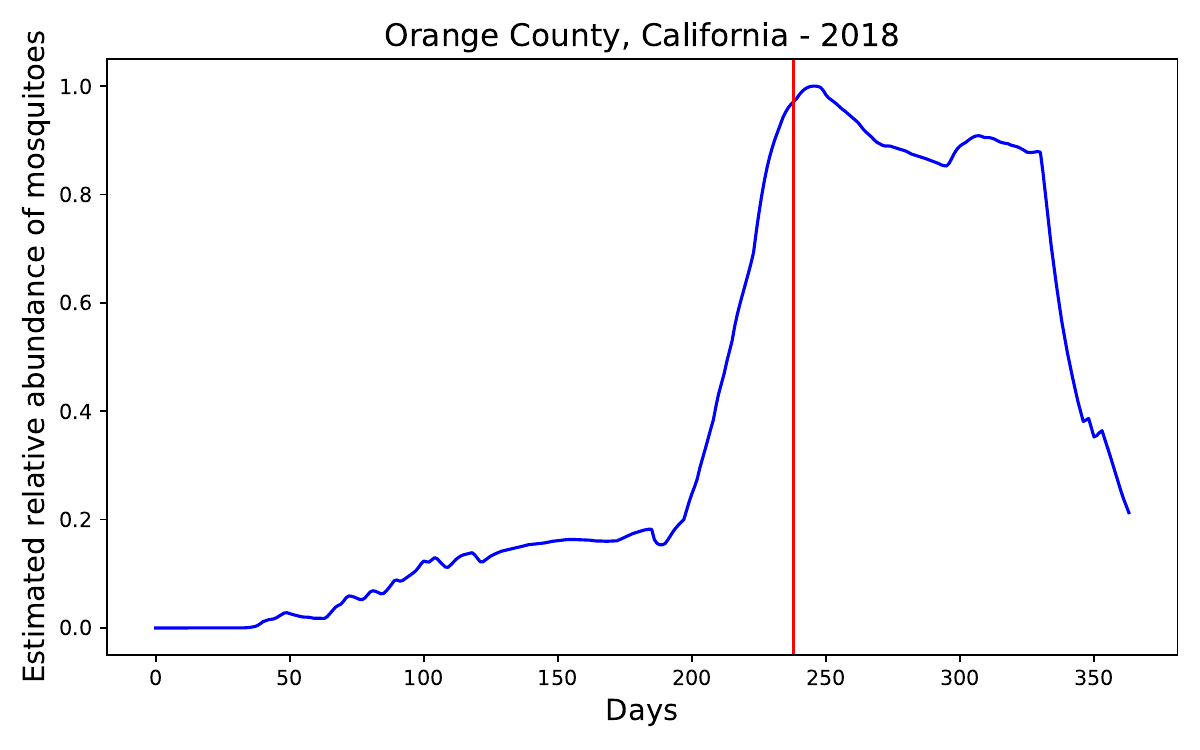}
        \caption{}
        \label{fig:june}
    \end{subfigure}
        \hfill
    \begin{subfigure}[t]{0.32\textwidth}
        \centering
        \includegraphics[width=\textwidth]{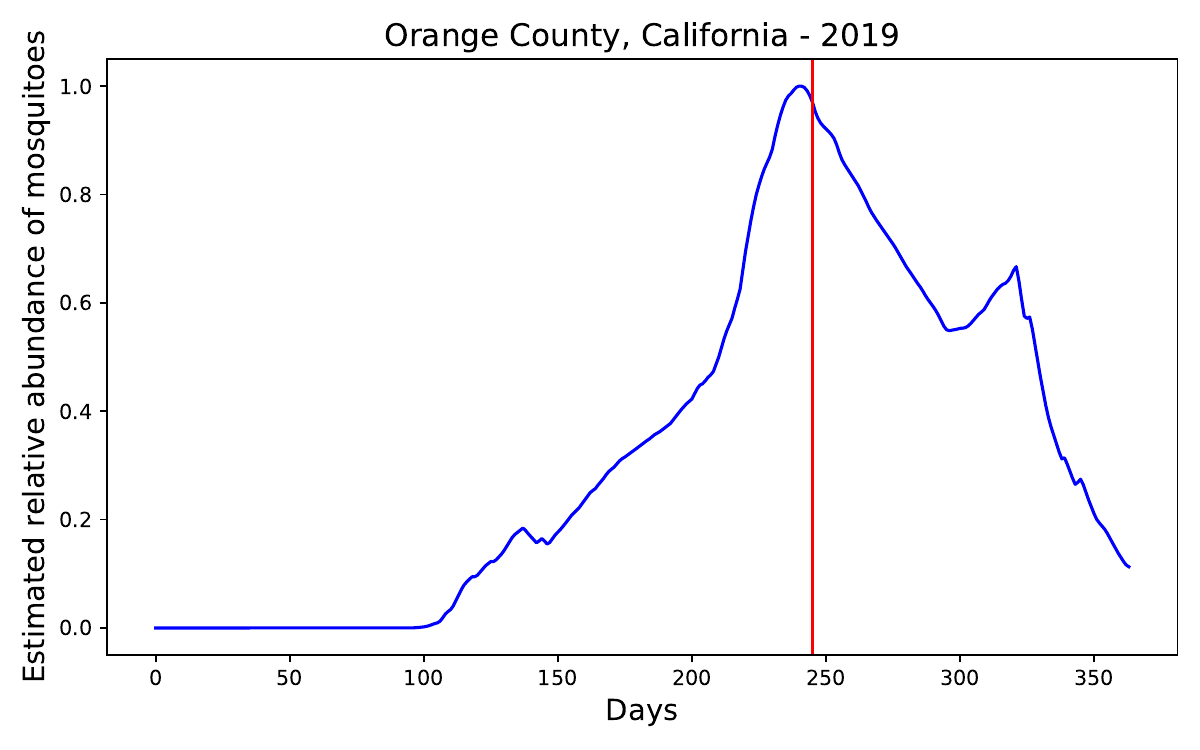}
        \caption{}
        \label{fig:june}
    \end{subfigure}
        \hfill
    \begin{subfigure}[t]{0.32\textwidth}
        \centering
        \includegraphics[width=\textwidth]{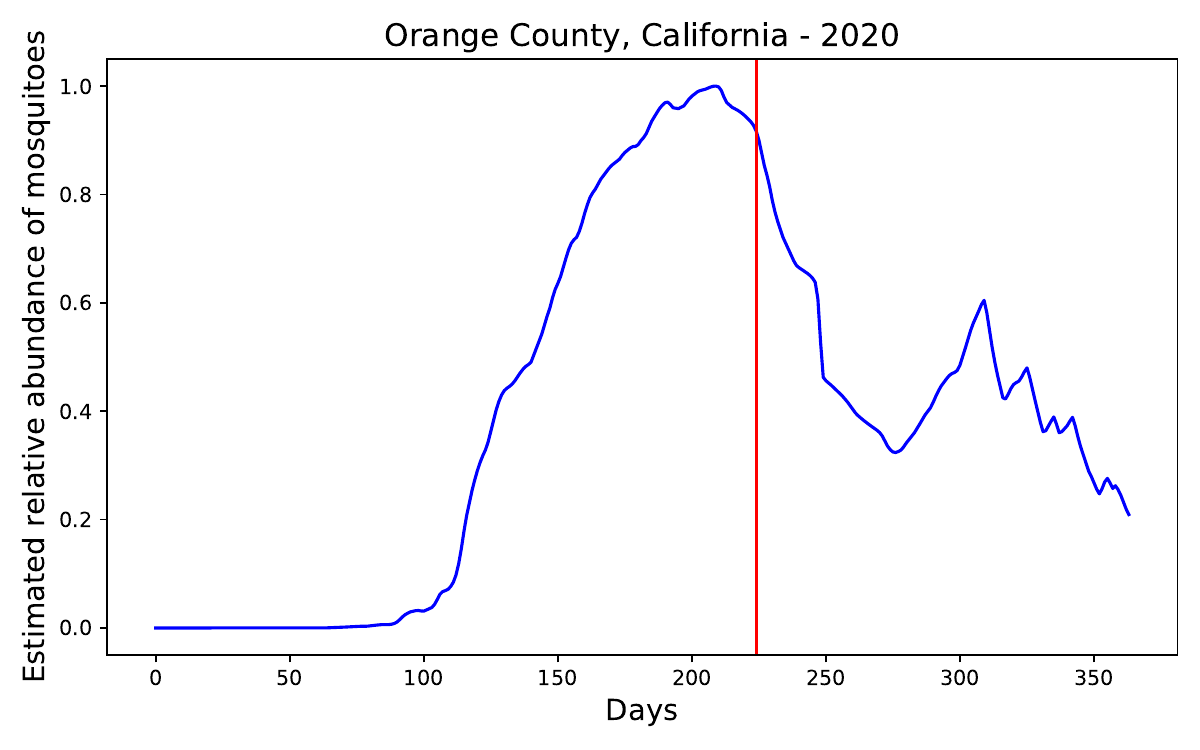}
        \caption{}
        \label{fig:june}
    \end{subfigure}
    \vspace{1em}
    \caption{Estimated profile of mosquito and times of onset of the spillover to human population.}
    \label{fig: Mosquito profiles-orange}
\end{figure*}
In the next step, a joint PDF is fitted to the coordinate sample collected from the time of the spillover onset and their weights which are the number of cases for each coordinate. Practically we have used the Kernel Density Estimation (KDE) method to fit this bivariate PDF. So, this PDF measures the density of the first time spillover to the human population based on the values of two parameters of the mosquito profile and $R0$. This PDF is called the spillover-onset PDF.\\

It should be noted that this is a predictive Bayesian PDF because it updates its densities based on new data. The PDF depends on the historical data, climate parameters and the number of new cases from past years. Therefore, if we update the dataset $Data{1}$ - which contains historical data for a certain period - by adding more years of information (or even more accurate data for the same period) to create $Data{2}$, the resulting PDFs are also updated. This process captures the details of recent years, reflecting the Bayesian approach of continuously refining predictions as new evidence becomes available. Statistically:\\
\begin{equation*}
    Dat{{a}_{1}}\to Dat{{a}_{2}}\Rightarrow {{f}_{(M,{{R}_{0}}\left| Dat{{a}_{1}} \right.)}}\to {{f}_{(M,{{R}_{0}}\left| Dat{{a}_{2}} \right.)}}
\end{equation*}

Figure (\ref{fig:Joint Copula pdf Orange County}) shows the joint PDF of the spillover-onset in Orange County. To create this PDF, historical data from 2006 to 2020 have been used. 
It is important to note that the joint PDF for each county (or each geographical location) has different distributions. After establishing the PDF of the spillover onset (for a given location), it will be used to predict the probability of spillover over time.

\subsection{Spillover Onset Risk Estimation }\label{Assessment of the Risk of Infection}
\begin{figure}
    \centering
    \includegraphics[width=0.45\linewidth]{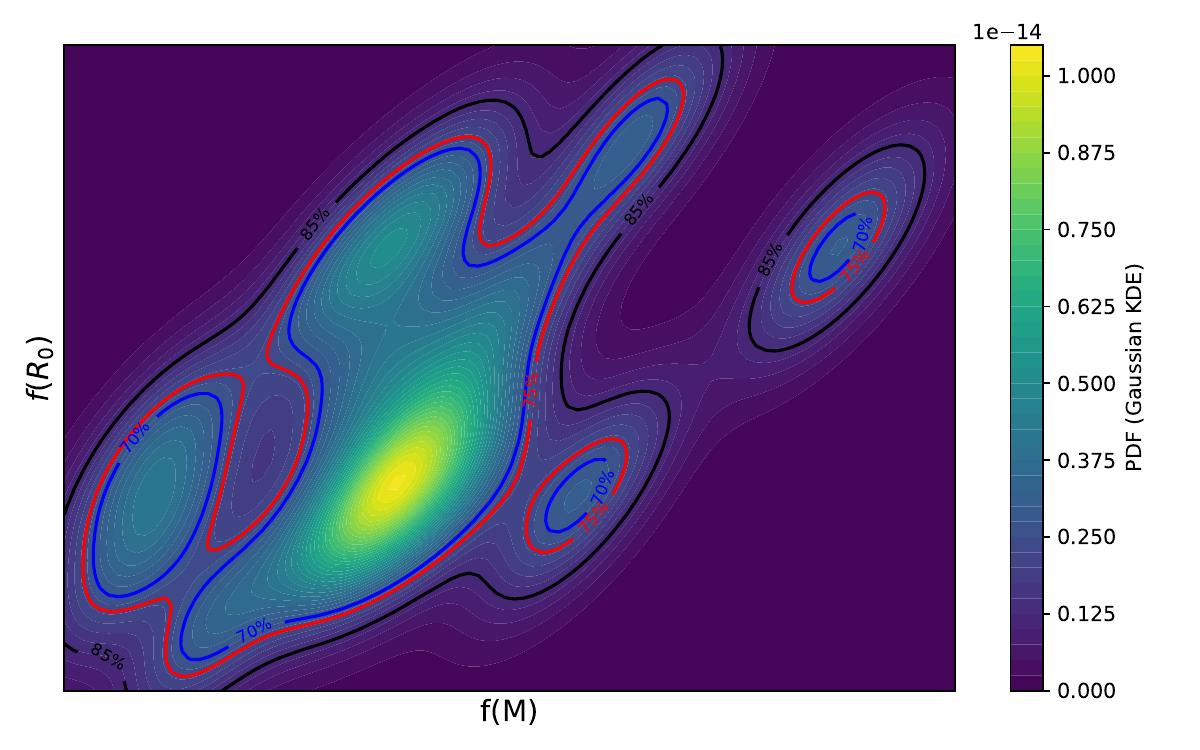}
    \caption{Joint PDF of the onset of spillover as a function of the Effective Number of Mosquitoes and $R_{0}$ for Orange County, California}
    \label{fig:Joint Copula pdf Orange County}
\end{figure}

\begin{figure*}[!ht]
    \centering
    \begin{subfigure}[t]{0.45\textwidth}
        \centering
        \includegraphics[width=\textwidth]{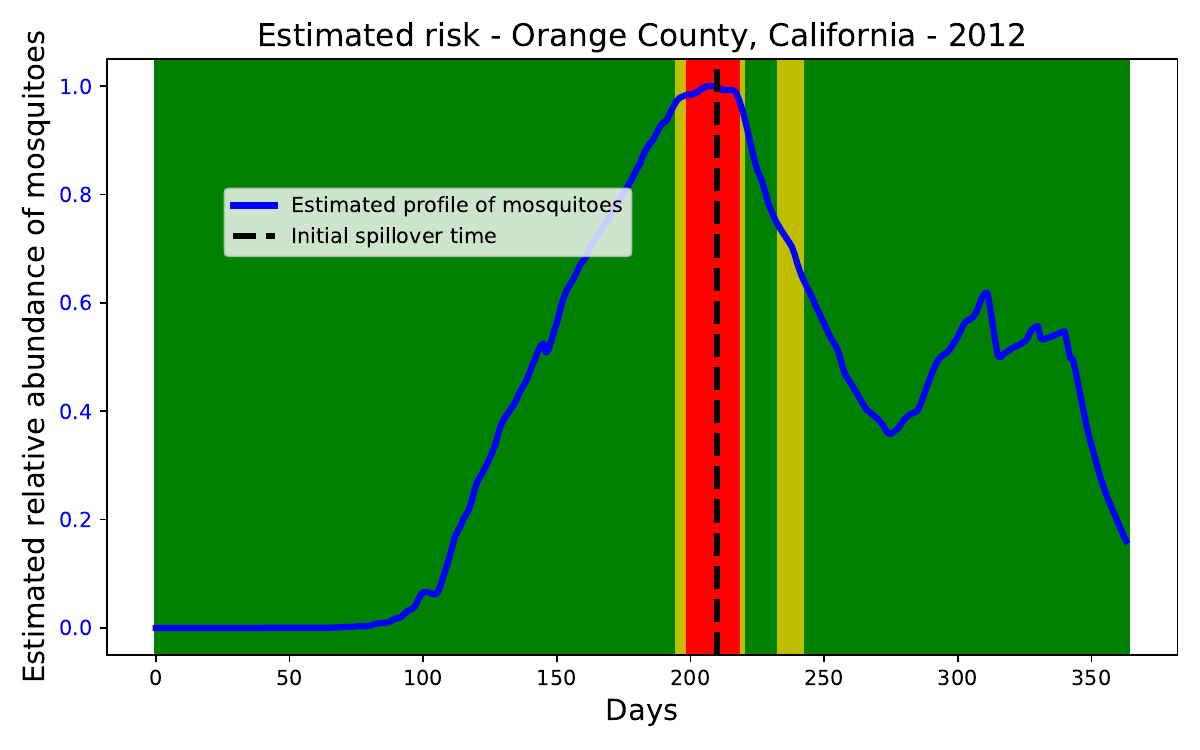}
        \caption{2012}
    \end{subfigure}
        \hfill
    \begin{subfigure}[t]{0.45\textwidth}
        \centering
        \includegraphics[width=\textwidth]{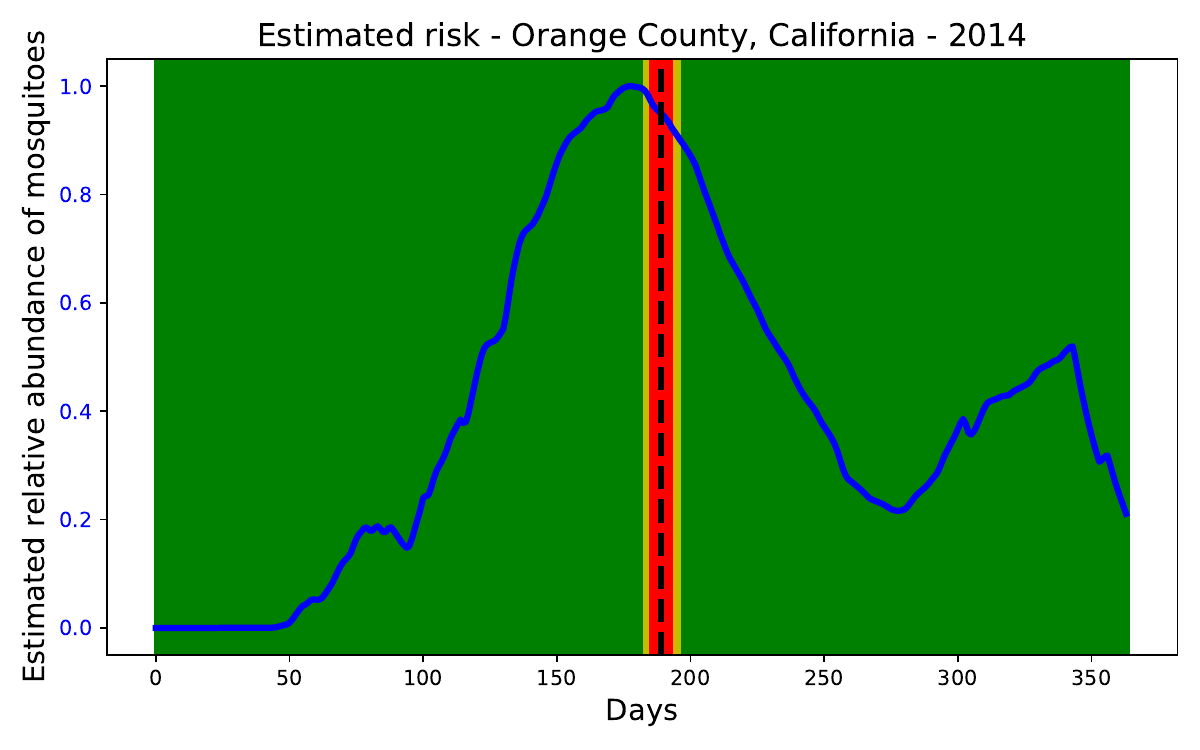}
        \caption{2014}
    \end{subfigure}
        \hfill
    \begin{subfigure}[t]{0.45\textwidth}
        \centering
        \includegraphics[width=\textwidth]{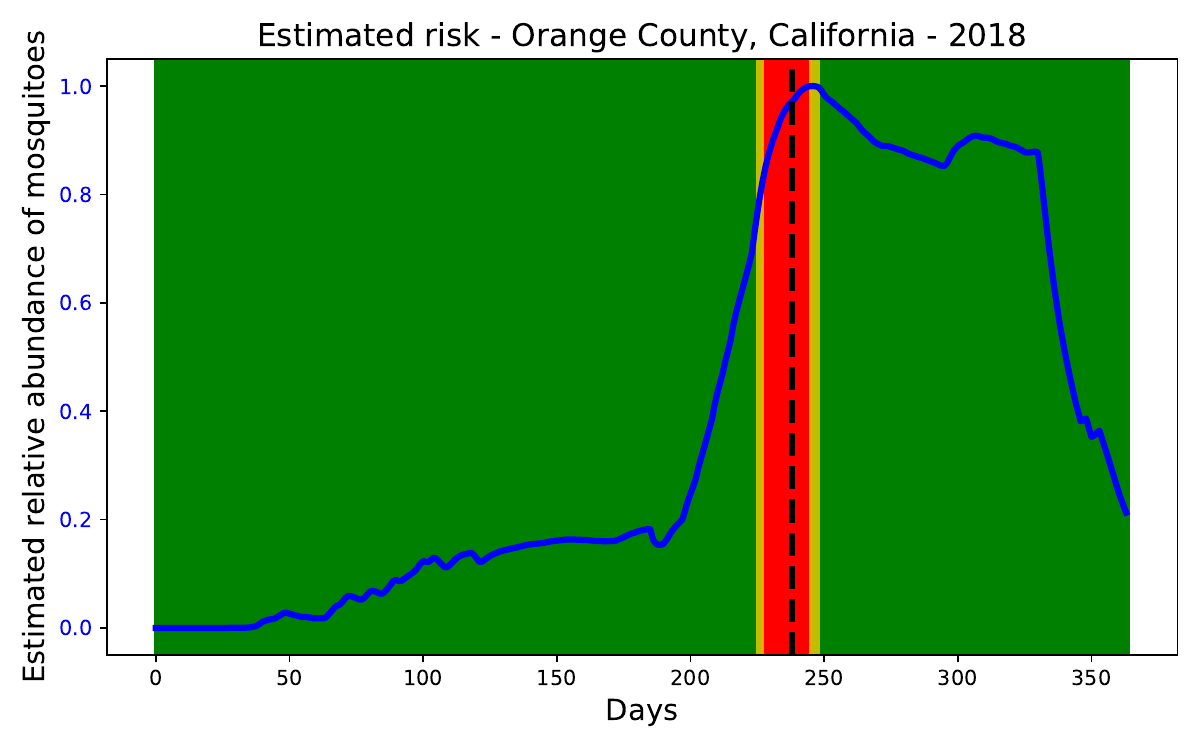}
        \caption{2018}
    \end{subfigure}
        \hfill
    \begin{subfigure}[t]{0.45\textwidth}
        \centering
        \includegraphics[width=\textwidth]{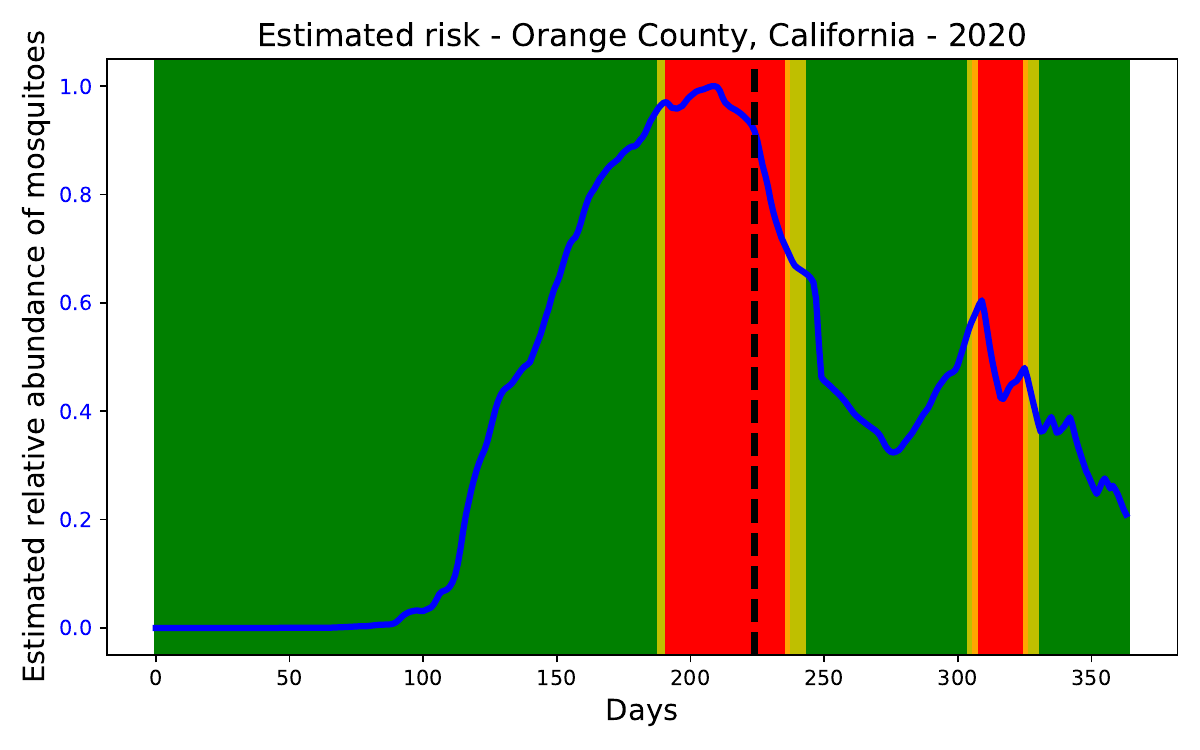}
        \caption{2020}
    \end{subfigure}

    \vspace{1em}
    \caption{Estimated Risk of the spillover onset to human population, Orange County, California, 2012, 2014, 2016, and 2020.}
    \label{fig: Risk-Orange county}
\end{figure*}

After finding the appropriate joint PDF for the risk of spillover onset, we can now define the risk range based on needs and strategies. We have defined three contours of the joint PDF at 0.8, 0.9, and 0.95. The contour of 0.88 (it means 88 percent of the density is within this contour) corresponds to the values with the highest frequency, indicating that spillover for the first time is more likely to occur within this contour (Figure \ref{fig:Joint Copula pdf Orange County}). Values between the contours of 0.88 and 0.9, as well as between the contours of 0.9 and 0.95, have lower frequencies. Consequently, the expectation of the spillover onset in these regions should be lower. If the coordinates of the effective $R_{0}$ and mosquito profile fall within the 88 percent contour (which has the highest frequency), the situation is classified as high risk (red). It is considered risky if it falls between the 80 and 90 percent contours (orange). It is low risk if it falls between the 90 and 95 percent contours (yellow). Finally, if it lies outside the 95 percent contour, it is considered a green situation.\\
Practically, after creating PDF for the onset of spillover, we can run the model using weather parameters to estimate the daily values of the mosquito profile and $R{0}$. By comparing these estimates with the PDF of the onset of the spillover and the defined contours (Figure \ref{fig:Joint Copula pdf Orange County}), we can determine the level of risk for each day. Figure (\ref{fig: Risk-Orange county}) illustrates the risk of spillover onset throughout the year in different years with the reported time of spillover onset.

\subsection{Spillover Onset Risk Prediction} \label{sunsection:Spillover Onset Risk Prediction}
\begin{figure}[!ht]
    \centering
    \begin{subfigure}[t]{0.45\textwidth}  
        \centering
        \includegraphics[width=\textwidth]{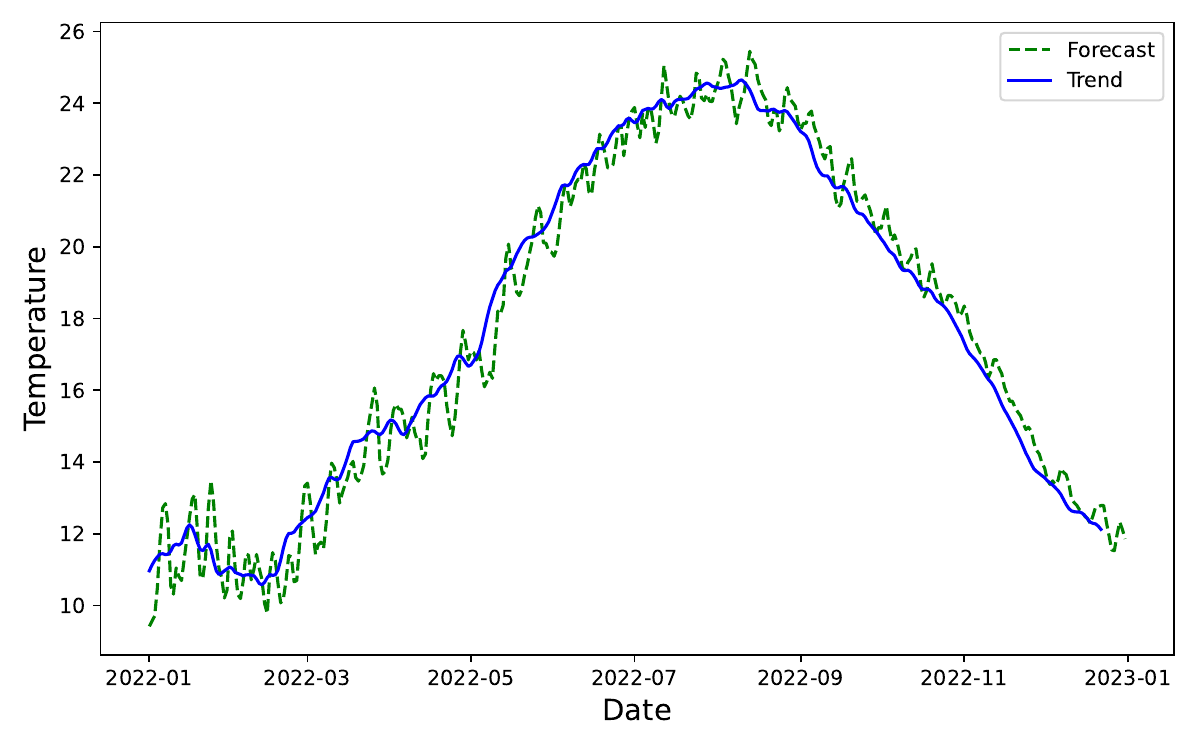}
        \caption{Temperature forecast}
        \label{fig:Tem}
    \end{subfigure}
    \hfill
    \begin{subfigure}[t]{0.45\textwidth}  
        \centering
        \includegraphics[width=\textwidth]{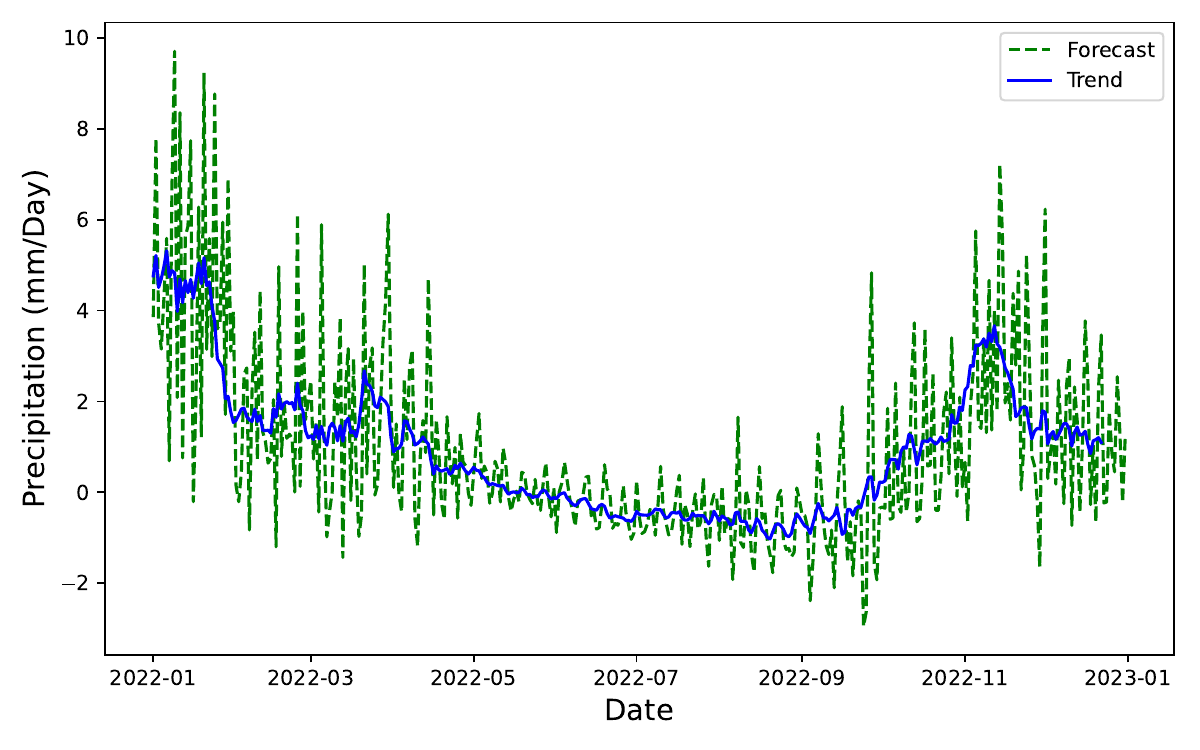}
        \caption{Precipitation forecast}
        \label{fig:Hum}
    \end{subfigure}
    \caption{Long term forecast of temperature and precipitation for Orange County, California, in 2022.}
    \label{fig:climate_forecast}
\end{figure}

\begin{figure*}
    \centering
    \includegraphics[width=1.1\linewidth]{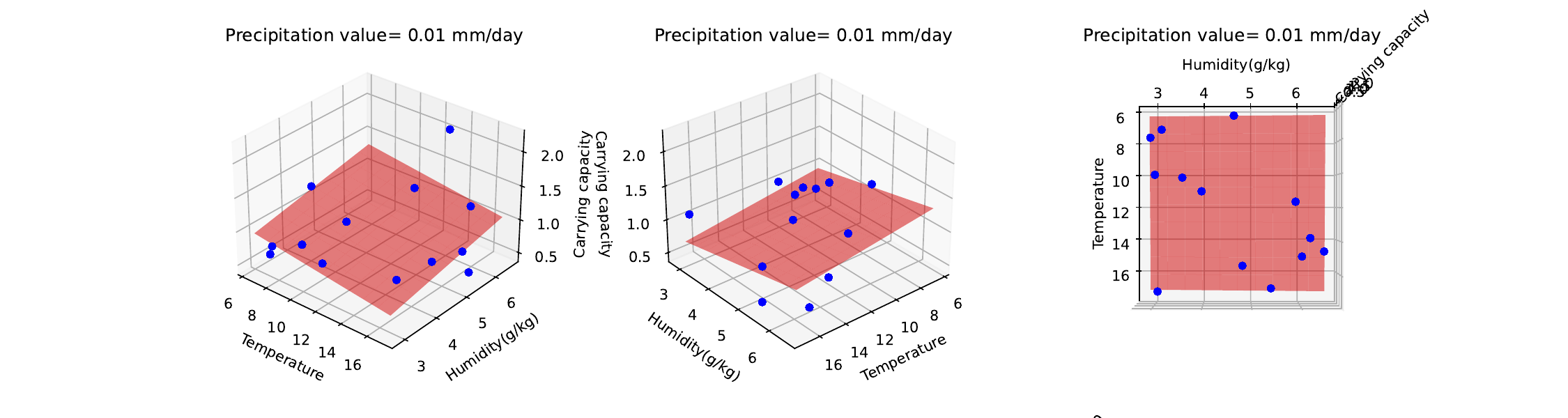}
    \caption{Predicted carrying capacity when precipitation value equals 0.01 mm/day.}
    \label{fig: predictionof carryin-orange}
\end{figure*}

The general approach to probabilistic risk assessment for the onset of spillover was discussed earlier in Subsection \ref{Assessment of the Risk of Infection}. However, in future scenarios where no weather data is available, alternative methods must be considered. This is important because related organizations require risk predictions in advance to define effective strategies to address potential epidemics. In the following section, we will explain how to make risk predictions for the onset of spillover for both long- and short-term lead times.\\ 
To effectively predict risk, it is essential to forecast both weather parameters and carrying capacity.
With regard to weather parameters, we have utilized time series methods for forecasting. Since time series techniques for weather prediction are well-established and extensively researched, they are not the focus of this work. Instead, we have applied well-known models for weather forecasting. However, it is important to design weather parameter forecasting in a way that creates an ideal scenario by offering long-term predictions, with lead times of one year (Figure \ref{fig:climate_forecast}), seasons or similar intervals, and short-term predictions, with lead times of one to two weeks, to continuously update conditions and reassess the risk. \\
To obtain a predictive function for the carrying capacity, we can take three approaches.
The first approach involves using the average estimated carrying capacity of the past years (2006-2021), which is useful when thinking about the long-term prediction of the carrying capacity.
Second, we used the time series technique for short-term prediction of the calculating capacity.
In the third approach, we leverage the full data set of estimated carrying capacities from 2006 to 2021. Each day is associated with the corresponding values for transport capacity, temperature, humidity, and precipitation. For each precipitation value, we collected the corresponding temperature, humidity, and carrying capacity samples. Subsequently, we fitted a plane to these three variables and the resulting equation is used to predict new carrying capacity values based on the forecasted precipitation, temperature, and humidity. Figure (\ref{fig: predictionof carryin-orange}) shows the plane fitted for a given value of precipitation. After predicting the carrying capacity, we can start predicting the probabilistic risk of spillover onset.
\clearpage

\subsubsection{Long-term Onset Risk Prediction}\label{subsection: long ter, for onset}
\begin{figure}[!ht]
    \centering
    \begin{subfigure}[t]{0.45\textwidth}  
        \centering
        \includegraphics[width=\textwidth]{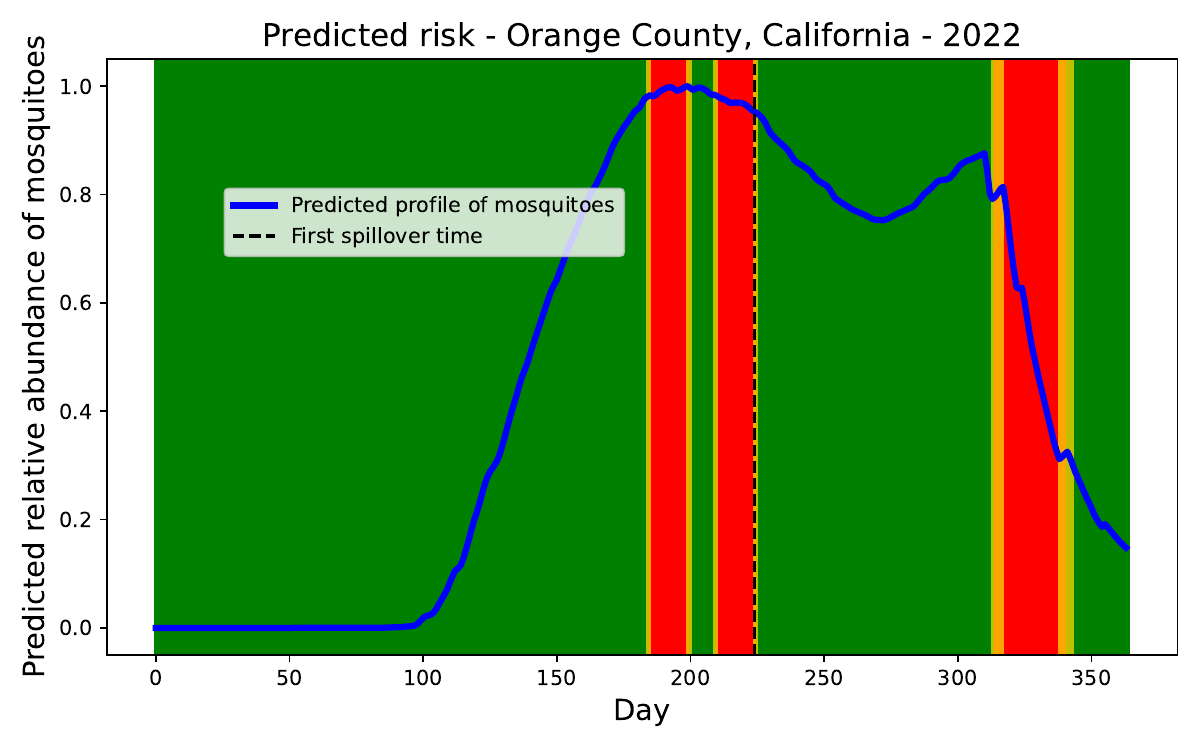}
        \caption{Long-term prediction for 2022}
        \label{fig:pred_2022}
    \end{subfigure}
    \hfill
    \begin{subfigure}[t]{0.45\textwidth}  
        \centering
        \includegraphics[width=\textwidth]{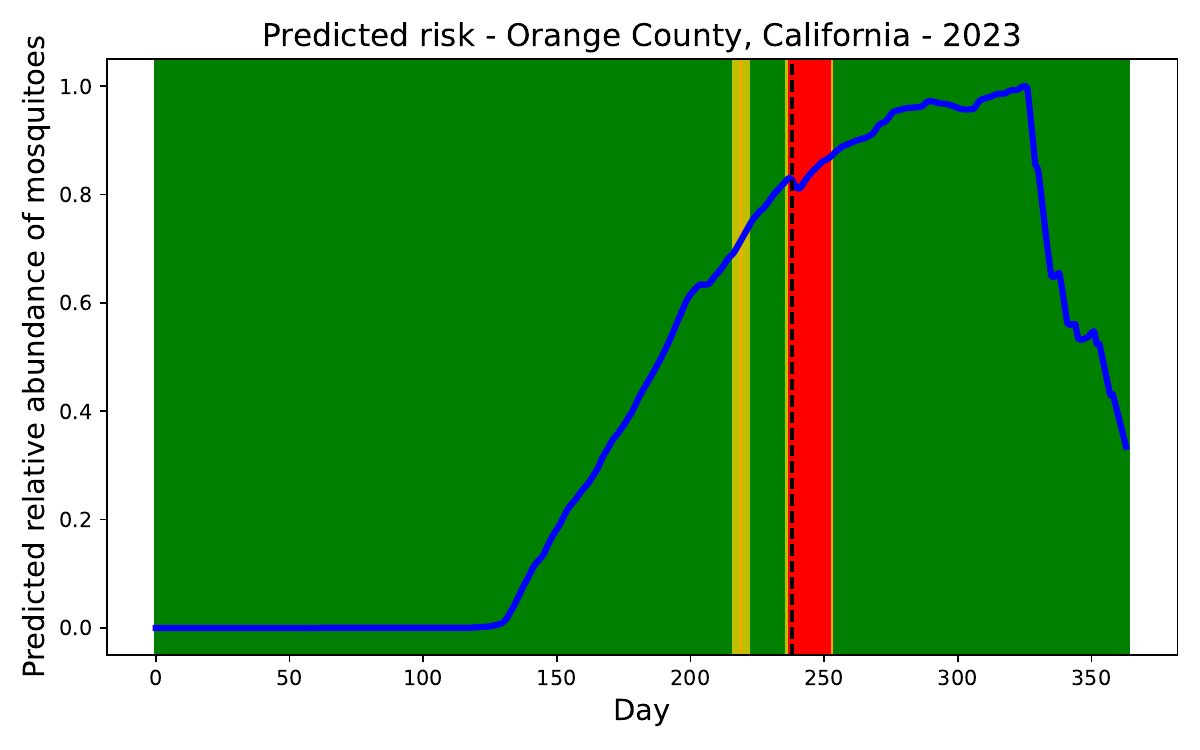}
        \caption{Long-term prediction for 2023}
        \label{fig:pred_2023}
    \end{subfigure}
    \hfill
    \begin{subfigure}[t]{0.45\textwidth}  
        \centering
        \includegraphics[width=\textwidth]{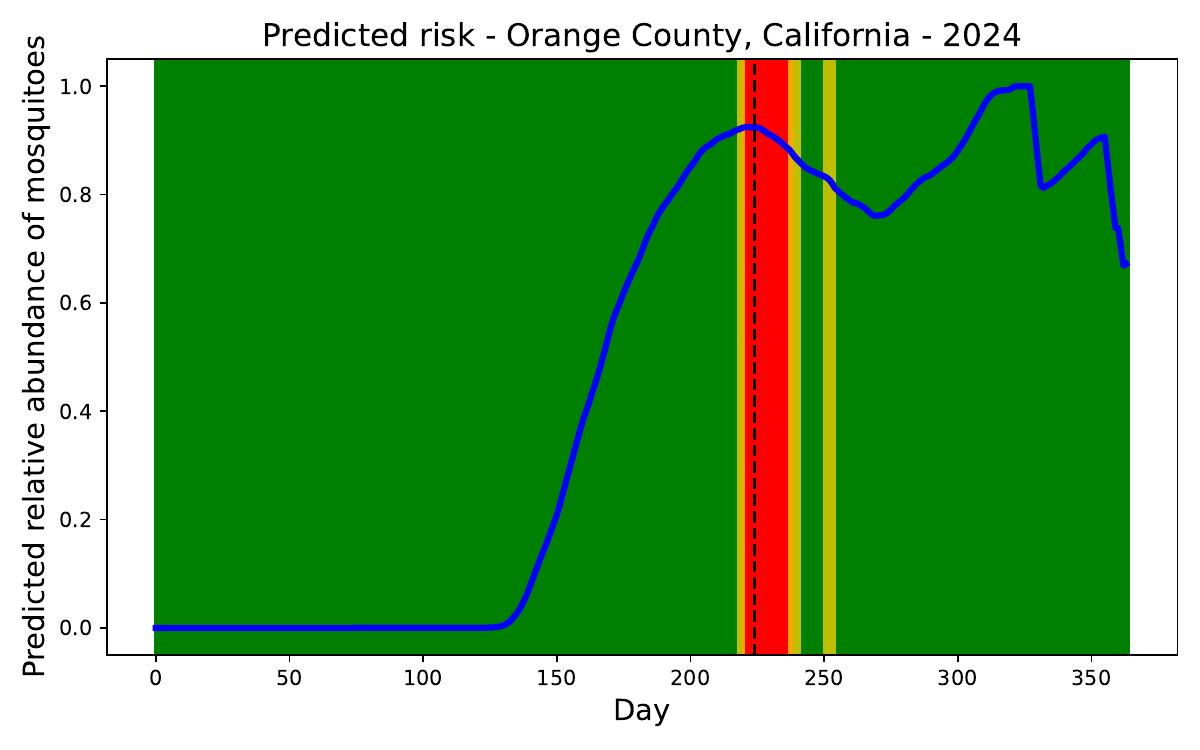}
        \caption{Long-term prediction for 2024}
        \label{fig:pred_2024}
    \end{subfigure}
    \caption{Long-term prediction of the risk of first-time spillover to the human population in Orange County, California for 2022, 2023, and 2024.}
    \label{fig:long_term_predictions}
\end{figure}

It is highly beneficial to understand the general trajectory of an epidemic at the beginning of the year (or a season such as summer), as this provides crucial information on the necessary actions and timings to mitigate the spread of the disease throughout the year. Although accurate predictions are impossible with such an extended lead time (365 days), this foresight can still be extremely valuable in shaping general policies for anticipated scenarios. \cite{hales1999nino} is a very good example of the effect of this cyclic pattern on vector-borne diseases.\\
One of the most significant benefits of long-term prediction is that analyzing historical weather data from recent years can reveal emerging trends in global warming. This improves our understanding of how climate change influences disease patterns and can alter the timing of initial spillover events.
Furthermore, long-term prediction can capture various cyclic weather patterns, such as the El Nio-Southern Oscillation (ENSO), the Quasi-Biennial Oscillation (QBO), and the Solar Cycle (Sunspot Cycle). These patterns have a substantial impact on weather parameters, which in turn affect the dynamics of vector-borne diseases such as WND.\\
To achieve an accurate prediction of the risk of onset, as mentioned earlier, time series techniques can be used to forecast weather parameters for an entire year at a specific location. In this study, an auto-regressive model with a degree of 365 has been used to forecast the annual trends (there are many methods to obtain the trend) of temperature, humidity, and precipitation for a given location. This model primarily provides trends and a little more than the trend about these variables, which are sufficient for our purposes. For example, Figure (\ref{fig:climate_forecast}) displays the forecast of the climate parameters (temperature, humidity, and precipitation) for the year 2021, using data from 2006 to 2020 with an auto-regressive model.
Using this predicted information, we have run the differential equation model, and using the model's results on the forecast parameters, we can predict mosquito profile and the values of $R{0}$. These predictions can then be compared with the PDF of the onset of the spillover to predict the level of risk.\\
Figure (\ref{fig:long_term_predictions}) shows the prediction results with a one-year lead time for 2022, 2023, and 2024. The carrying capacity has been predicted for these years using the third approach mentioned in (\ref{sunsection:Spillover Onset Risk Prediction}). The vertical line indicates the first spillover time, based on reported data from the California Department of Public Health.
It is evident that in all three years, the spillover onset time coincided with the high-risk days predicted by the model. In other words, the first cases occurred on the days identified as having a high likelihood of initial spillover. This is particularly remarkable, as the prediction was made at the beginning of the year. Also, it should be mentioned that the number of high risk days of 2022, 2023 and 2024 is 23, 32 and 23 days, respectively.\\

\subsubsection{Short-term Onset Risk Prediction}
\begin{figure*}[!ht]
    \centering
    \begin{subfigure}[t]{0.32\textwidth}
        \centering
        \includegraphics[width=\textwidth]{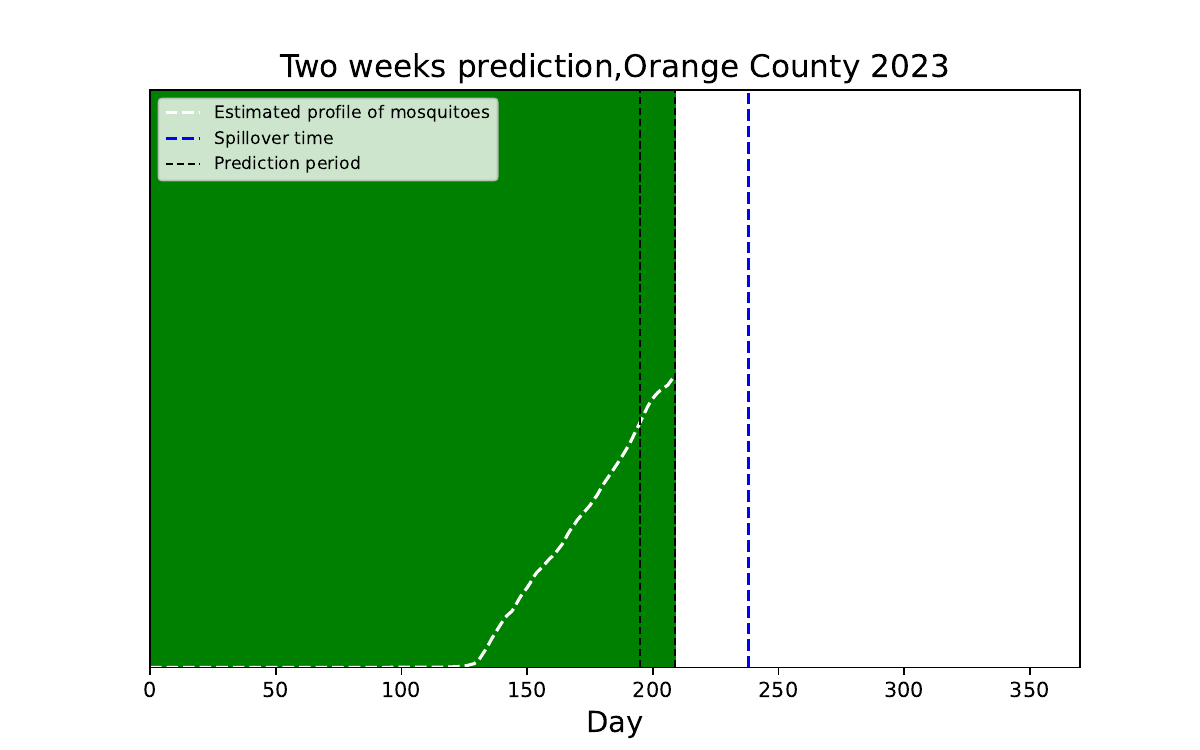}
        \caption{}
        \label{fig:january}
    \end{subfigure}
    \hfill
    \begin{subfigure}[t]{0.32\textwidth}
        \centering
        \includegraphics[width=\textwidth]{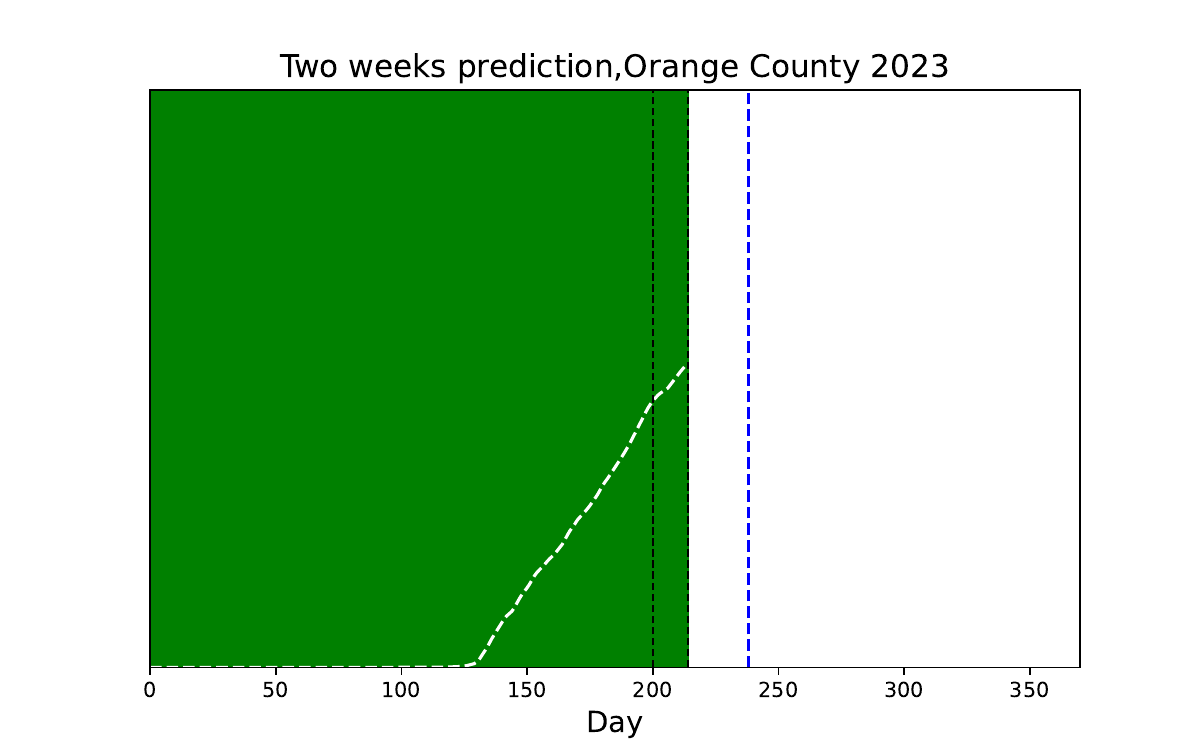}
        \caption{}
        \label{fig:february}
    \end{subfigure}
    \hfill
    \begin{subfigure}[t]{0.32\textwidth}
        \centering
        \includegraphics[width=\textwidth]{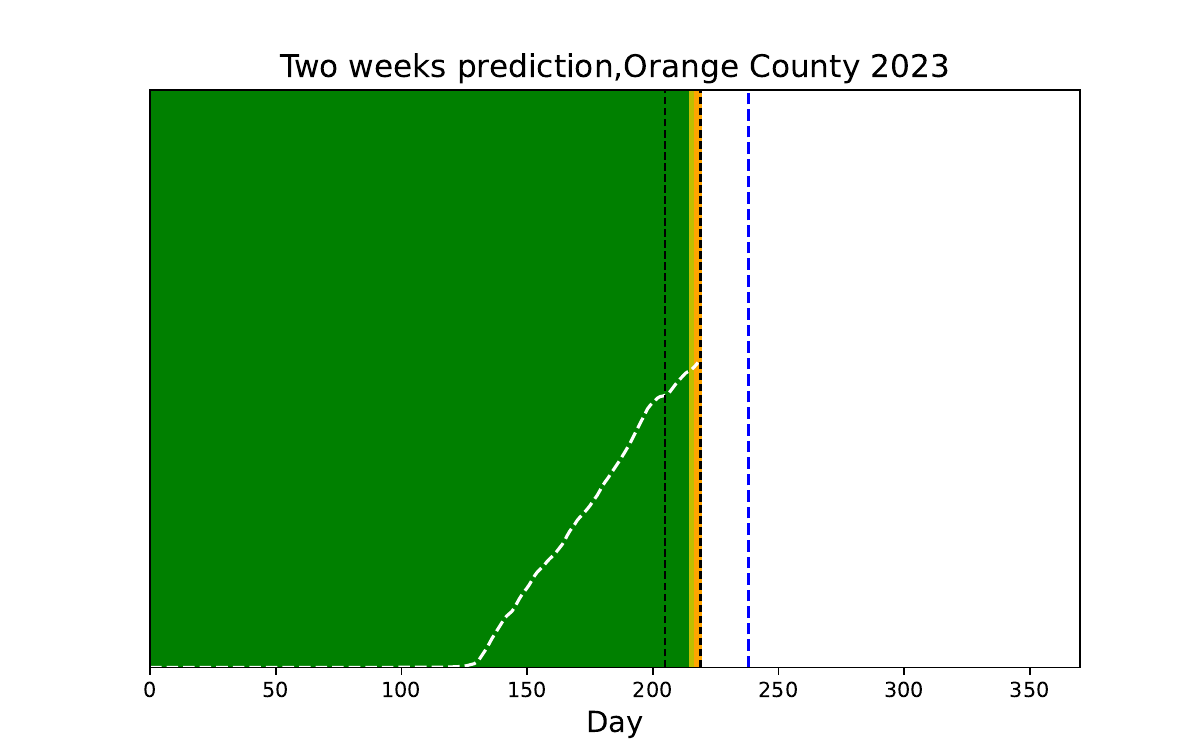}
        \caption{}
        \label{fig:april}
    \end{subfigure}
    \hfill
    \begin{subfigure}[t]{0.32\textwidth}
        \centering
        \includegraphics[width=\textwidth]{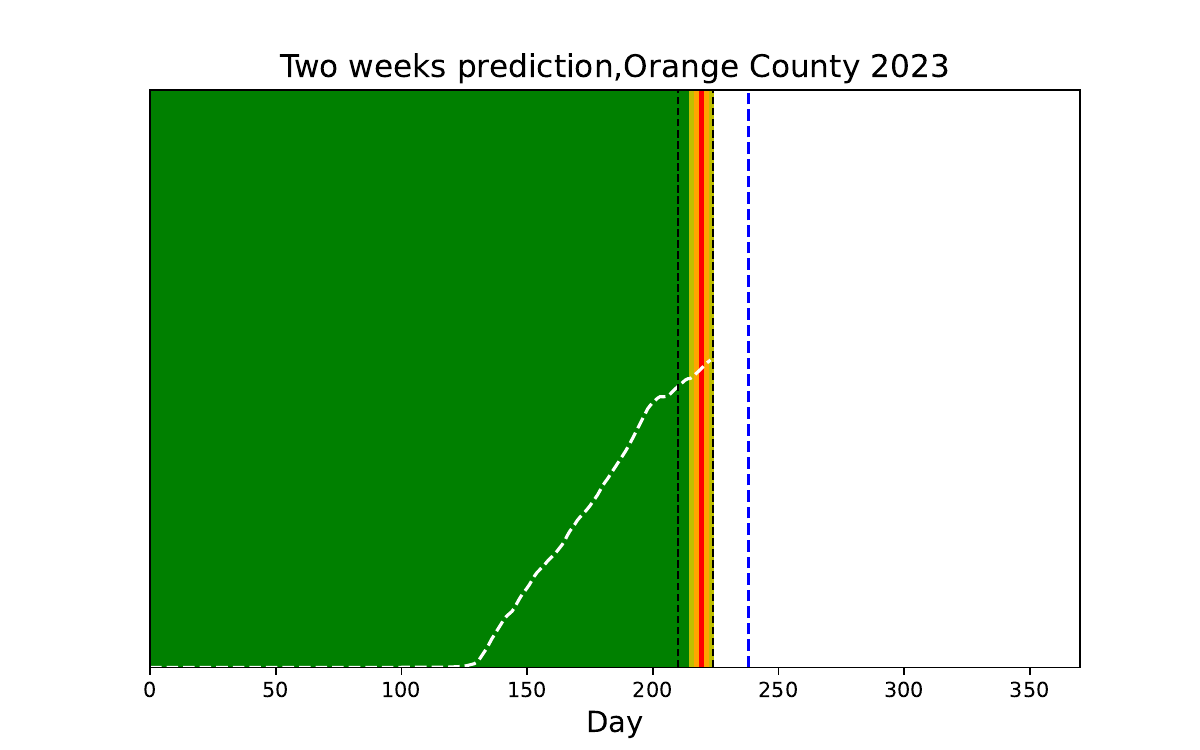}
        \caption{}
        \label{fig:may}
    \end{subfigure}
    \hfill
    \begin{subfigure}[t]{0.32\textwidth}
        \centering
        \includegraphics[width=\textwidth]{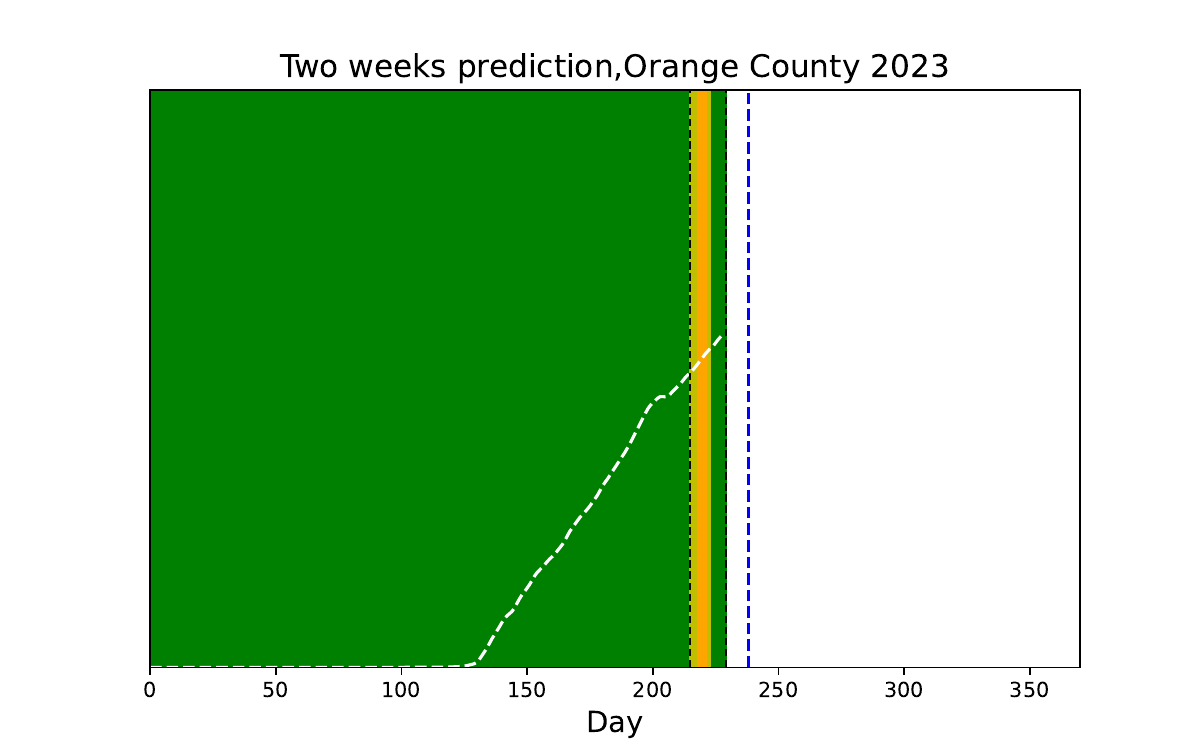}
        \caption{}
        \label{fig:june}
    \end{subfigure}
        \hfill
    \begin{subfigure}[t]{0.32\textwidth}
        \centering
        \includegraphics[width=\textwidth]{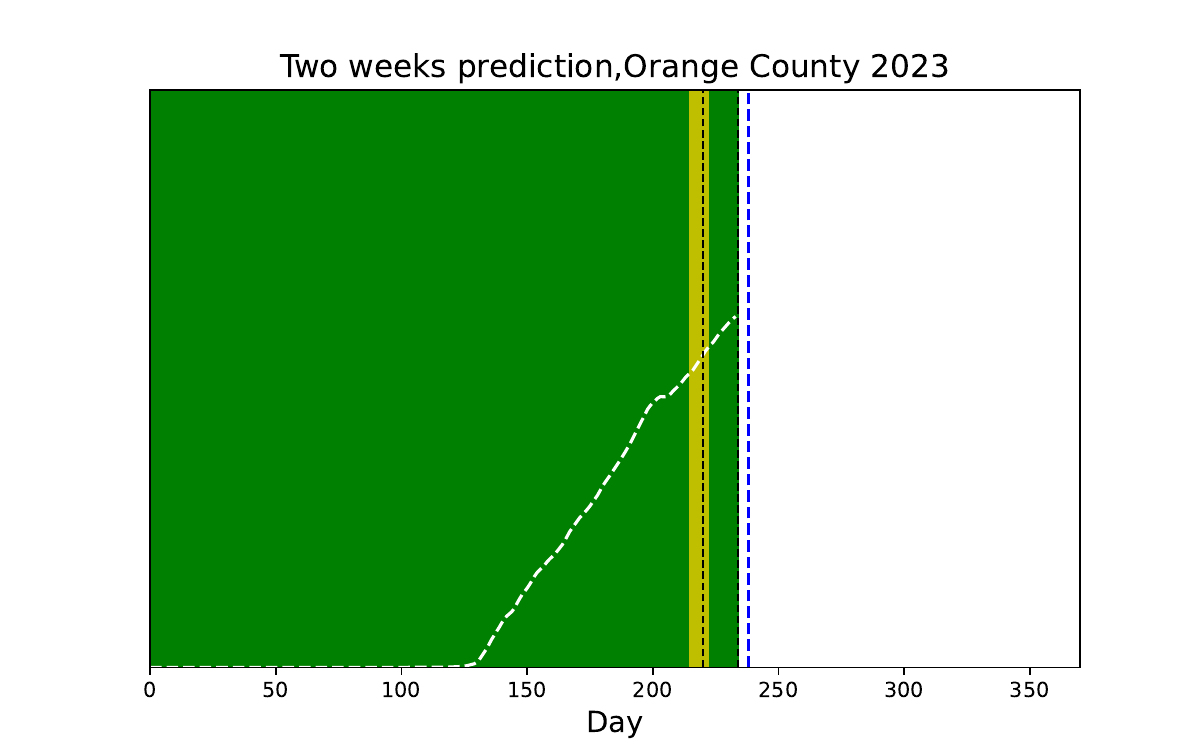}
        \caption{}
        \label{fig:june}
    \end{subfigure}
        \hfill
    \begin{subfigure}[t]{0.32\textwidth}
        \centering
        \includegraphics[width=\textwidth]{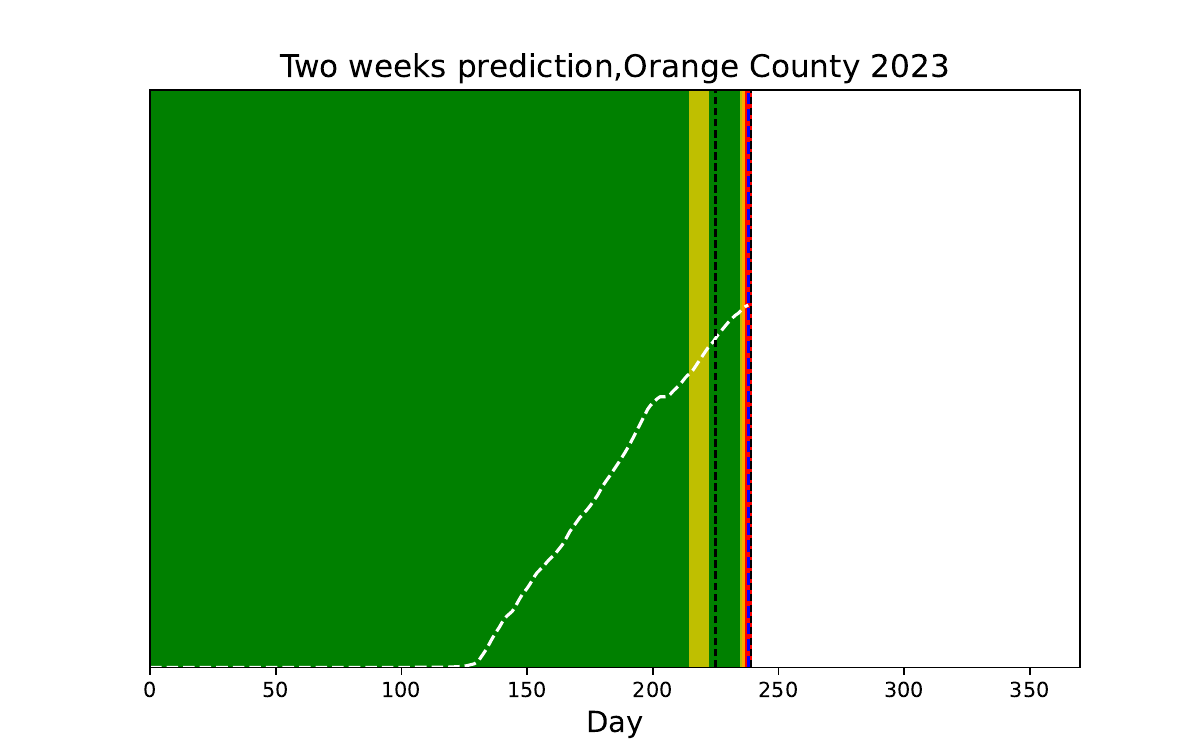}
        \caption{}
        \label{fig:june}
    \end{subfigure}
        \hfill
    \begin{subfigure}[t]{0.32\textwidth}
        \centering
        \includegraphics[width=\textwidth]{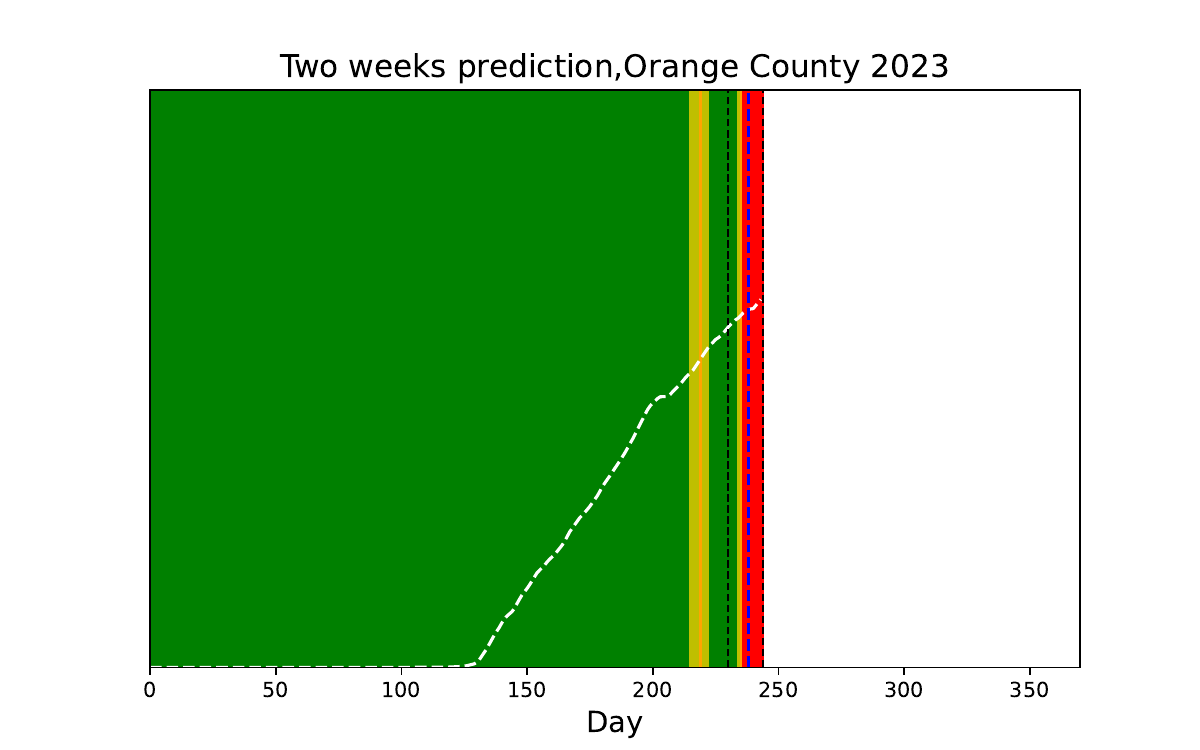}
        \caption{}
        \label{fig:june}
    \end{subfigure}
        \hfill
    \begin{subfigure}[t]{0.32\textwidth}
        \centering
        \includegraphics[width=\textwidth]{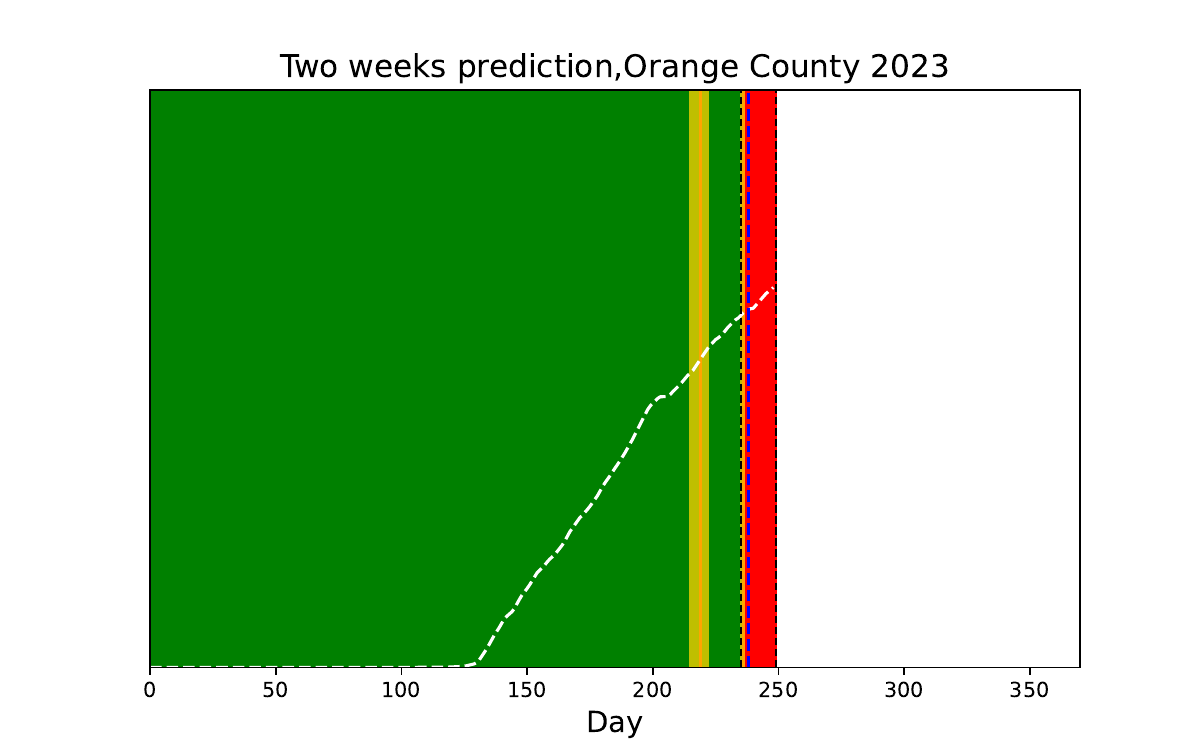}
        \caption{}
        \label{fig:june}
    \end{subfigure}
        \hfill
    \begin{subfigure}[t]{0.32\textwidth}
        \centering
        \includegraphics[width=\textwidth]{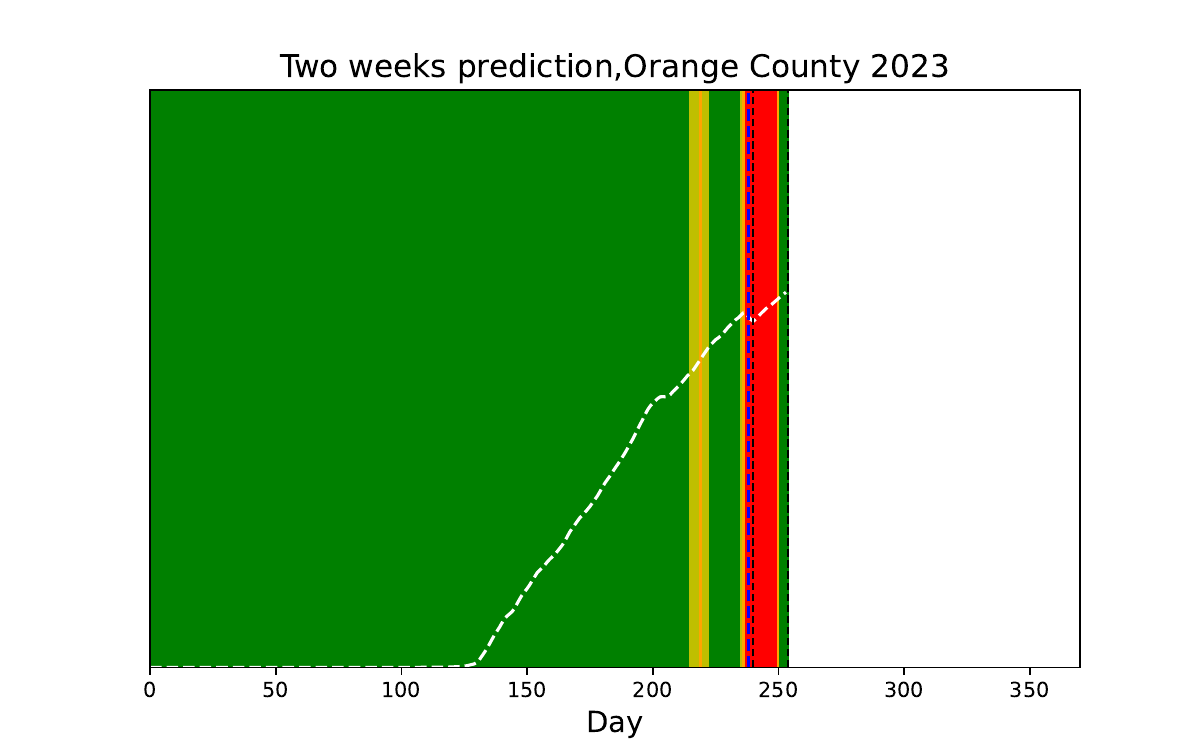}
        \caption{}
        \label{fig:june}
    \end{subfigure}
        \hfill
    \begin{subfigure}[t]{0.32\textwidth}
        \centering
        \includegraphics[width=\textwidth]{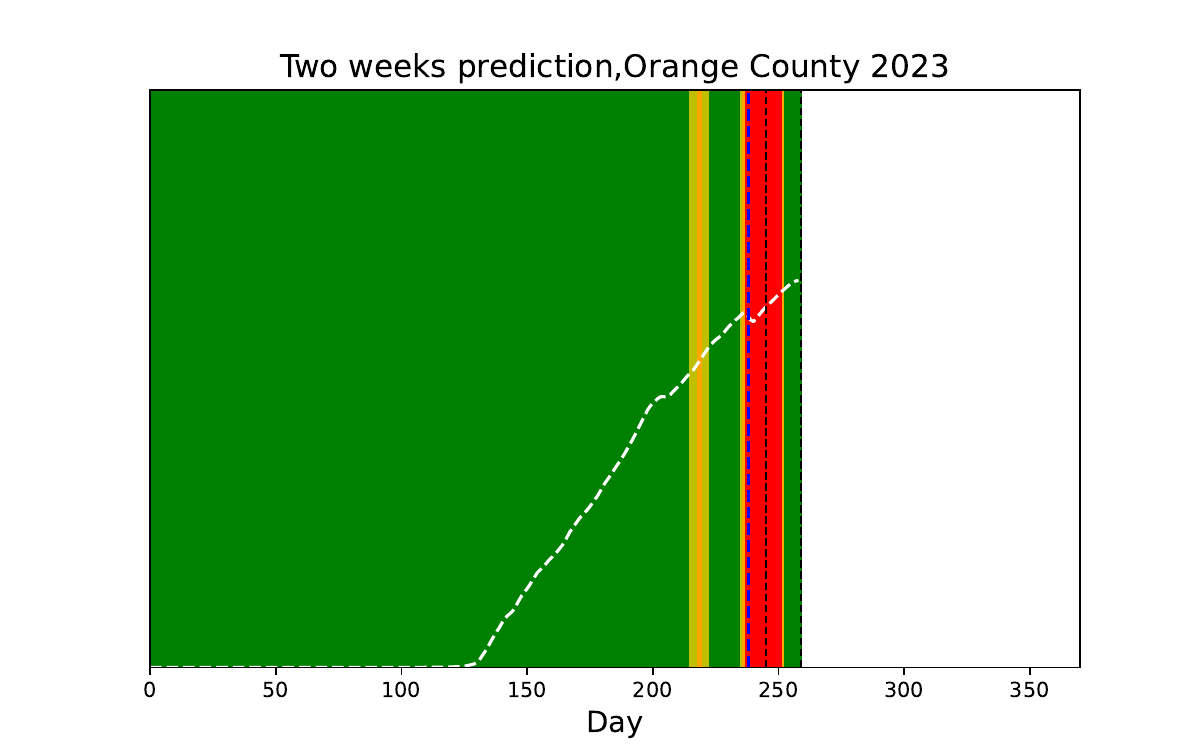}
        \caption{}
        \label{fig:june}
    \end{subfigure}
        \hfill
    \begin{subfigure}[t]{0.32\textwidth}
        \centering
        \includegraphics[width=\textwidth]{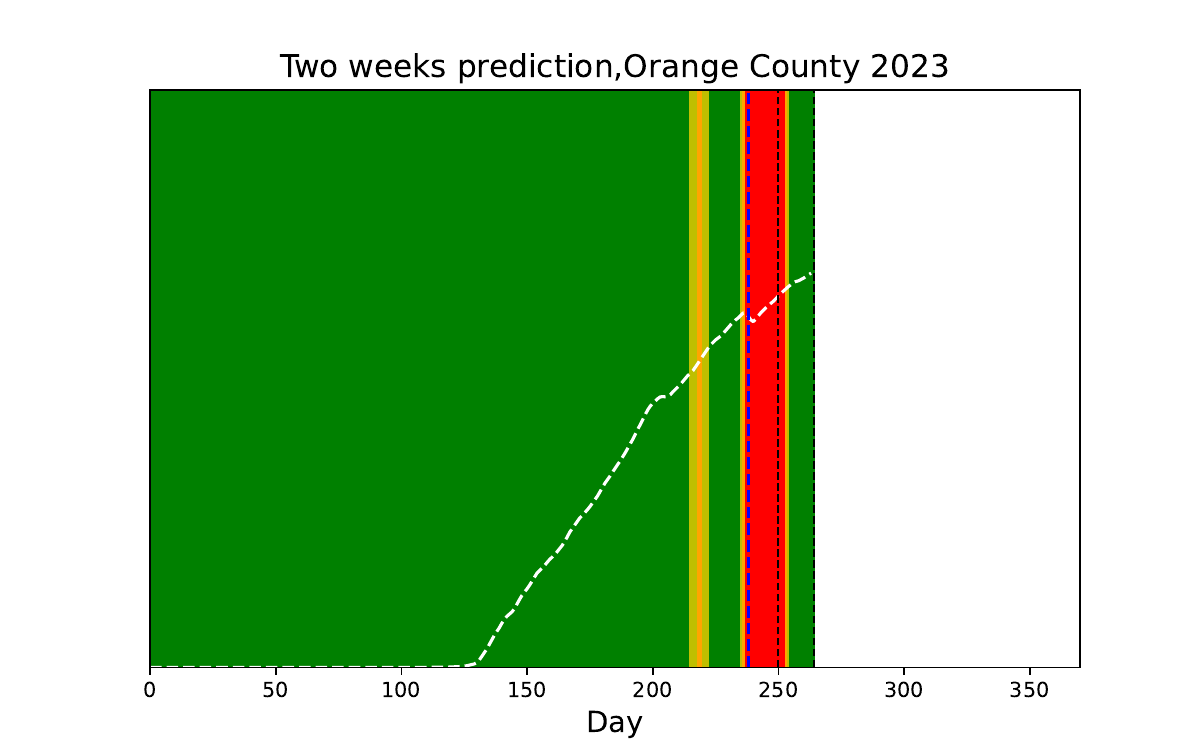}
        \caption{}
        \label{fig:june}
    \end{subfigure}
    \vspace{1em}
    \caption{Short-term risk assessment for the onset of spillover to human population, Orange County, California.}
    \label{fig: short term Risk-Orange county}
\end{figure*}

During certain periods, such as the onset of the hot season, increased vigilance is required regarding potential disease outbreaks and preparations must be made to address them. In these instances, our strategy involves employing a more precise model that uses current-year climate data instead of historical climate data, allowing for shorter prediction leads and more immediate and accurate forecasts. This approach is used to predict the weather parameters over a brief period of 10 to 14 days. The resulting data are then used to run the model, helping to evaluate the short-term risk of initial spillover to humans.\\
Figure \ref{fig: short term Risk-Orange county} illustrates the step-by-step process of predicting short-term risk. The initial spillover time is indicated by vertical blue dashed lines, while the period for the two-week prediction is shown by the area between two vertical black dashed lines. Up to the date of the first black dashed line, we used actual values of temperature, humidity, and precipitation. For the next two weeks, we forecasted these parameters, ran the model based on these forecasts, and obtained the mosquito profile values (M) and $R{0}$, and for each forecast day we compared the coordinated with the PDF of the spillover onset and find the level of the risk as we discussed before.\\
The animation file illustrating this process has been included as supplementary material that visually demonstrates the entire procedure ($Shortterm2023.mp4$.)

\begin{figure}[!ht]
    \centering
    \begin{subfigure}[t]{0.45\textwidth}
        \centering
        \includegraphics[width=\textwidth]{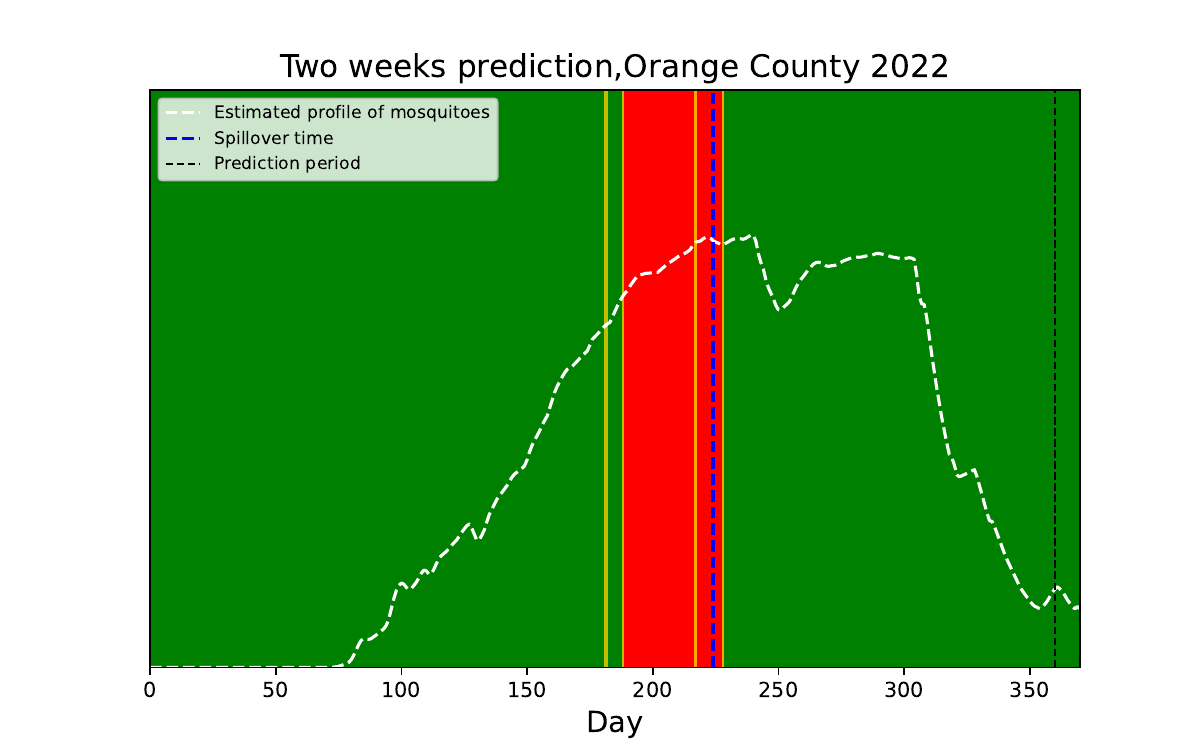}
        \caption{Short-term prediction for 2022}
        \label{fig:short-term-2022}
    \end{subfigure}
    \hfill
    \begin{subfigure}[t]{0.45\textwidth}
        \centering
        \includegraphics[width=\textwidth]{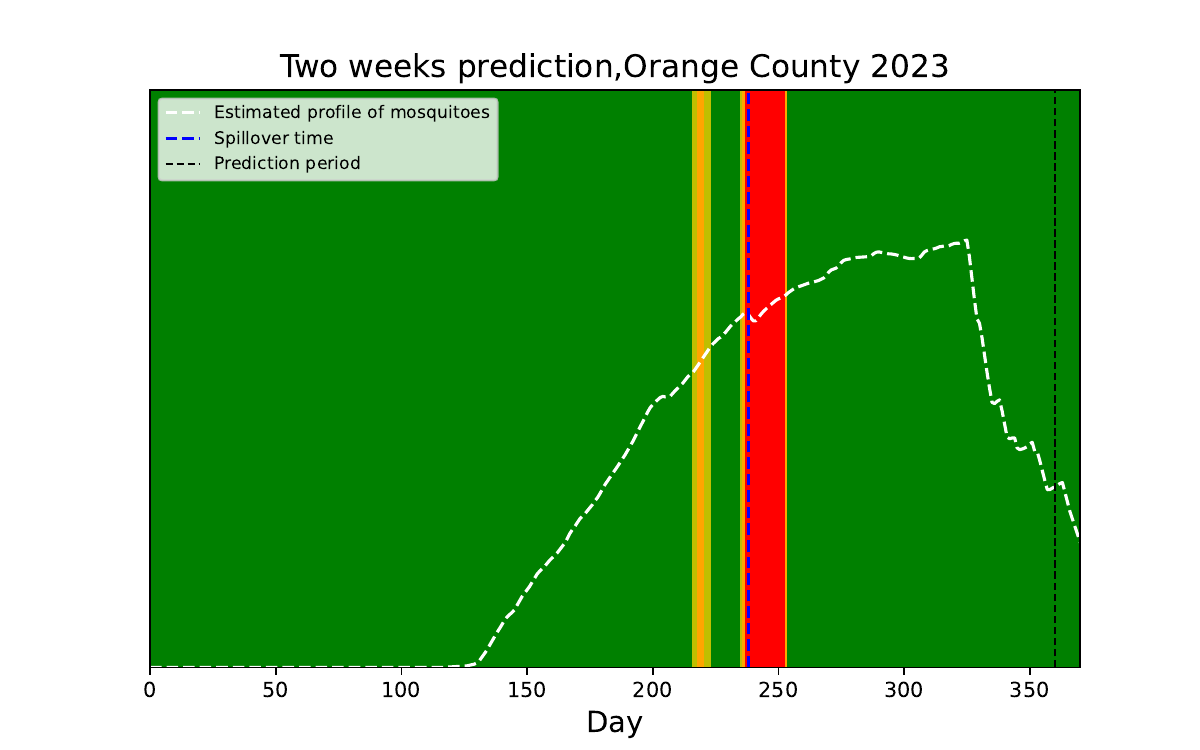}
        \caption{Short-term prediction for 2023}
        \label{fig:short-term-2023}
    \end{subfigure}
    \hfill
    \begin{subfigure}[t]{0.45\textwidth}
        \centering
        \includegraphics[width=\textwidth]{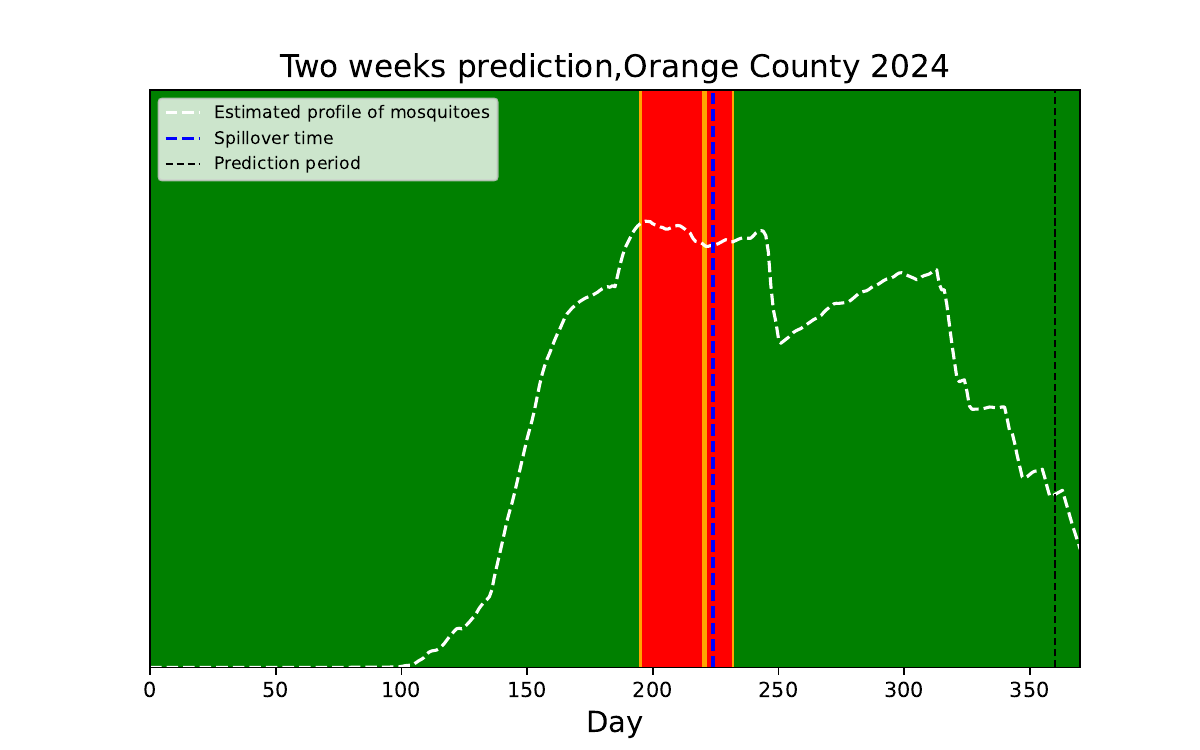}
        \caption{Short-term prediction for 2024}
        \label{fig:short-term-2024}
    \end{subfigure}
    \caption{Short-term prediction of the risk of the spillover onset to humans in Orange County, California, for years 2022–2024.}
    \label{fig:short-term-predictions}
\end{figure}

A good example demonstrating the advantage of short-term predictions is evident when comparing them with long-term predictions (\ref{fig:short-term-2022}). In particular, short-term predictions significantly reduce the uncertainty around the duration of high-risk periods. For example, the long-term prediction identified a total of 46 high-risk (and 65 risky) days, spread over three separate intervals, whereas the short-term prediction reduced this number to 38 days (and 65 risky) days in a continuous interval, thus providing a clearer and more actionable risk assessment.

\section{Bayesian Assessment of WND Spillover Severity Risk}
After assessing the risk of spillover onset, we now turn to assessing the possible severity of the epidemic following this timeline. Given the inherent uncertainty in situations such as the transmission of diseases such as WND, we again adopt a probabilistic framework for assessment. 
The first step is to establish a PDF to quantify the severity of the situation. To achieve this, we collected data on effective mosquito profiles (M) and effective values of weather parameters, such as temperature, humidity, and precipitation, at the time of each spillover event (reported cases after the onset of the spillover). We then used the number of reported cases as a weight (or frequency) for each (M, W) coordinate (actually, we have used a function of these parameters). 

\begin{figure}
    \centering
    \includegraphics[width=0.45\linewidth]{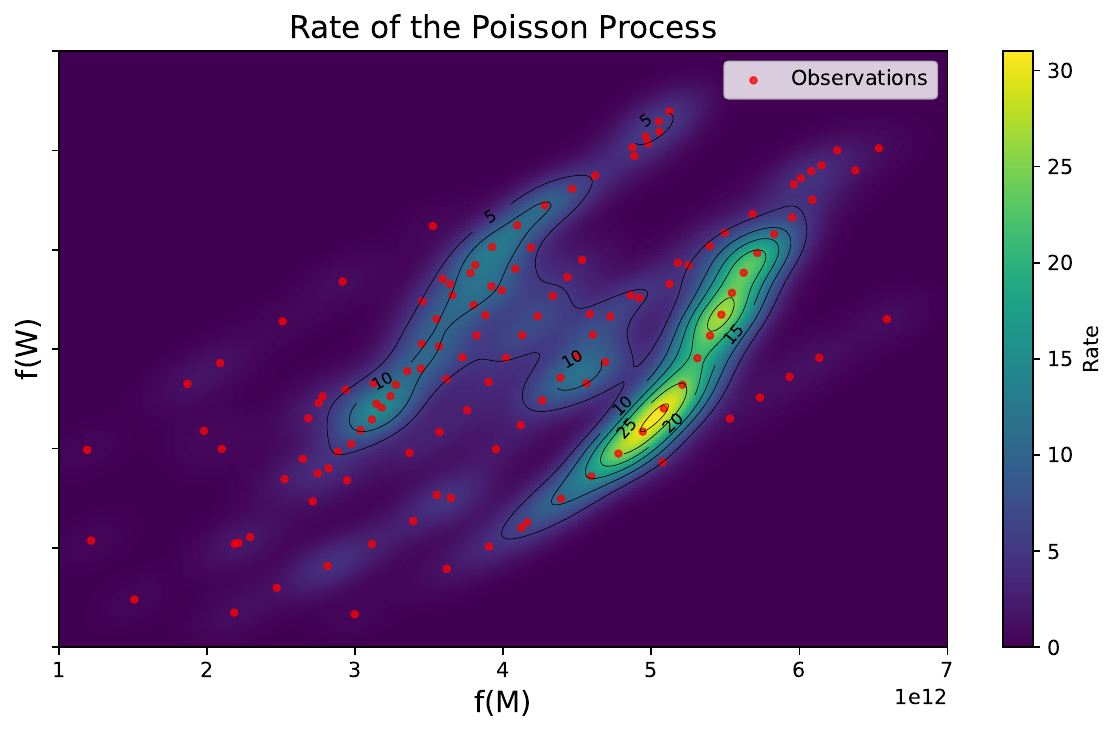}
    \caption{Fitted 2D function with contour lines for the Poisson rate as a function of (M, W)}
    \label{fig:rate_plot}
\end{figure}

In the next step, we fitted a bivariate histogram and a density using the Kernel Density Estimation (KDE) \cite{chen2017tutorial} method for these coordinates, where the height of each coordinate represents the number of cases associated with that specific combination (M, W). This results in a three-dimensional curve with axes corresponding to the mosquito profile (M), the weather parameters (W), and the number of reported cases. 
In doing this, we effectively have derived a function for the rate of a Poisson process, where the rate parameter $\lambda$ is a function of $(M, W)$. This means that we have assumed that the value of the number of reported cases is the mean of the Poisson process, which is a function (M,W), and the fitted curve defines this functionality (Figure\ref{fig:rate_plot}). \\\\
Using this Poisson rate function, we can define the likelihood function for the Poisson process as follows:
\begin{equation}
    f(x \mid (M, W)) = \frac{e^{-\lambda_{(M, W)}} \cdot \lambda^{x}_{(M, W)}}{x!},
\end{equation}
in which $x$ is the number of infected cases. Consequently based on the Bayes rule:
\begin{equation*}
    \pi{((M,W) \mid X=x)}\propto f(x \mid (M, W))\pi(M,W)
\end{equation*}
\clearpage

As an initial approach, it is possible to consider the prior as a bivariate uniform PDF to explore the regions of the parameter space $(M, W)$ associated with different values. Figure(\ref{fig:posteriorcomparison}) illustrates the posterior PDFs for various observation values. 
\begin{figure*}[htbp]  
    \centering
    \begin{subfigure}{0.33\linewidth}
        \centering
        \includegraphics[width=\linewidth]{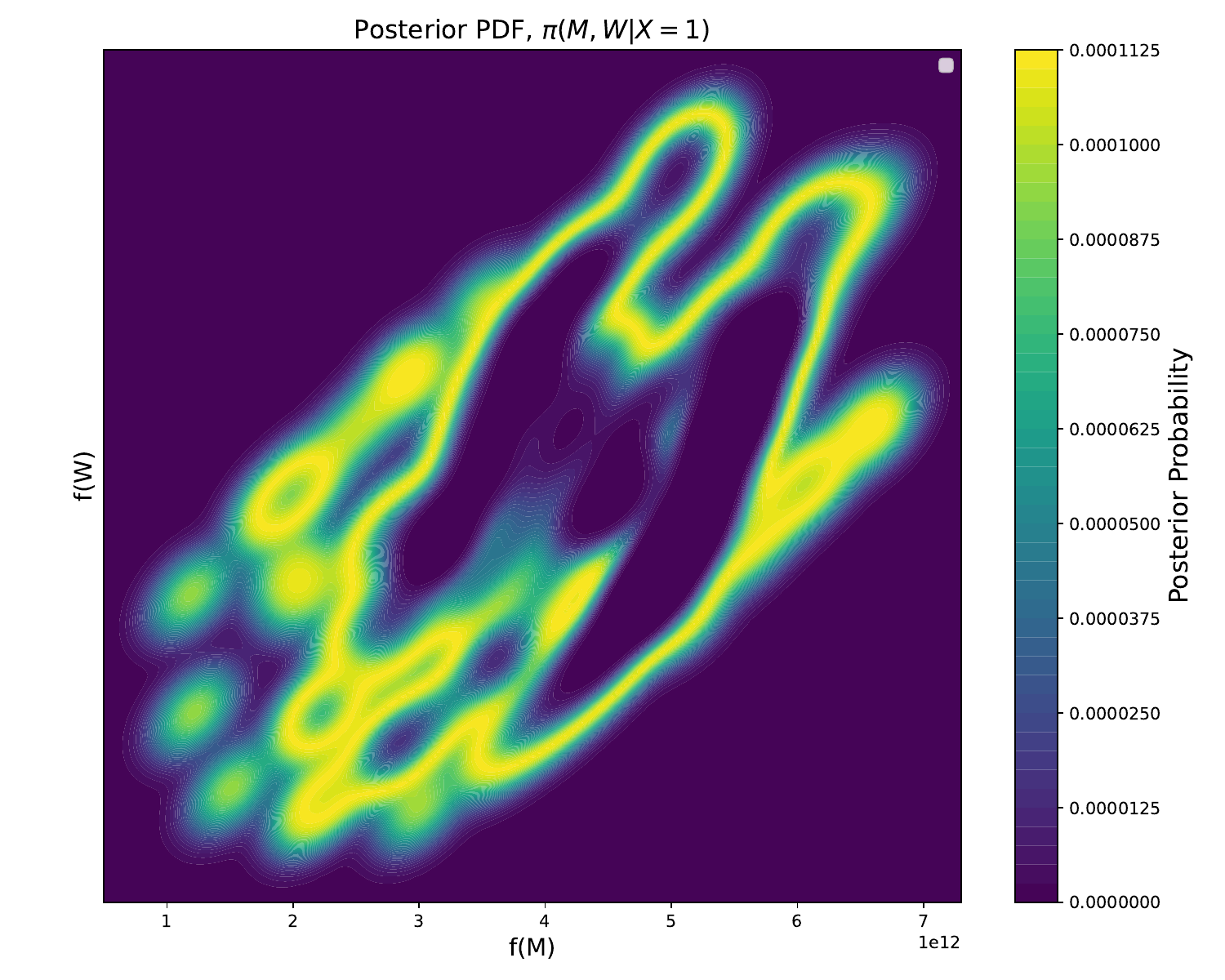}
        \caption{Posterior for $X=1$}
        \label{fig:posterior1}
    \end{subfigure}
    \hfill
    \begin{subfigure}{0.33\linewidth}
        \centering
        \includegraphics[width=\linewidth]{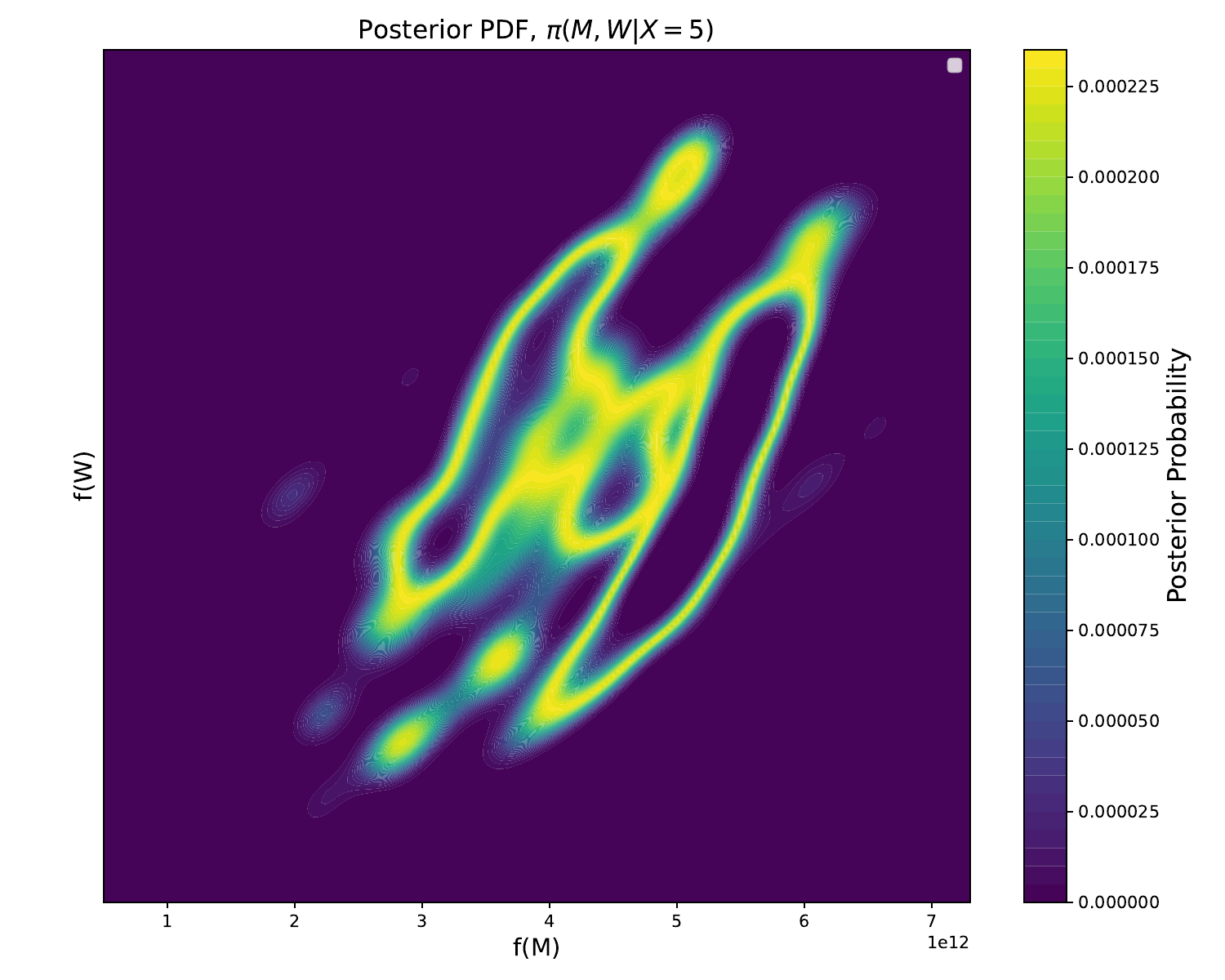}
        \caption{Posterior for $X=5$}
        \label{fig:posterior5}
    \end{subfigure}
    \hfill
    \begin{subfigure}{0.33\linewidth}
        \centering
        \includegraphics[width=\linewidth]{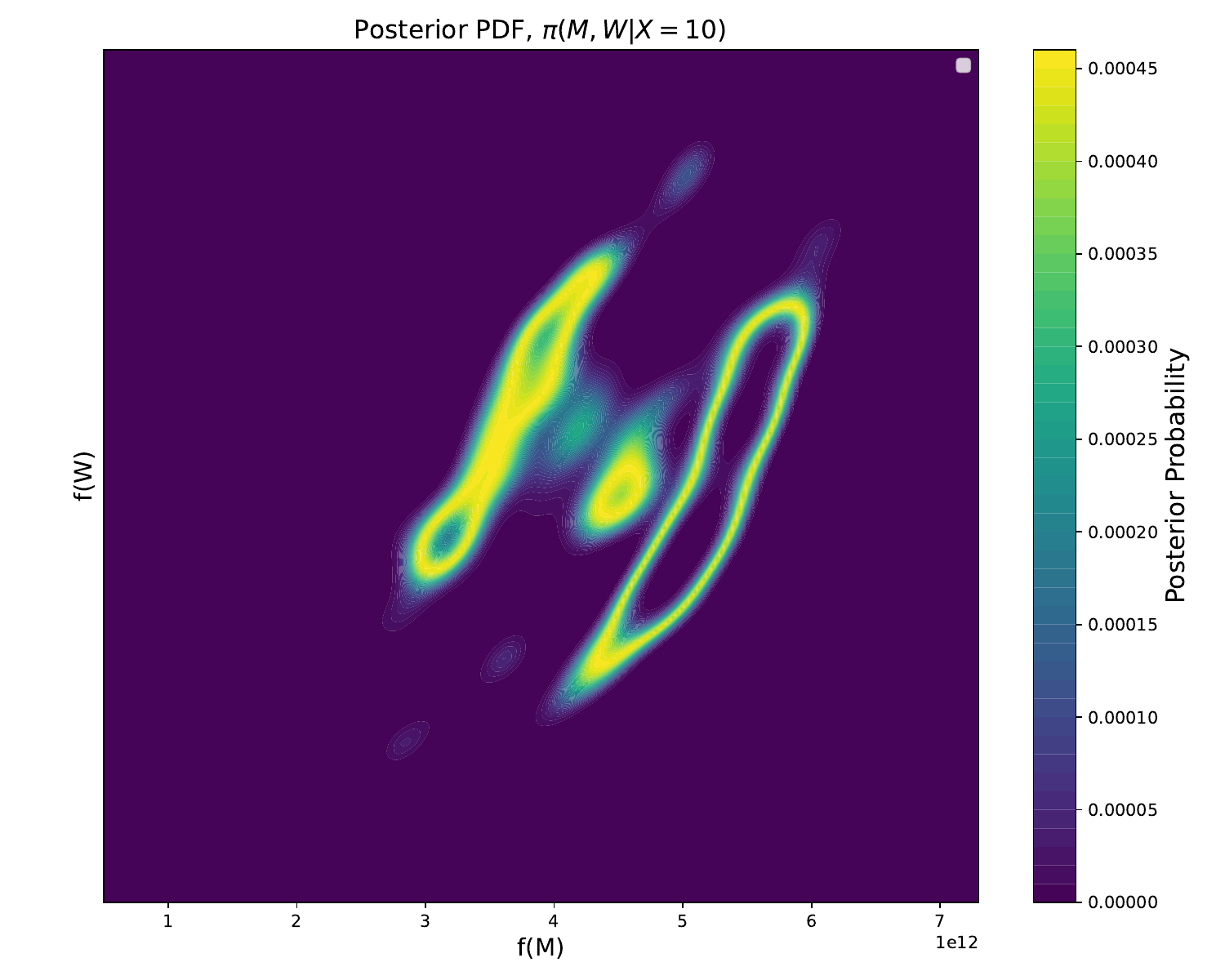}
        \caption{Posterior for $X=10$}
        \label{fig:posterior10}
    \end{subfigure}
    \\[1em]
    \begin{subfigure}{0.33\linewidth}
        \centering
        \includegraphics[width=\linewidth]{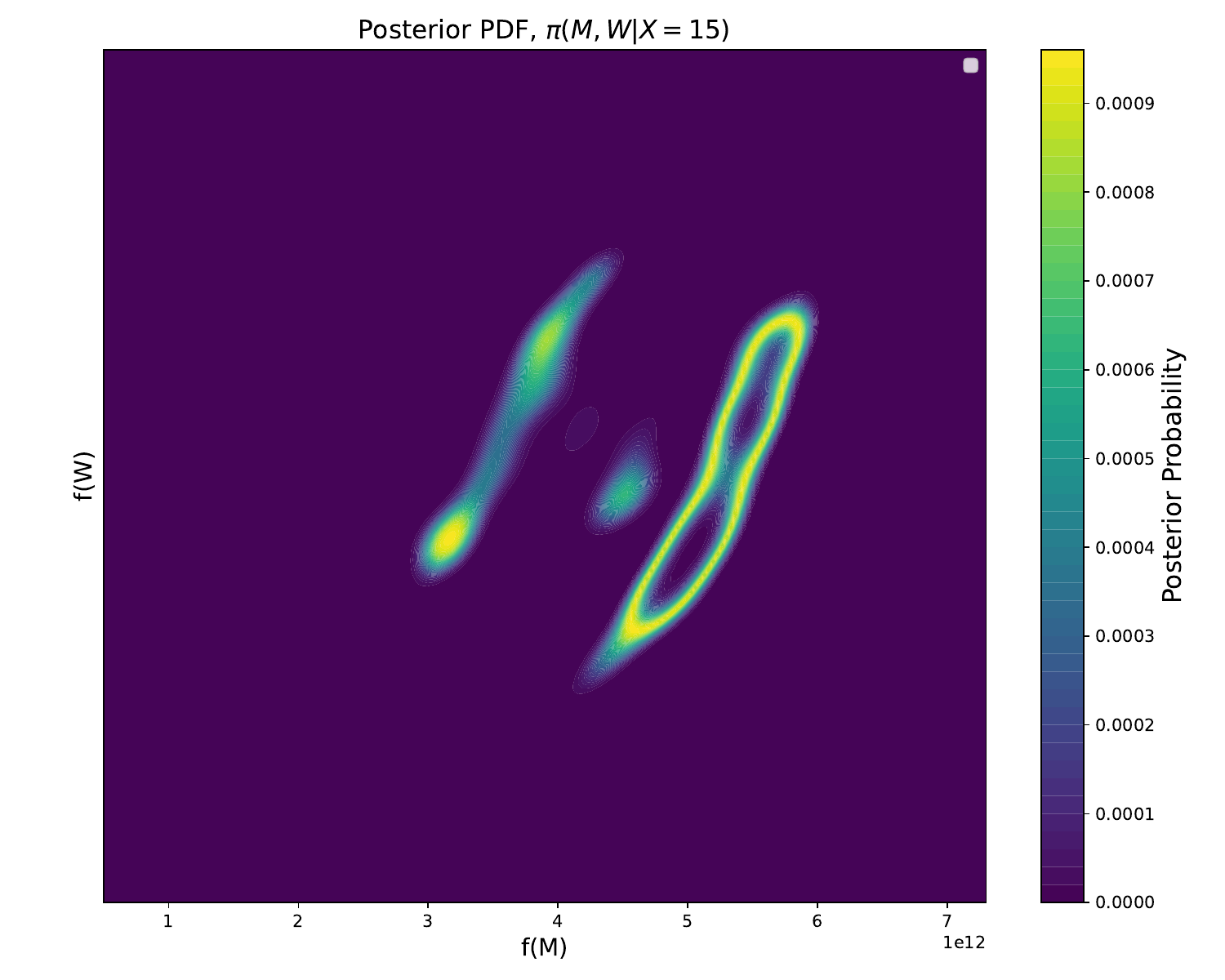}
        \caption{Posterior for $X=15$}
        \label{fig:posterior15}
    \end{subfigure}
    \hfill
    \begin{subfigure}{0.33\linewidth}
        \centering
        \includegraphics[width=\linewidth]{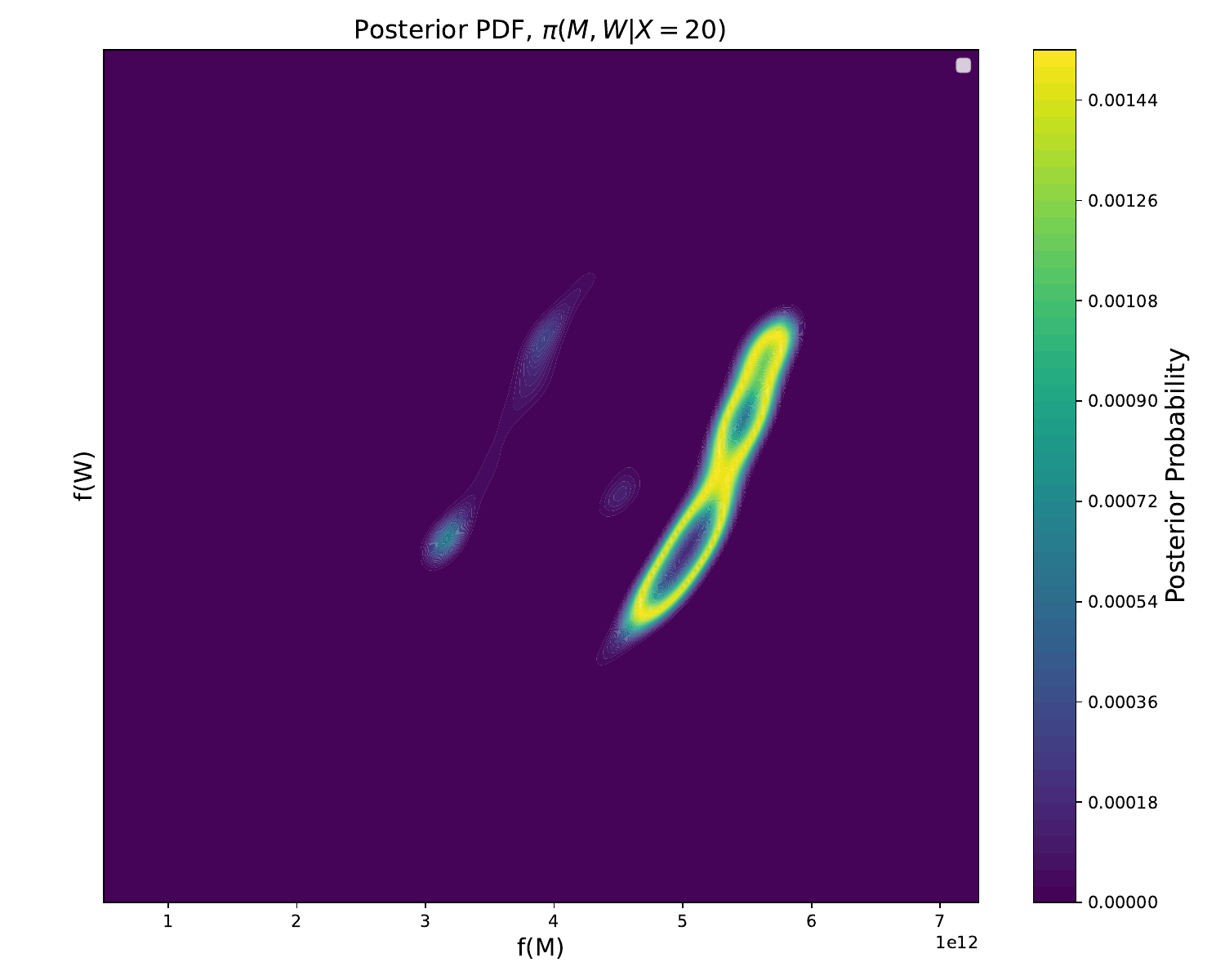}
        \caption{Posterior for $X=20$}
        \label{fig:posterior20}
    \end{subfigure}
    \hfill
    \begin{subfigure}{0.33\linewidth}
        \centering
        \includegraphics[width=\linewidth]{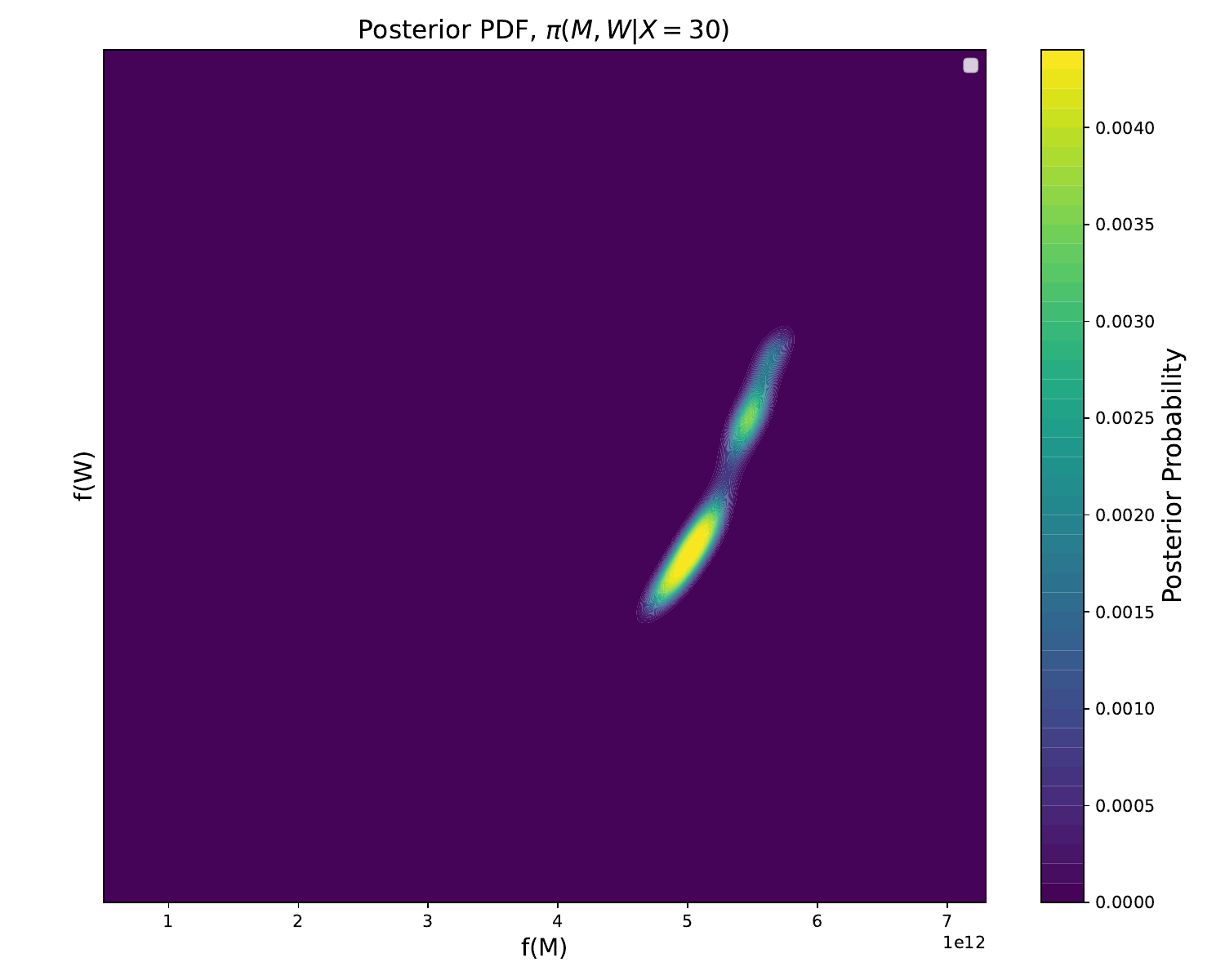}
        \caption{Posterior for $X=30$}
        \label{fig:posterior30}
    \end{subfigure}
    \caption{Comparison of posterior PDFs for different observation values.}
    \label{fig:posterior-comparison}
\end{figure*}

\begin{figure*}
    \centering
    \begin{subfigure}{0.45\linewidth}
        \centering
        \includegraphics[width=1\linewidth]{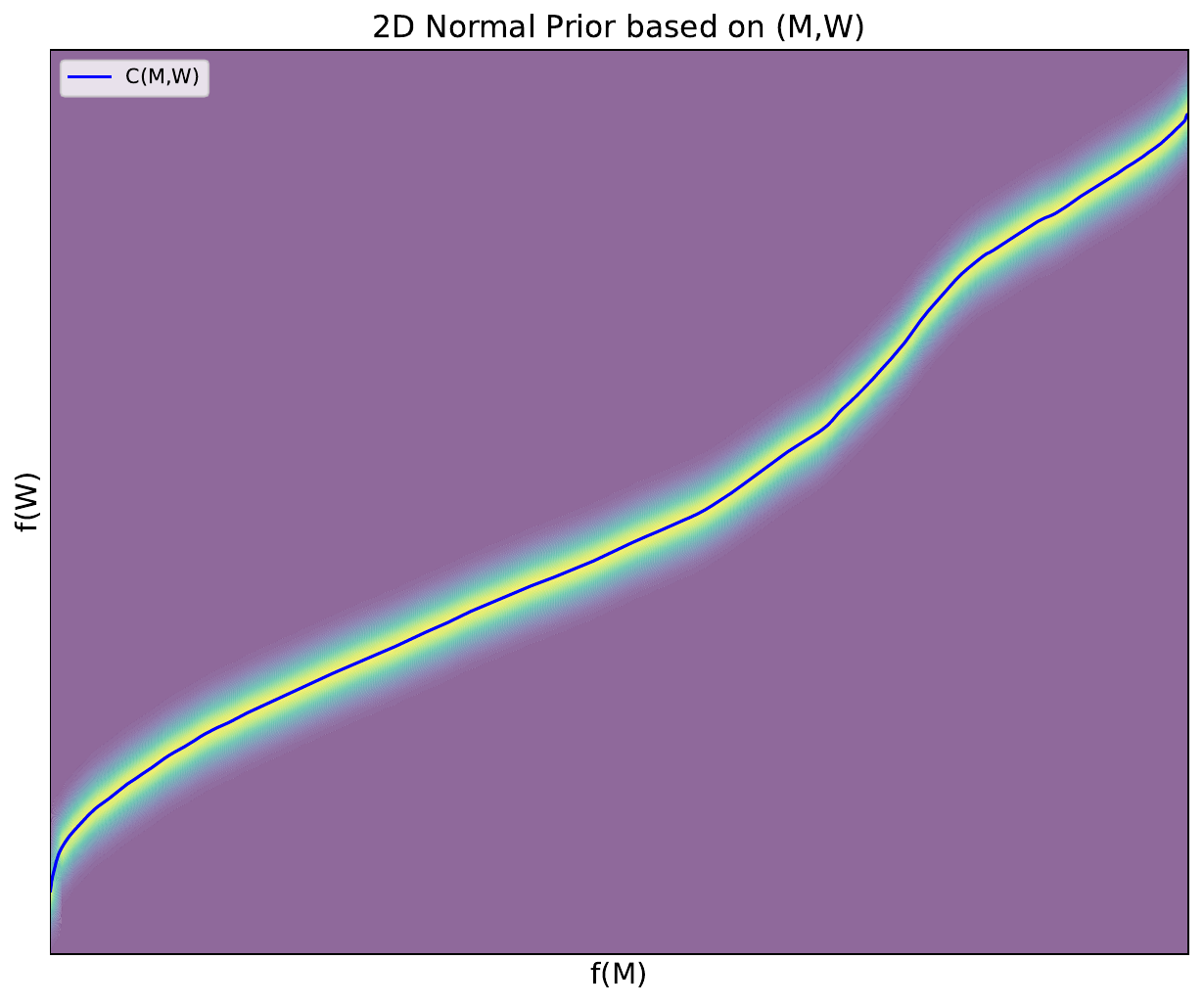}
        \caption{Normal prior}
        \label{fig:normal}
    \end{subfigure}
    \begin{subfigure}{0.45\linewidth}
        \centering
        \includegraphics[width=1\linewidth]{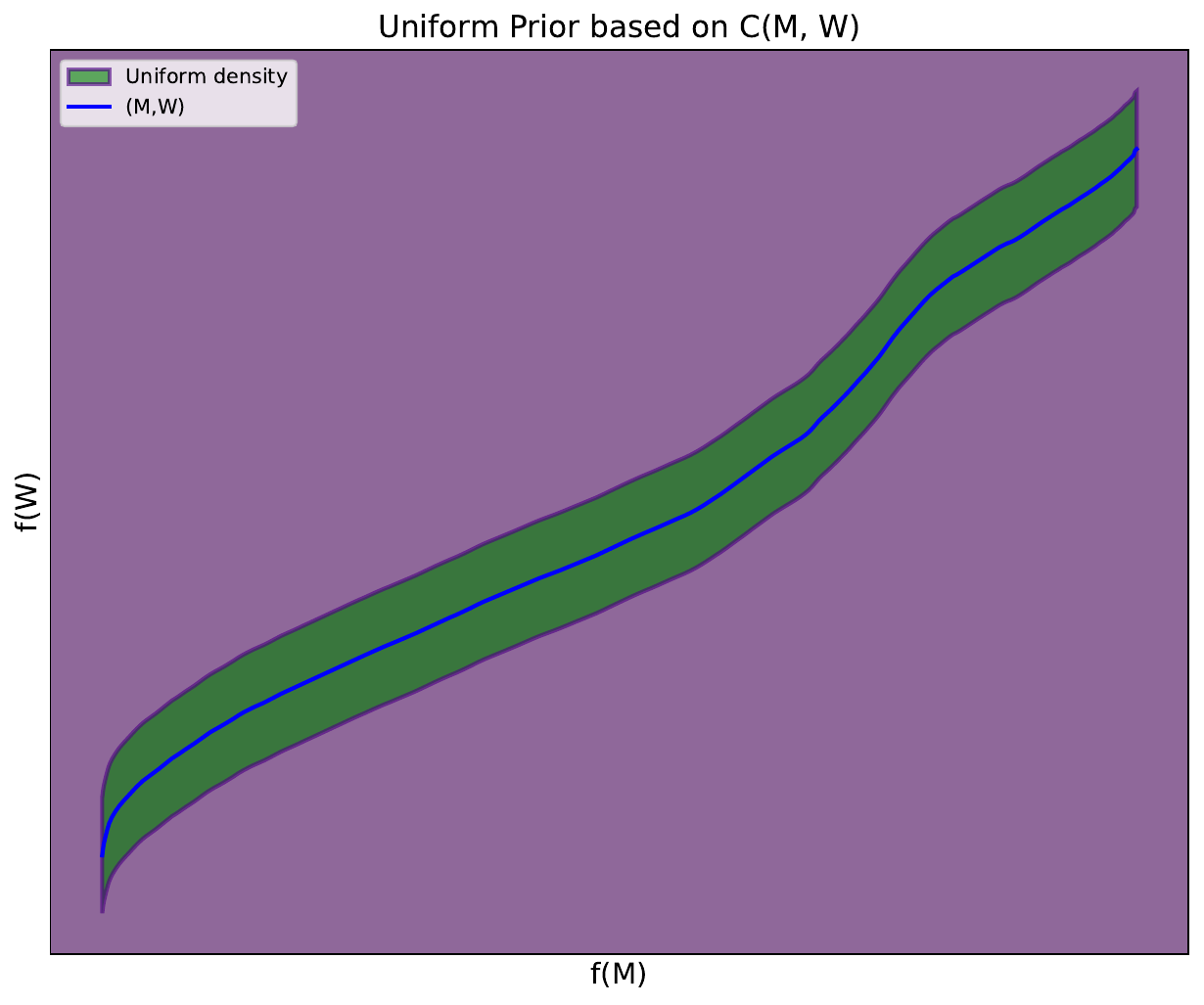}
        \caption{Uniform prior}
        \label{fig:uniform}
    \end{subfigure}
    
    \caption{Two Types of Prior PDFs Derived from the (M,C) Curve Generated by the Compartmental Model for Orange County, California, 2020}
    \label{fig:priors}
\end{figure*}

In order to estimate or even prediction for the future, we can define informative prior. This process can be obtained by using the compartmental differential equations model we have used to get the curve of (M,W). It can be considered as a Gaussian density with the mean of the curve of(M,W), or even a uniform shape around the curve of (M,W). This means that based on this approach, we believe that the forecast for (M,W) gives us values in the prior probability density function around the curve of (M,W) Figure (\ref{fig:priors}). In this work, we use uniform priors for inference to prevent distortions in the results, as it is known that priors can sometimes influence the outcome by incorporating additional, albeit unbiased, information into the model.\\

\subsection{Spillover Severity Risk Estimation }
\begin{figure*}
    \centering
    \begin{subfigure}{0.45\linewidth}
        \centering
        \includegraphics[width=1\linewidth]{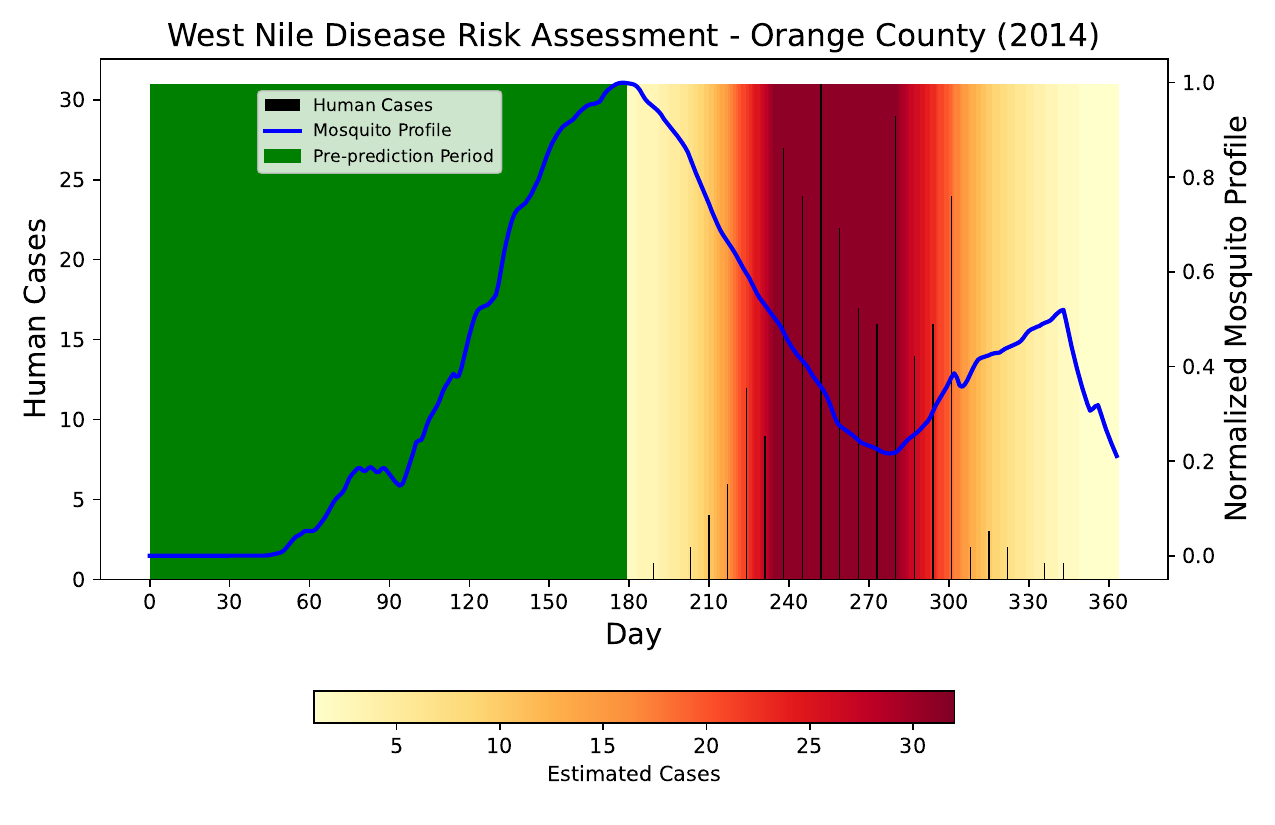}
        \caption{2014}
        \label{fig:severity2014}
    \end{subfigure}
    \begin{subfigure}{0.45\linewidth}
        \centering
        \includegraphics[width=1\linewidth]{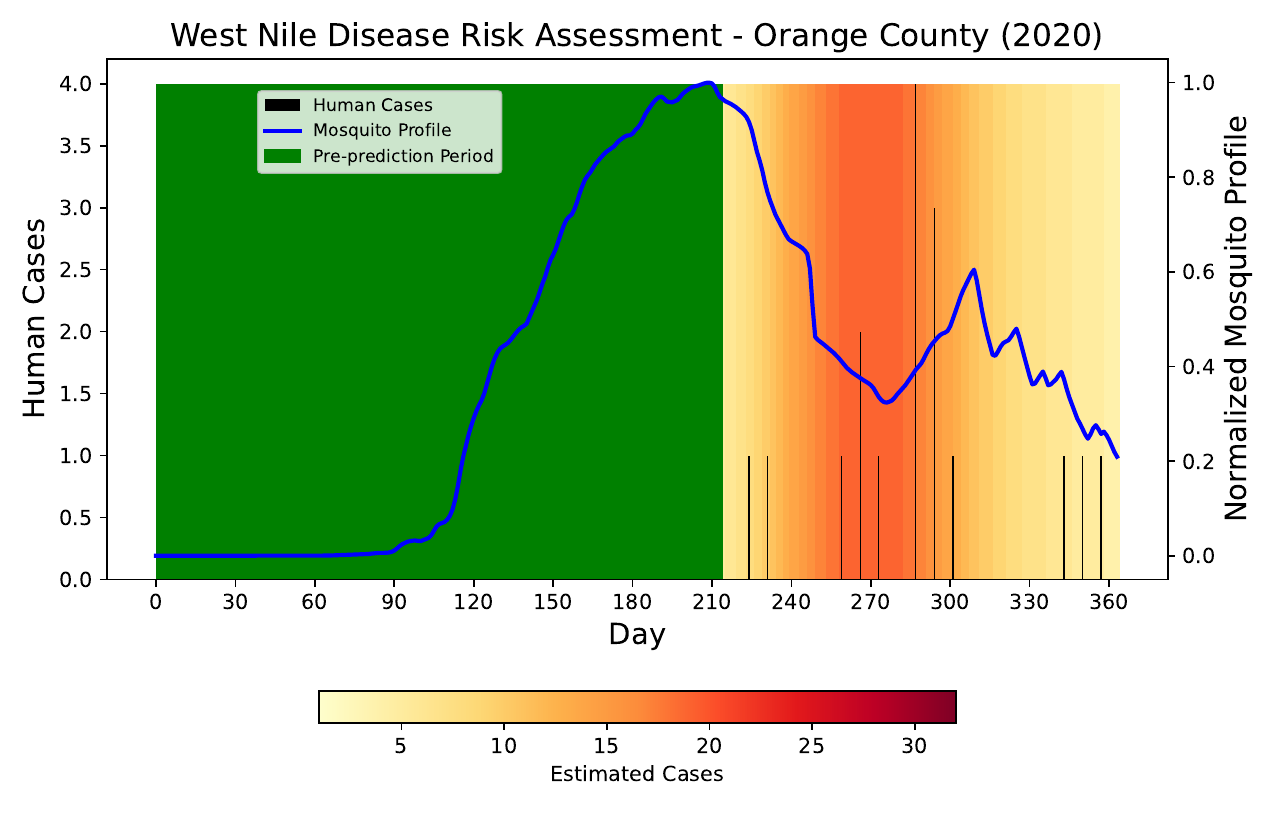}
        \caption{2020}
        \label{fig:severity2020}
    \end{subfigure}
    
    \caption{Estimated severity of outbreak for 2014 and 2020, Orange County, California}
    \label{fig:severity}
\end{figure*}

To estimate the severity of the outbreak for a given day, we use the posterior distributions generated. Specifically, we create posteriors for different values of $ X = x $, where $ X $ represents the number of cases. In the next step, we run the model using real weather parameters. For each day, the compartmental model provides a estimated coordinate $ (M_t, W_t)$, which we compare against all posteriors to determine which assigns the highest density to that coordinate. Based on the optimal posterior, the corresponding value of $ X $ is estimated. Figure (\ref{fig:severity}) shows the risk estimation based on this method.

\subsection{Spillover Severity Risk Prediction}
Prediction of risk severity is performed using posterior PDFs, following a maximum posterior predictive approach (MPP). Specifically, we estimate the value of $X$ by evaluating the posterior predictive density of the observed data $(W_0, M_0)$ in each candidate posterior distribution corresponding to $X = 1, 2, \dots, 30$. The chosen estimate for $X$ is the one that produces the highest posterior predictive density at the observation.

We used historical data up to time $t_0$ to forecast weather parameters at time $t_1$. The compartmental model is then run over this forecasted period to compute the curve $C(M,W)$. From this curve, we derive prior distributions (either uniform or normal) and subsequently obtain the posterior distributions for each candidate value of $X$. Finally, predictions are made for all predicted pairs $(M,W)$, allowing us to determine the corresponding days.
\subsubsection{Long-term Prediction of Risk Severity}
\begin{figure*}[htbp]
    \centering
    \begin{subfigure}{0.45\textwidth}
        \centering
        \includegraphics[width=\linewidth]{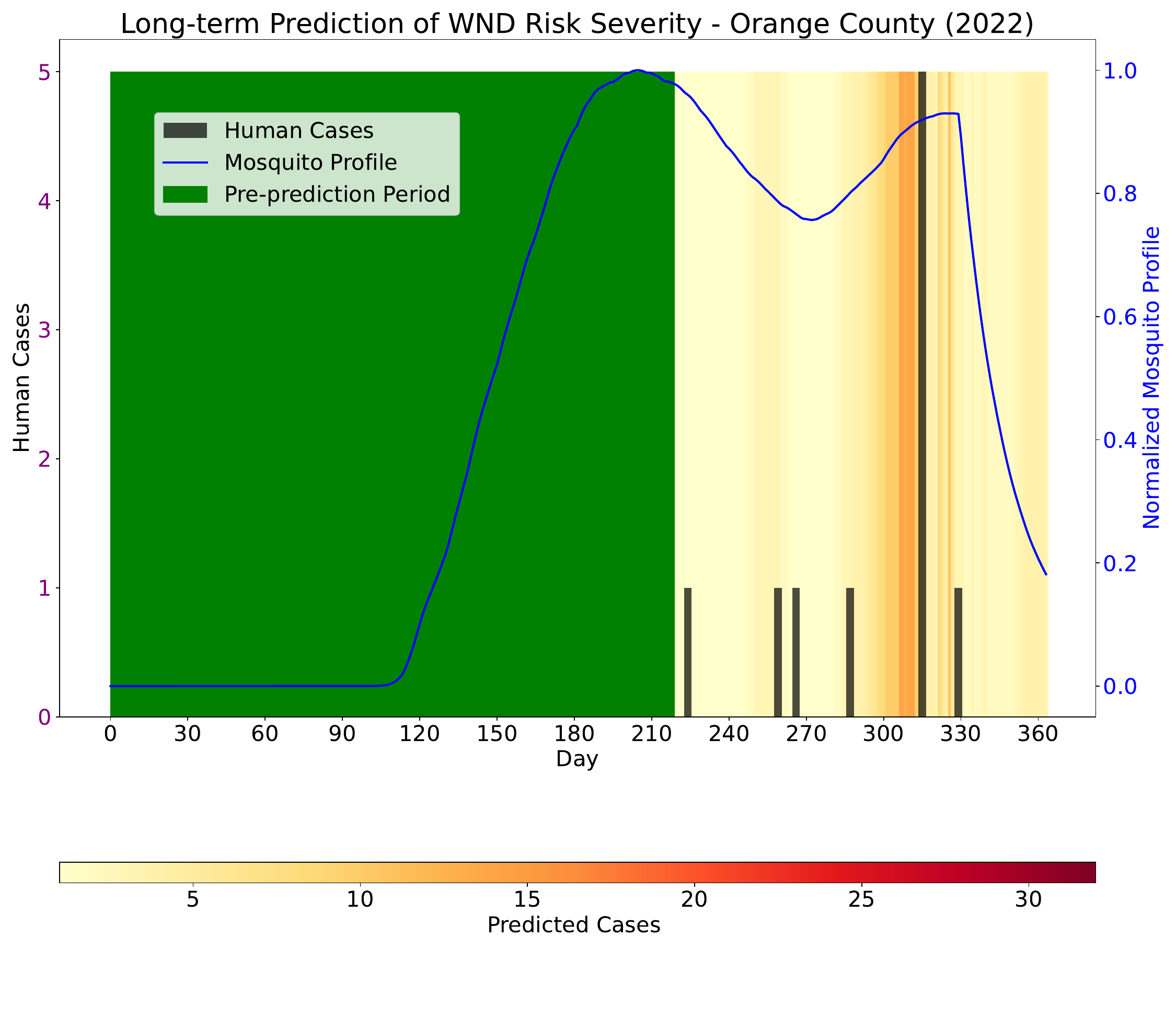}
        \caption{2022}
        \label{fig:2022}
    \end{subfigure}
    \hfill
    \begin{subfigure}{0.45\textwidth}
        \centering
        \includegraphics[width=\linewidth]{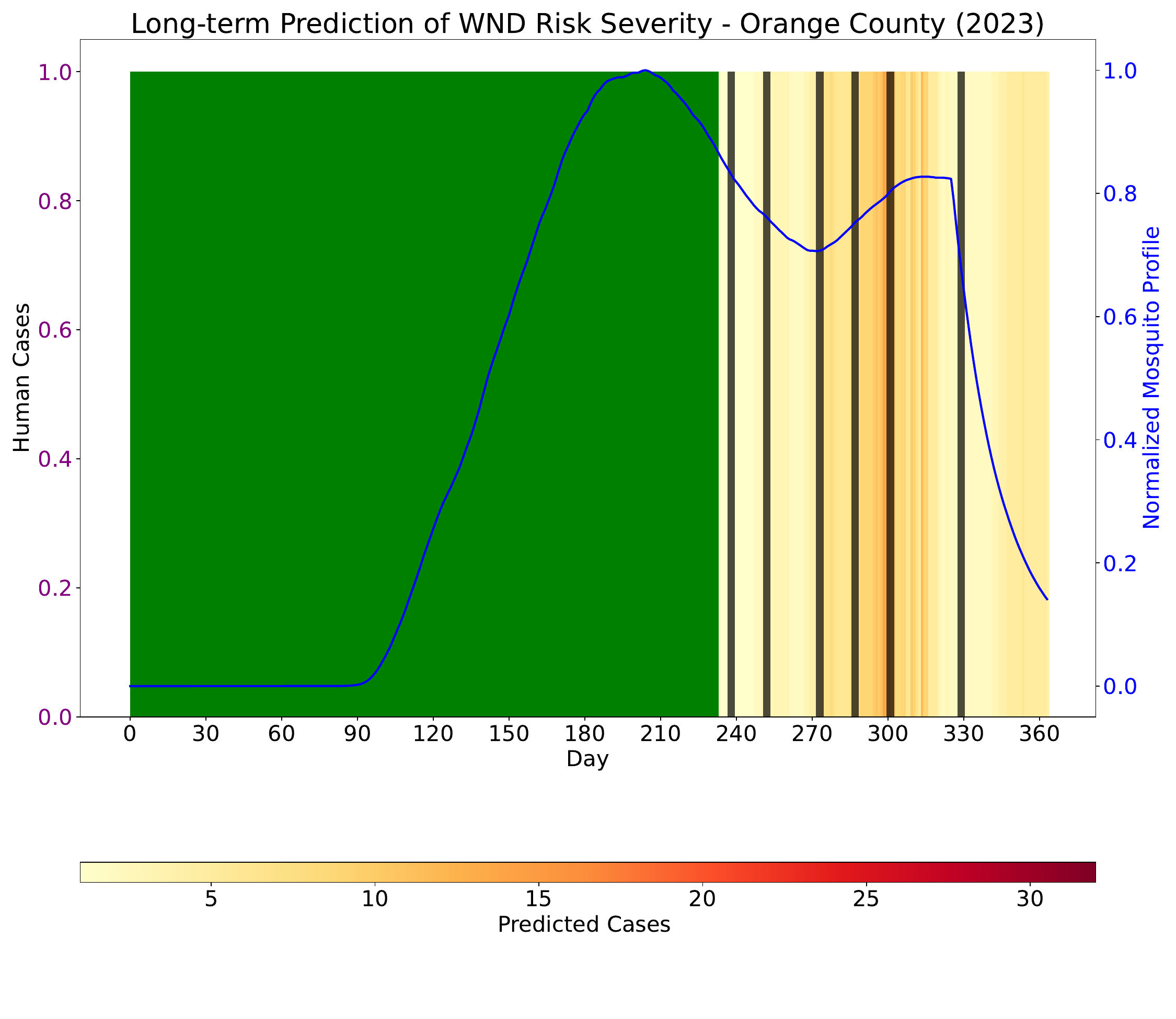}
        \caption{2023}
        \label{fig:2023}
    \end{subfigure}
    \hfill
    \begin{subfigure}{0.45\textwidth}
        \centering
        \includegraphics[width=\linewidth]{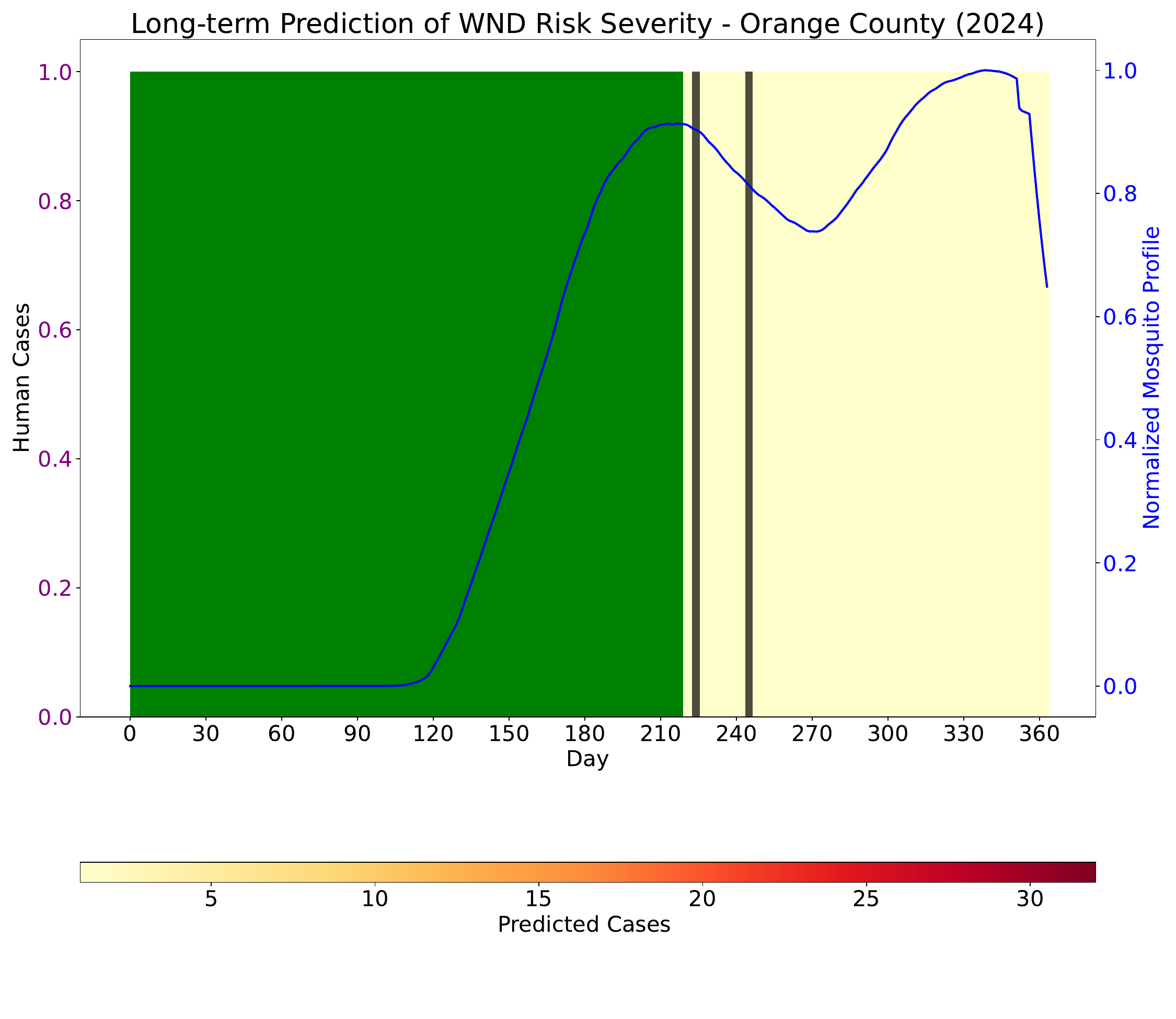}
        \caption{2024}
        \label{fig:2024}
    \end{subfigure}
    
    \caption{Long-term prediction for the severity of the risk for 2022, 2023, and 2024, Orange County, California}
    \label{fig:severity-longterm}
\end{figure*}

For long-term risk severity prediction, we employed an auto-regressive model (discussed in \ref{sunsection:Spillover Onset Risk Prediction}) to capture the trend of the weather parameters over a long period.
At a given time (starting point in time), the model was initialized using the predicted temperature (for a period that exceeds one month) to generate the corresponding disease trend during this time frame. In practice, we also incorporate a portion of the actual weather parameters of the target year, which means that we used \ref{subsection: long ter, for onset} to approximate the start date of the long-term severity prediction. Using the forcatsted parameter, the prior and consequently posterior is made.
The predicted disease trend was then used as input for the Bayesian model, which estimated the severity of the outbreak for each day within the predicted period.  In other words, for each day, we obtained a coordinate (M, W), which was then compared with the posterior distributions to predict the number of cases. 
Figure (\ref{fig:severity-longterm}) shows the plot related to this long-term prediction for 2022,2023, and 2024. 
\cleardoublepage



\subsubsection{Short-term Prediction of Risk Severity}
\begin{figure*}[htbp]
    \centering
    \begin{subfigure}{0.45\textwidth}
        \centering
        \includegraphics[width=\linewidth]{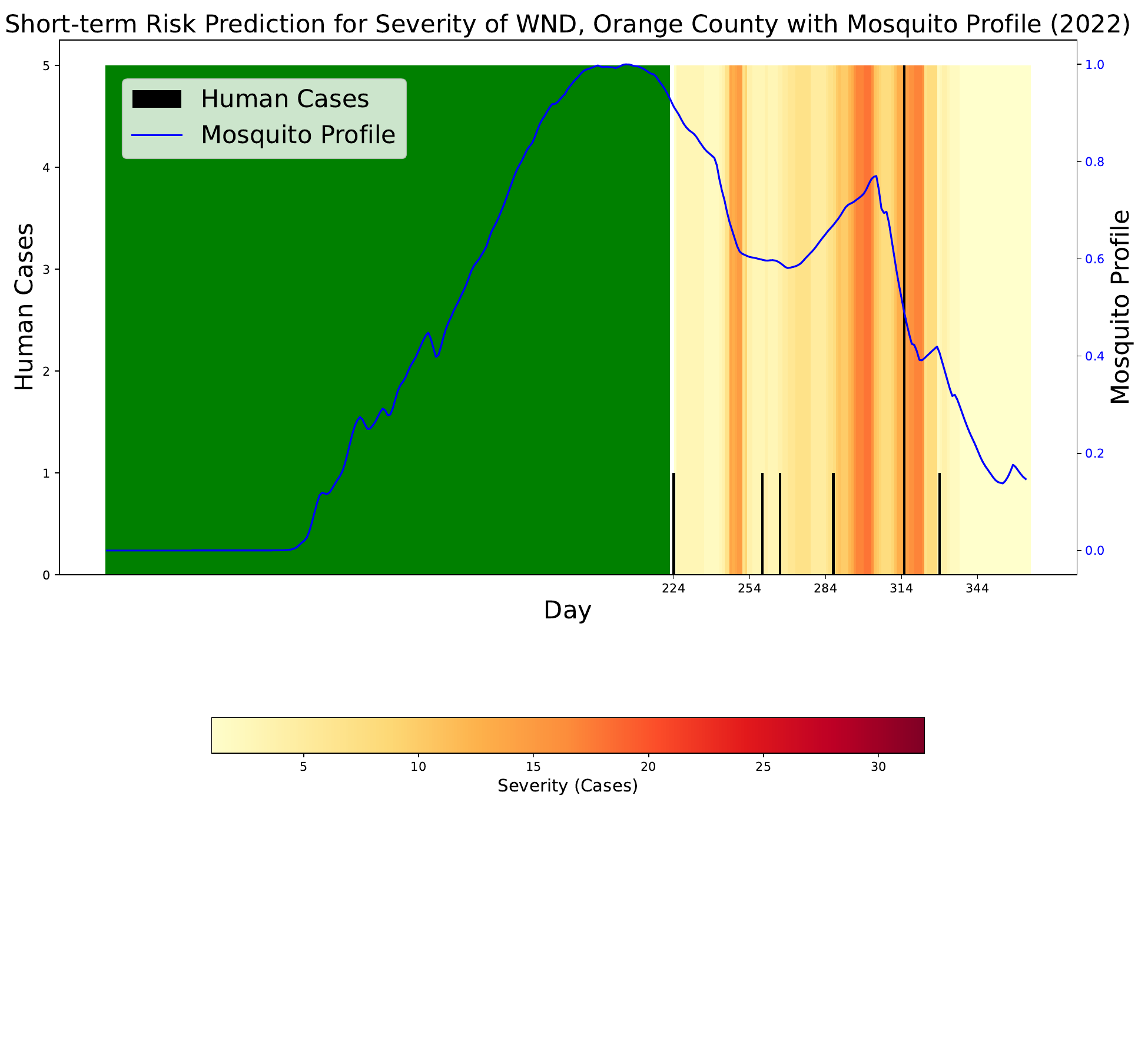}
        \caption{2022}
        \label{fig:2022}
    \end{subfigure}
    \hfill
    \begin{subfigure}{0.45\textwidth}
        \centering
        \includegraphics[width=\linewidth]{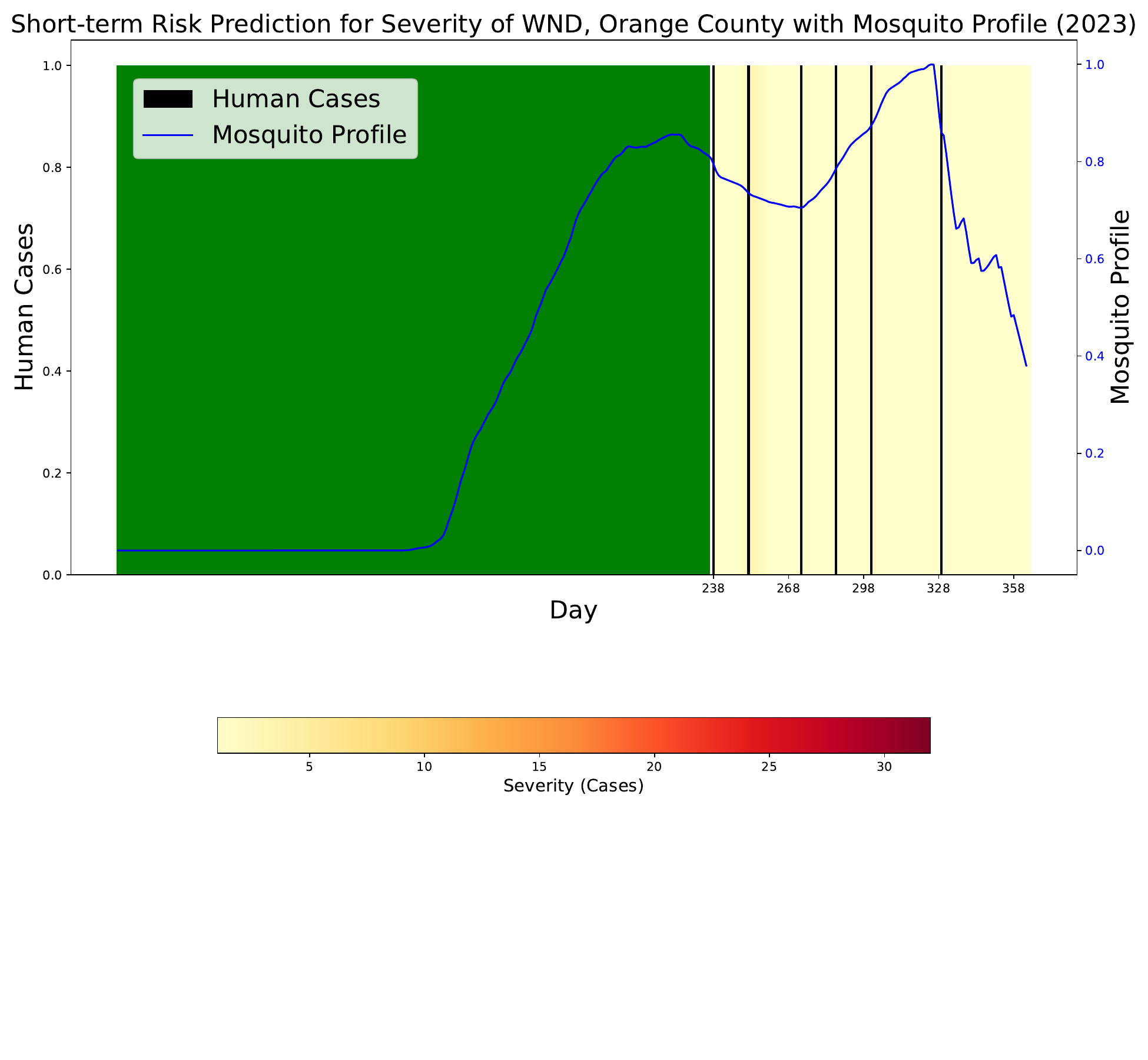}
        \caption{2023}
        \label{fig:2023}
    \end{subfigure}
    \hfill
    \begin{subfigure}{0.45\textwidth}
        \centering
        \includegraphics[width=\linewidth]{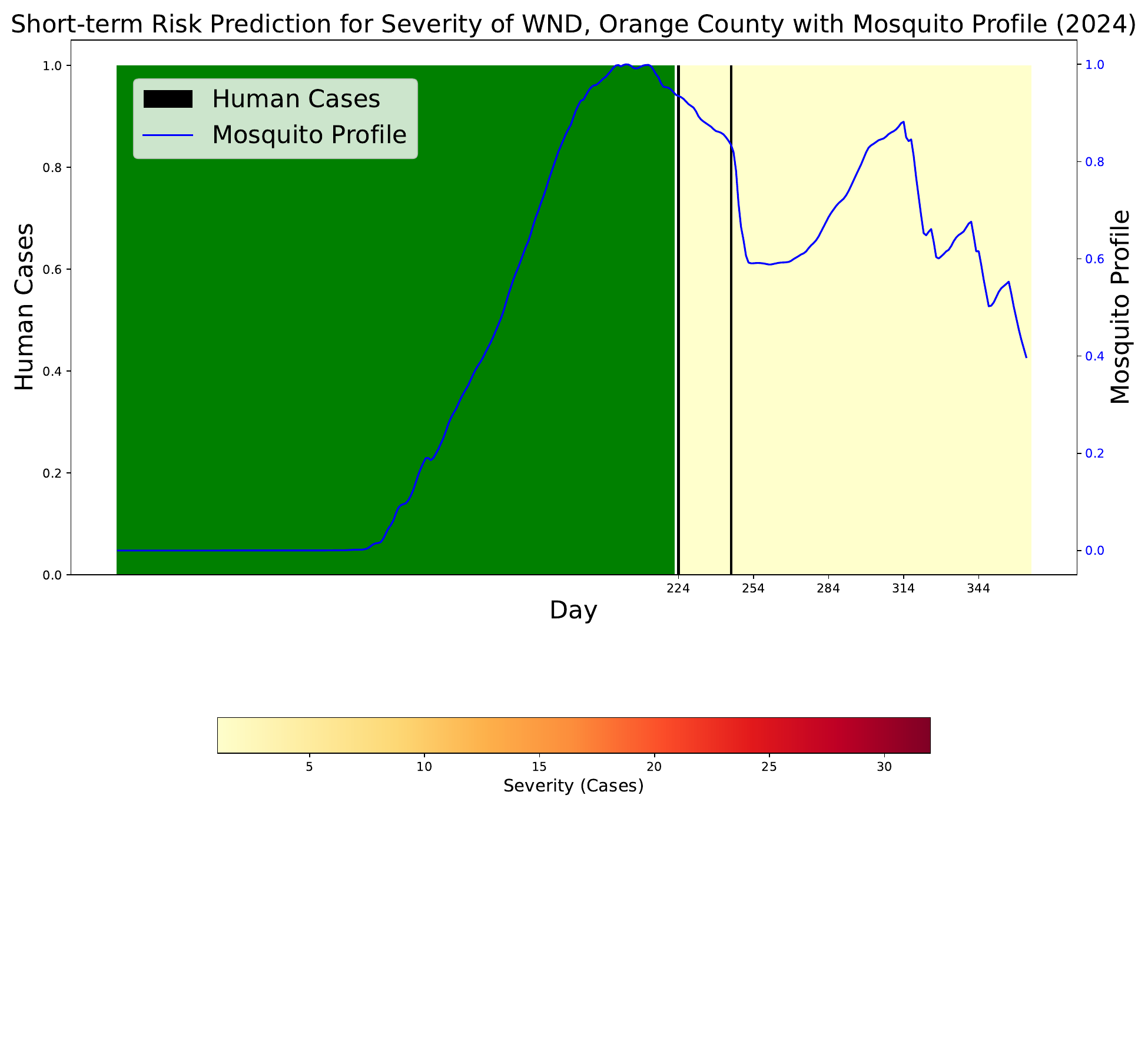}
        \caption{2024}
        \label{fig:2024}
    \end{subfigure}
    
    \caption{Short-term prediction for the severity of the risk for 2022, 2023, and 2024, Orange County, California}
    \label{fig:severity-short}
\end{figure*}

For short term prediction of the risk severity we have used an auto-regressive (degree=time lead ($l_t$)) model and used the weather data from the target year to time $t_0$ and predicted the parameters for day $t_{0}+l_t$, run the model and obtain the prior and posterior to $t_{0}+l_t$, then compare the values of ($M_{t_{0}+l_t},W_{t_{0}+l_t}$) with the posteriors and predict the prediction value $X=x$, then update the likelihood with date to time $t_0+l_t$ and so on. Figure (\ref{fig:severity-short}) shows the result of the one-step prediction of the severity of the risk for three consecutive years. 

\section{Model Evaluation}
So far, we have evaluated our model by comparing its predictions with data observed (new reported cases) from authoritative sources, such as the California Department of Public Health. 
To thoroughly assess the performance and validity of our model, we should also benchmark it against other existing models. However, comparing against every available model individually is impractical. Therefore, an effective strategy is to select a highly accurate and well-established model identified from prior research and use it as a benchmark to represent the performance of established high-quality models. This approach simplifies comparisons while maintaining rigor by ensuring that our model is tested against a reliable and representative baseline.
To do this comparison, we have chosen a Negative Binomial-based model that has had a high accuracy in CDC open forecasting challenge to assess skill of West Nile virus neuroinvasive disease prediction \cite{holcomb2023evaluation}.
The Negative Binomial model has been trained on historical case data (using the built-in function of "NegativeBinomialP" from the "statsmodels" library) to compare its results with our model. The Negative Binomial model performs one-step predictions, meaning it uses historical data up to the current time in a given year to predict the next time step (one week ahead in this case, because historical data are reported weekly) and continues iteratively.

We performed this comparison for Orange County, Kern County and Los Angeles County in California, as well as Dallas County and Harris County in Texas. Dallas County has the highest rate of West Nile Disease (WND) in Texas, and Harris County recorded a historically high number of cases of WND in 2014. WND data for Texas was obtained from the Texas Department of State Health Services (DSHS). For building the models for Harris and Dallas counties, we used data from 2014 to 2024, which is a shorter period compared to the data used for California counties (2006–2024).

In addition, to compare these two models, we have used the logarithmic score for each week that has been used in the open forecast challenge of the CDC to assess the performance of the prediction of neuroinvasive diseases of the West Nile virus \cite{holcomb2023evaluation} to score the results of the models of the participants. It is evident that, on the basis of this scoring system, negative values closer to zero indicate higher accuracy.
Regarding our model, we utilized weather data up to time $t_0$. Using AR(7) or SARIMA, we forecasted the weather parameters for time $t_{0}+7$ (one week ahead) and continued this process iteratively to generate weekly predictions. 
Although long-term risk severity predictions typically have a lead time of months rather than a week, making comparisons with the Negative Binomial model less relevant, we have still calculated the logarithmic score for this case to assess whether accuracy is maintained in short-term predictions.

\begin{figure*}
    \centering
    \begin{subfigure}{0.45\textwidth}
        \centering
        \includegraphics[width=\linewidth]{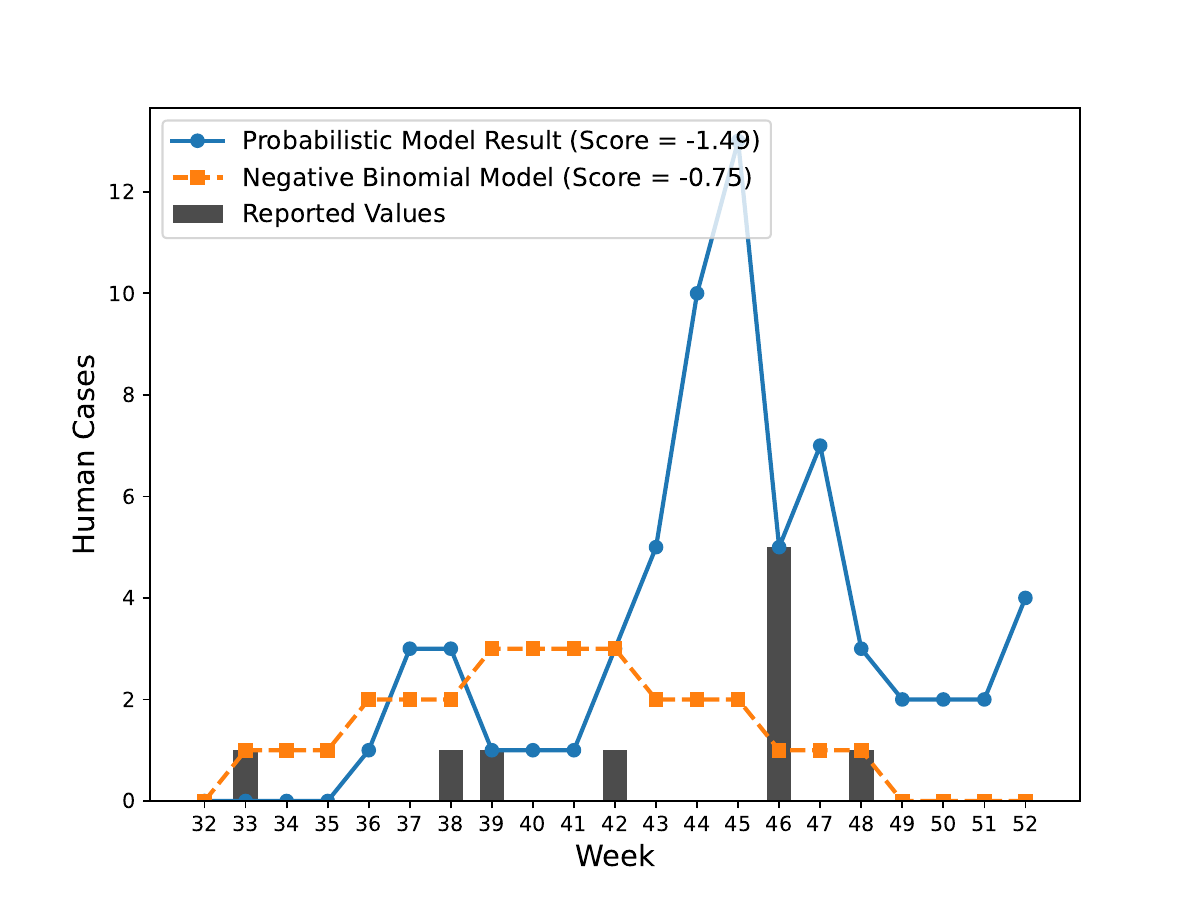}
        \caption{2022}
        \label{fig:model_2022}
    \end{subfigure}%
    \begin{subfigure}{0.45\textwidth}
        \centering
        \includegraphics[width=\linewidth]{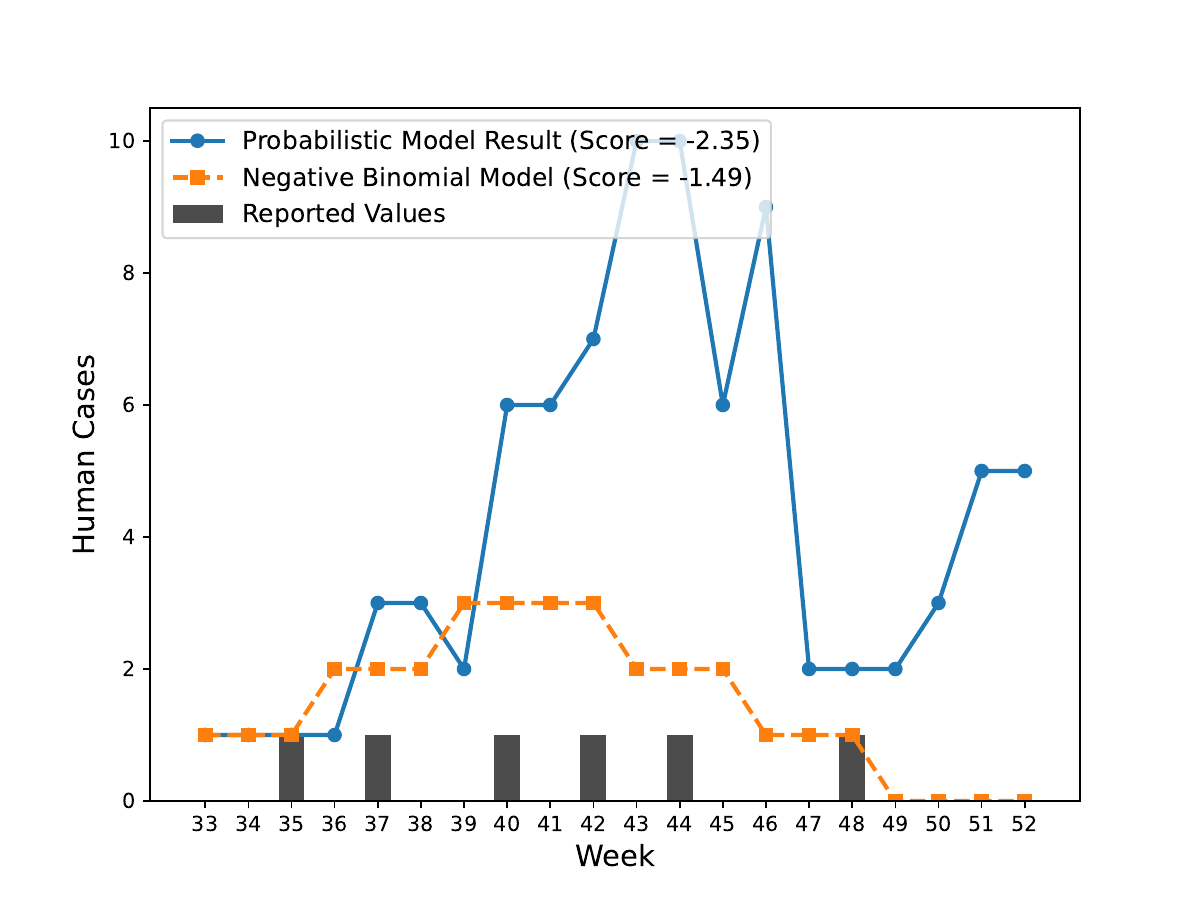}
        \caption{2023}
        \label{fig:model_2023}
    \end{subfigure}
    
    \begin{subfigure}{\textwidth}
        \centering
        \includegraphics[width=0.45\linewidth]{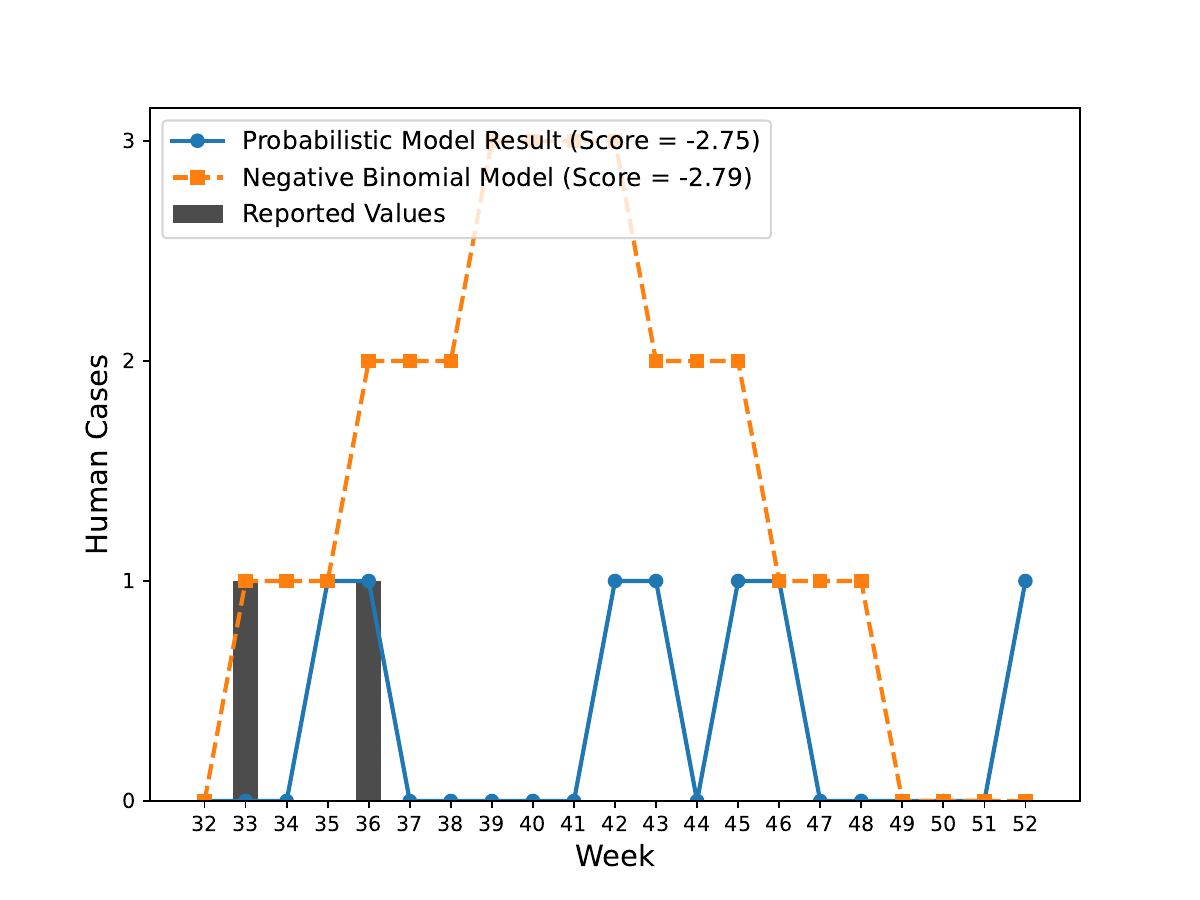}
        \caption{2024}
        \label{fig:model_2024}
    \end{subfigure}
    \caption{Comparison of the Bayesian model, long-term results, and Negative Binomial results for short-term prediction based on the logarithmic scoring system for 2022, 2023, and 2024, Orange County, California.}
    \label{fig:model_comparison}
\end{figure*}
\clearpage
\begin{figure}
    \centering
    \begin{subfigure}{0.48\linewidth}
        \centering
        \includegraphics[width=\linewidth]{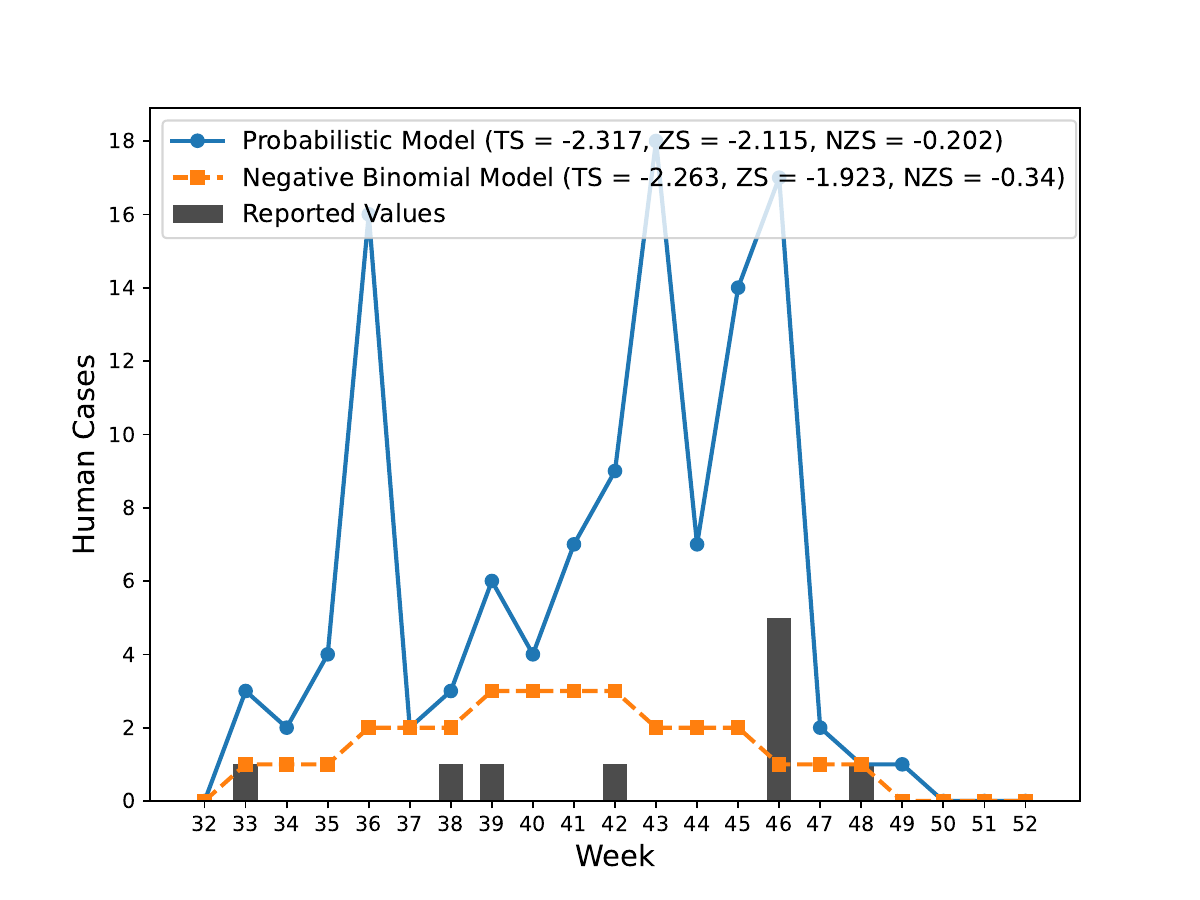}
        \caption{2022}
        \label{fig:model-comparison-2022}
    \end{subfigure}
    \hfill
    \begin{subfigure}{0.48\linewidth}
        \centering
        \includegraphics[width=\linewidth]{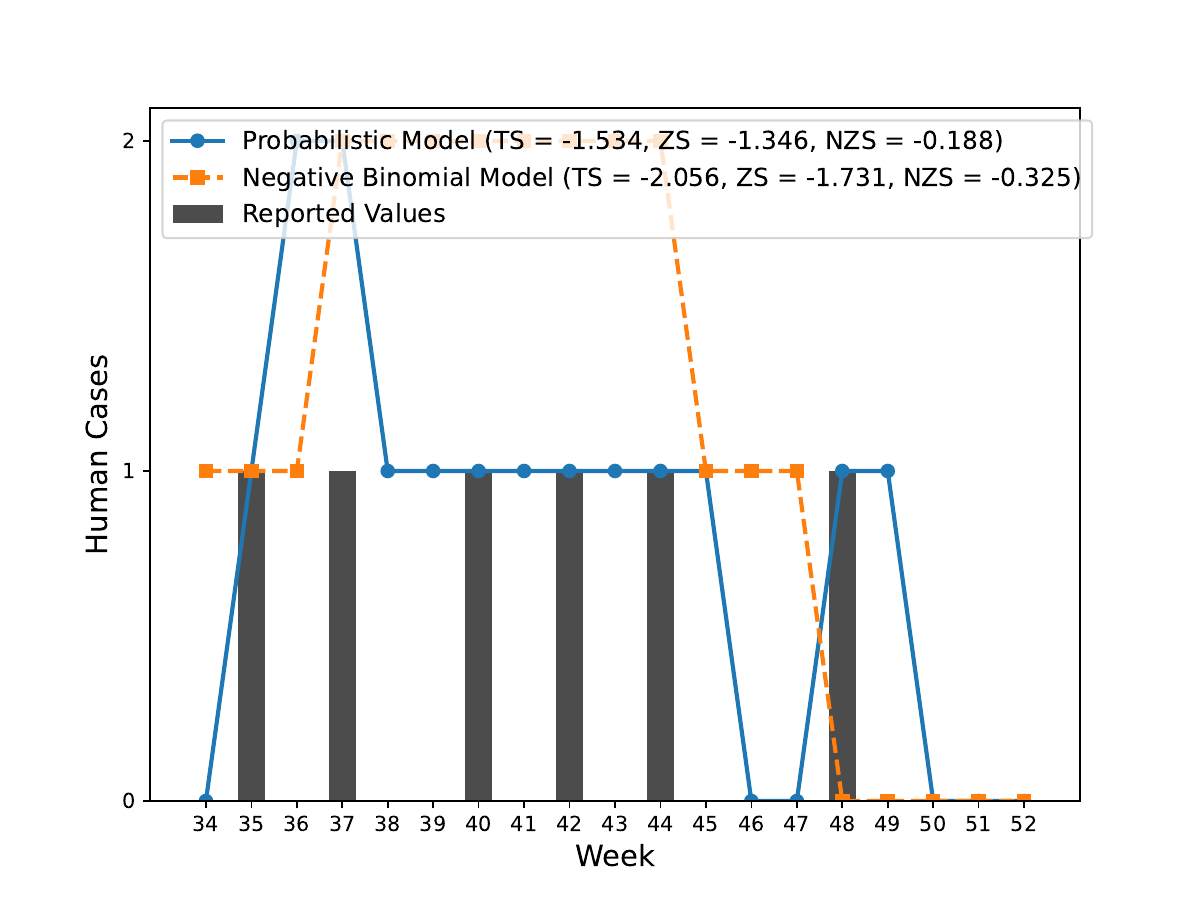}
        \caption{2023}
        \label{fig:model-comparison-2023}
    \end{subfigure}
    \hfill
    \begin{subfigure}{0.48\linewidth}
        \centering
        \includegraphics[width=\linewidth]{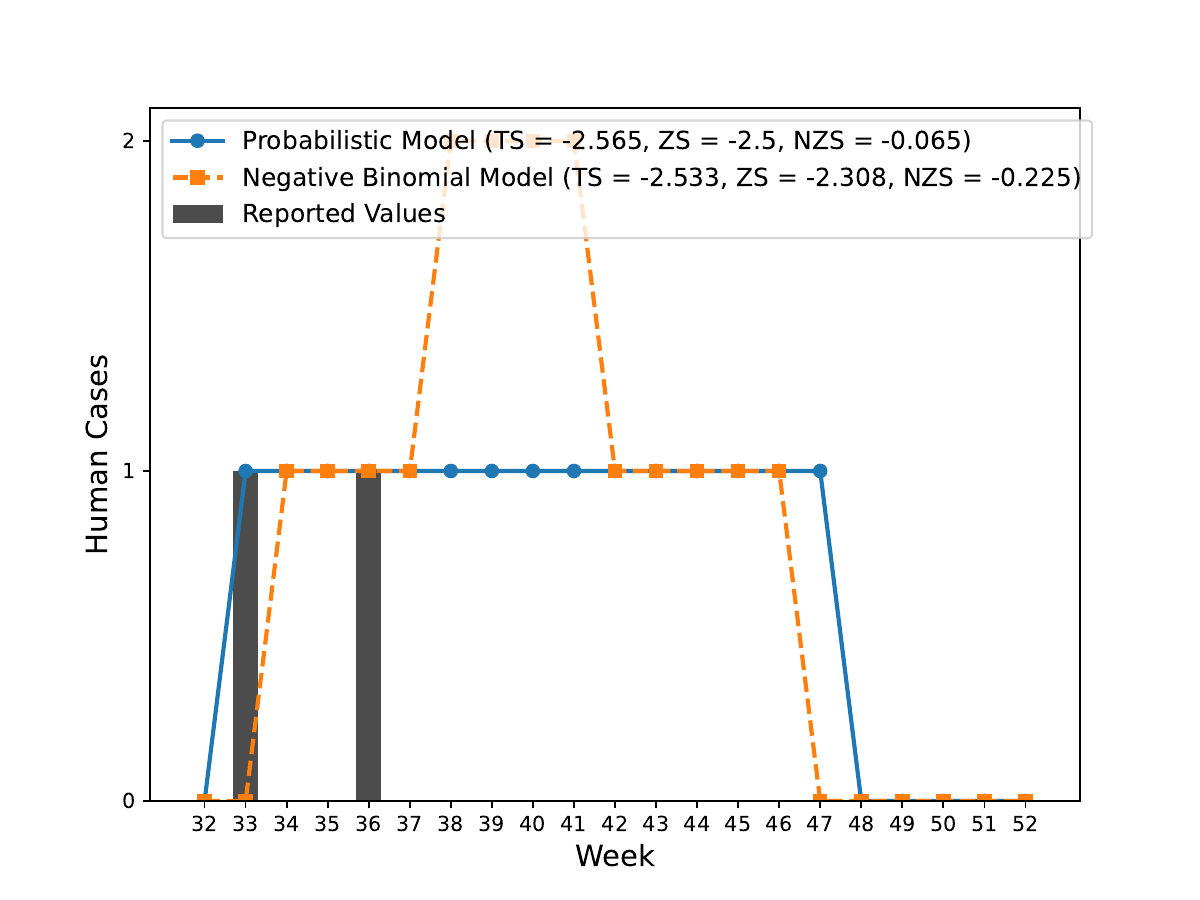}
        \caption{2024}
        \label{fig:model-comparison-2024}
    \end{subfigure}
    \caption{Comparison of Bayesian and Negative Binomial model results for short-term prediction using the logarithmic scoring system for 2022, 2023, and 2024, Orange County California.}
    \label{fig:short-term-model-comparison}
\end{figure}

\begin{figure}[htbp]
    \centering
    \begin{subfigure}[b]{0.45\linewidth}
        \centering
        \includegraphics[width=\linewidth]{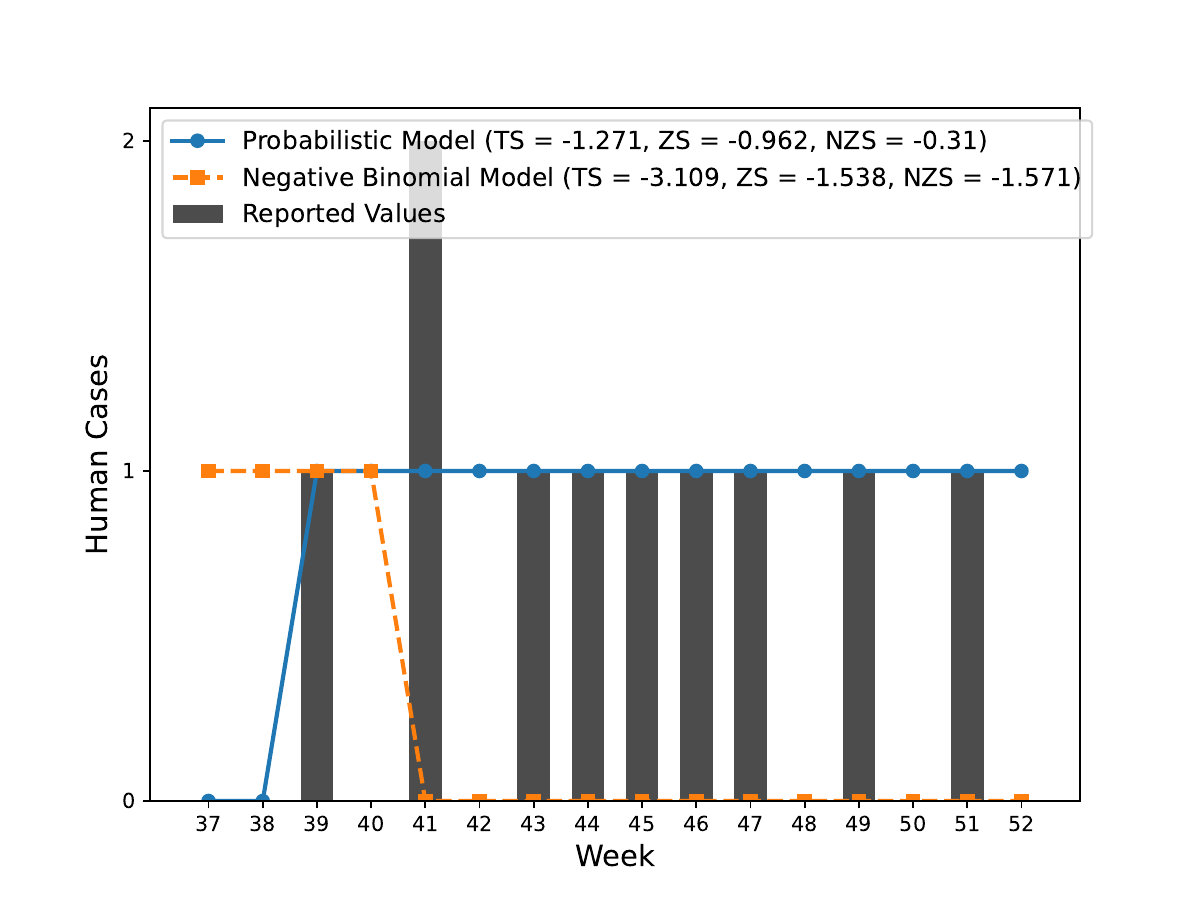}
        \caption{Dallas County}
        \label{fig:dallas2024}
    \end{subfigure}
    \hfill
    \begin{subfigure}[b]{0.45\linewidth}
        \centering
        \includegraphics[width=\linewidth]{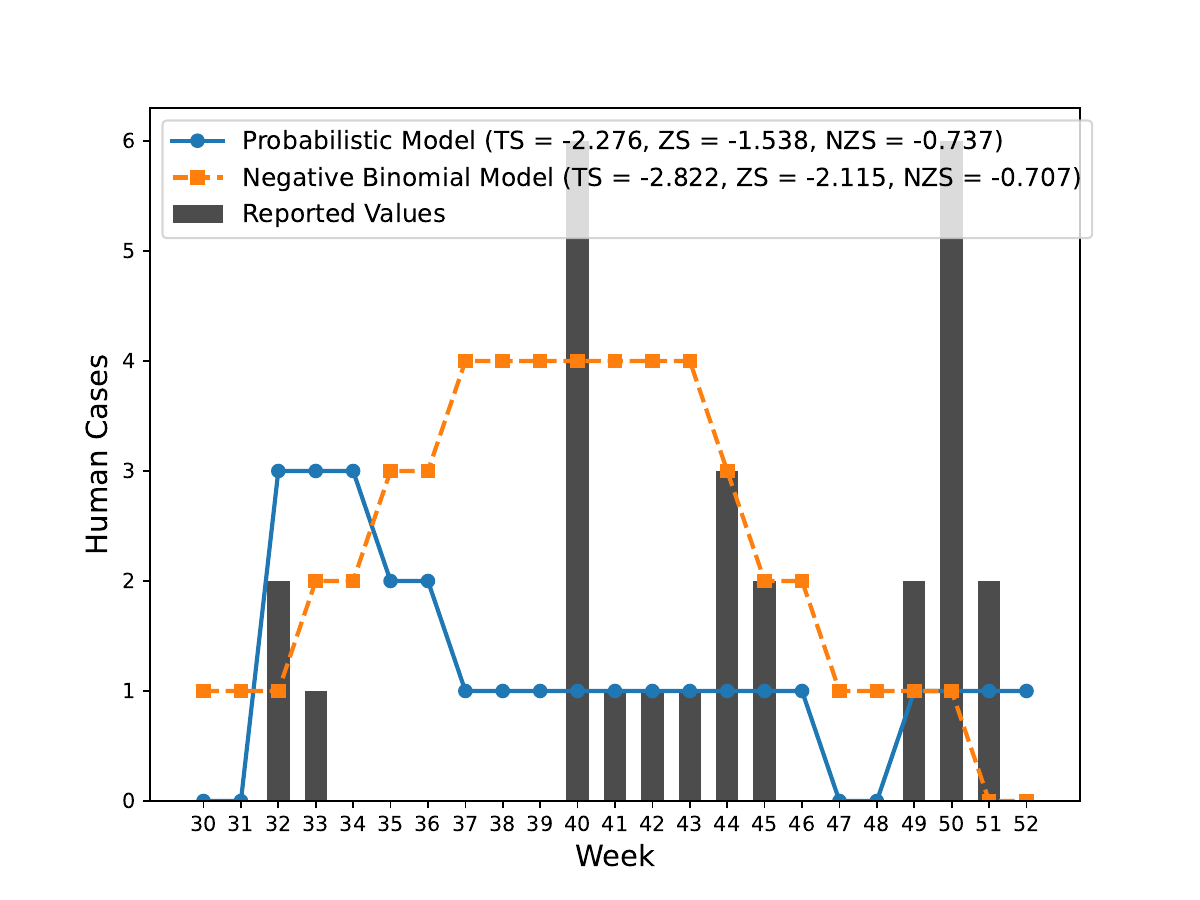}
        \caption{Harris County}
        \label{fig:harris2024}
    \end{subfigure}
    \caption{Comparison of short-term models for West Nile Disease (WND) predictions in Dallas and Harris Counties, Texas, for the year 2024.}
    \label{fig:short-term-model-comparison-Dallas-Harris-2024}
\end{figure}
To better understand the model's performance, we evaluated its score separately for zero and non-zero cases. This approach demonstrates how effectively the model predicts both zero and nonzero observations, and by considering the nonzero score, zero score, and total score together, we gain a comprehensive understanding of its overall performance.\\
Figure (\ref{fig:short-term-model-comparison}) shows the Orange County, Los Angeles County and Kern County results, for the total score (TS), the total score of zero cases (ZS), and the total score for non-zero cases. 
In addition, (\ref{fig:short-term-model-comparison-Dallas-Harris-2024}) shows the results for Dallas and Harris County, Texas, for 2024. 

Furthermore, comparing Figures (\ref{fig:short-term-model-comparison}) with (\ref{fig:model_comparison}) shows that in the three years (for Orange, Los Angeles and Kern County, California), the short-term prediction of our model has a higher score than the long-term prediction, which we expect.
\begin{figure}
    \centering
    \begin{subfigure}{0.45\linewidth}
        \centering
        \includegraphics[width=\linewidth]{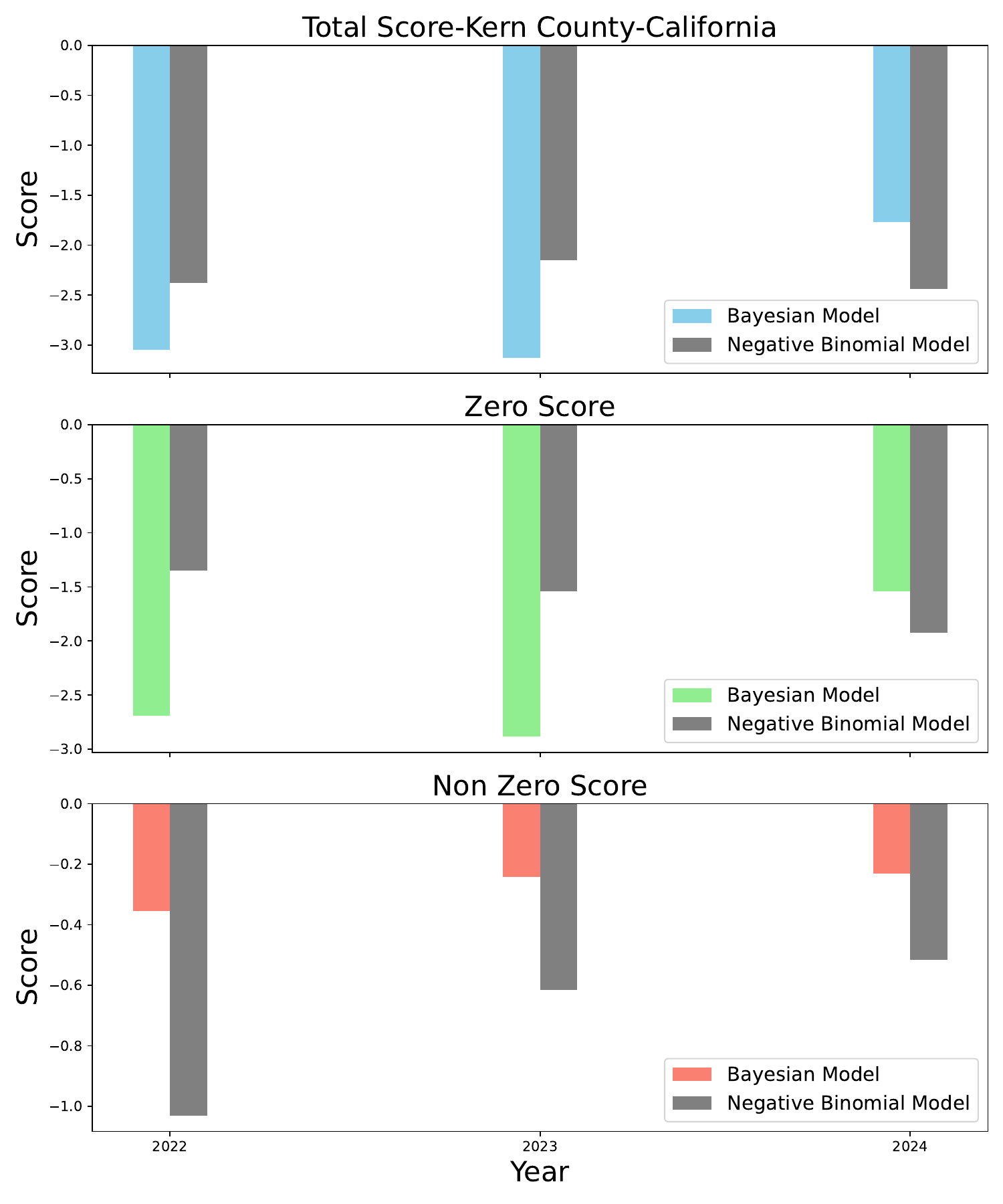}
        \caption{Kern County}
        \label{fig:kern}
    \end{subfigure}
    \hfill
    \begin{subfigure}{0.45\linewidth}
        \centering
        \includegraphics[width=\linewidth]{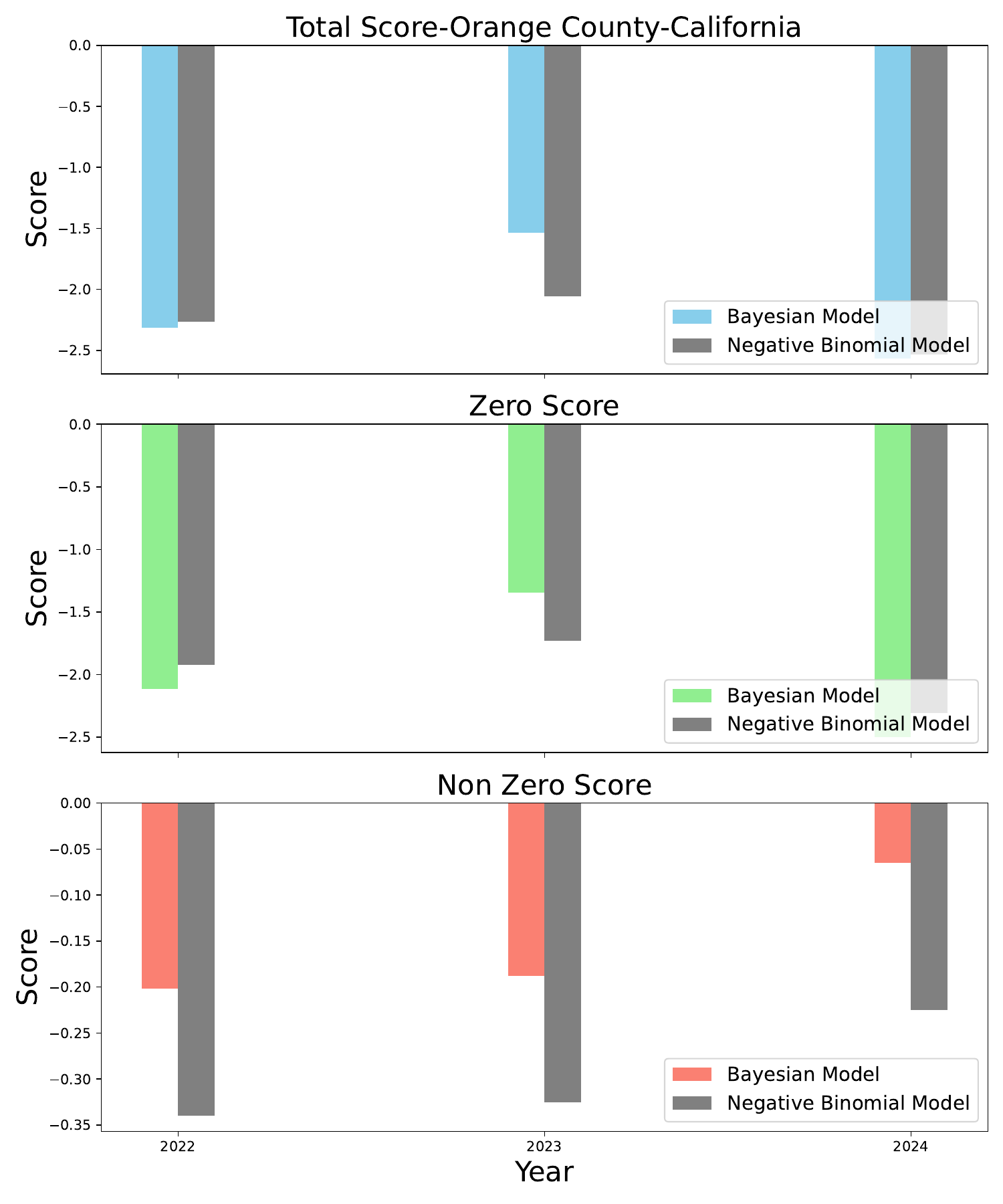}
        \caption{Orange County}
        \label{fig:orange}
    \end{subfigure}
    \hfill
    \begin{subfigure}{0.45\linewidth}
        \centering
        \includegraphics[width=\linewidth]{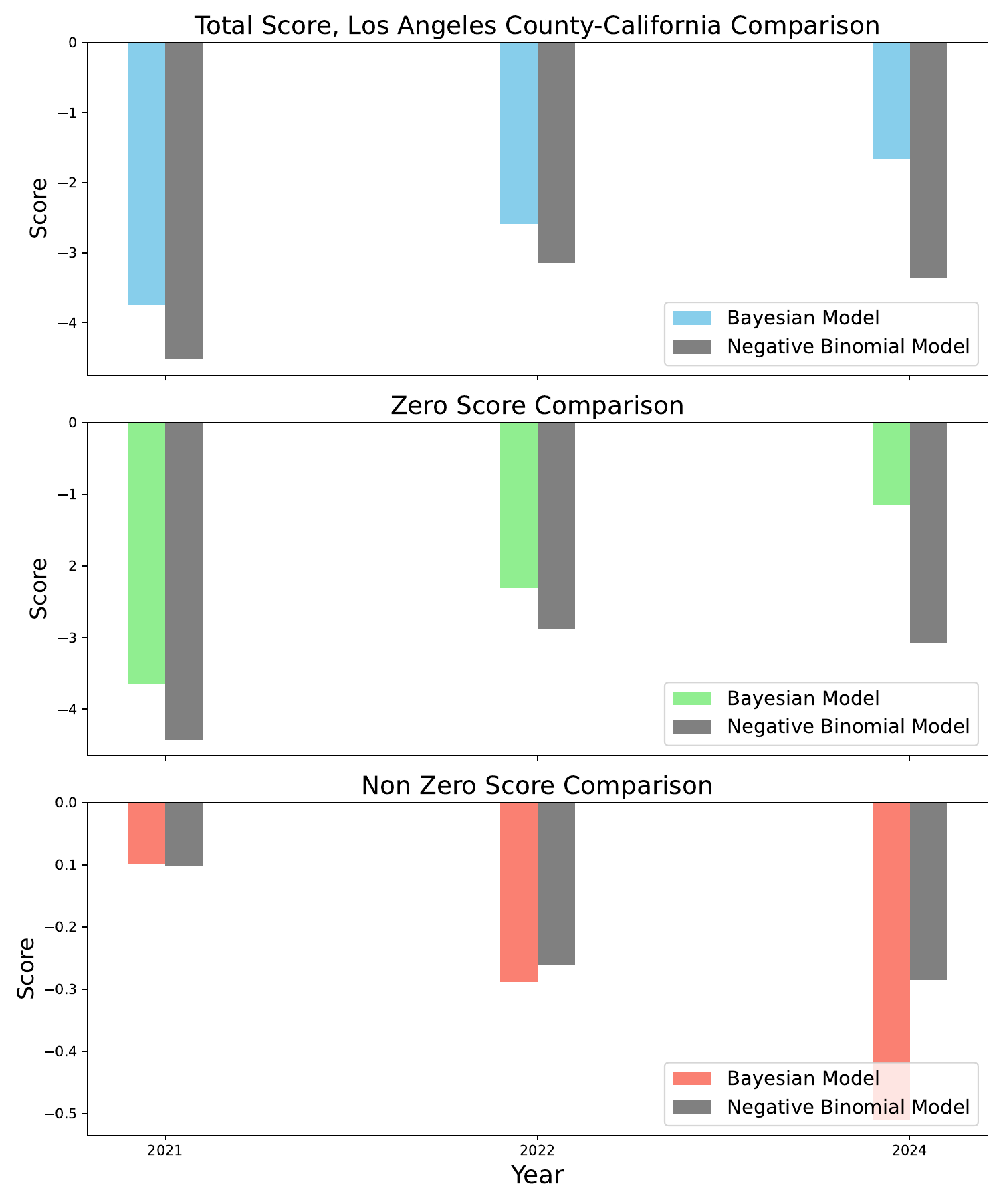}
        \caption{Los Angeles County}
        \label{fig:la}
    \end{subfigure}
    
    \caption{Model Score Comparisons Across Kern, Orange, and Los Angeles Counties.}
    \label{fig:model_comparison-three_counties}
\end{figure}

\begin{figure}[ht]
    \centering
    \begin{subfigure}[b]{0.45\textwidth}
        \centering
        \includegraphics[width=\linewidth]{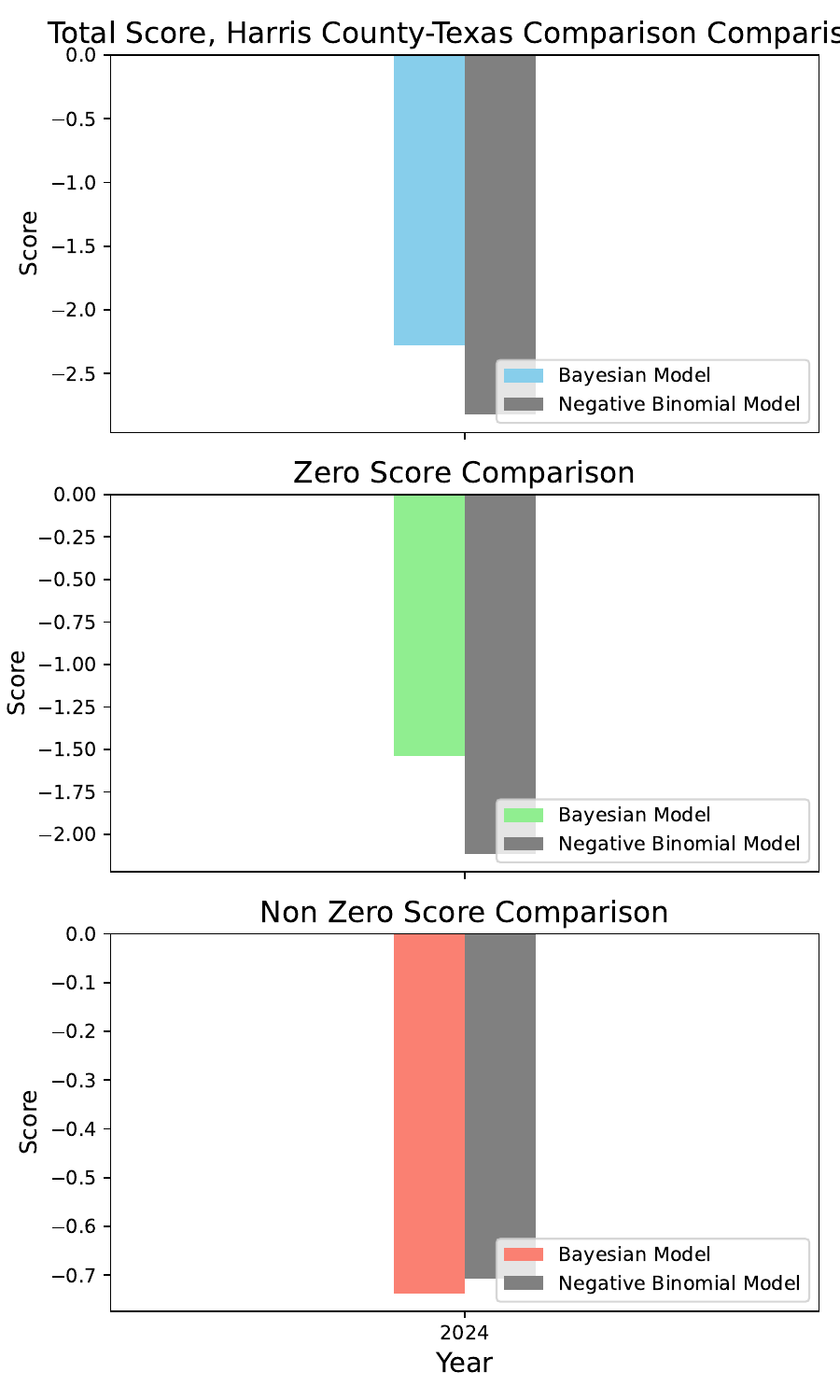}
        \caption{Harris County Comparison}
        \label{fig:harris-comparison}
    \end{subfigure}
    \hfill
    \begin{subfigure}[b]{0.45\textwidth}
        \centering
        \includegraphics[width=\linewidth]{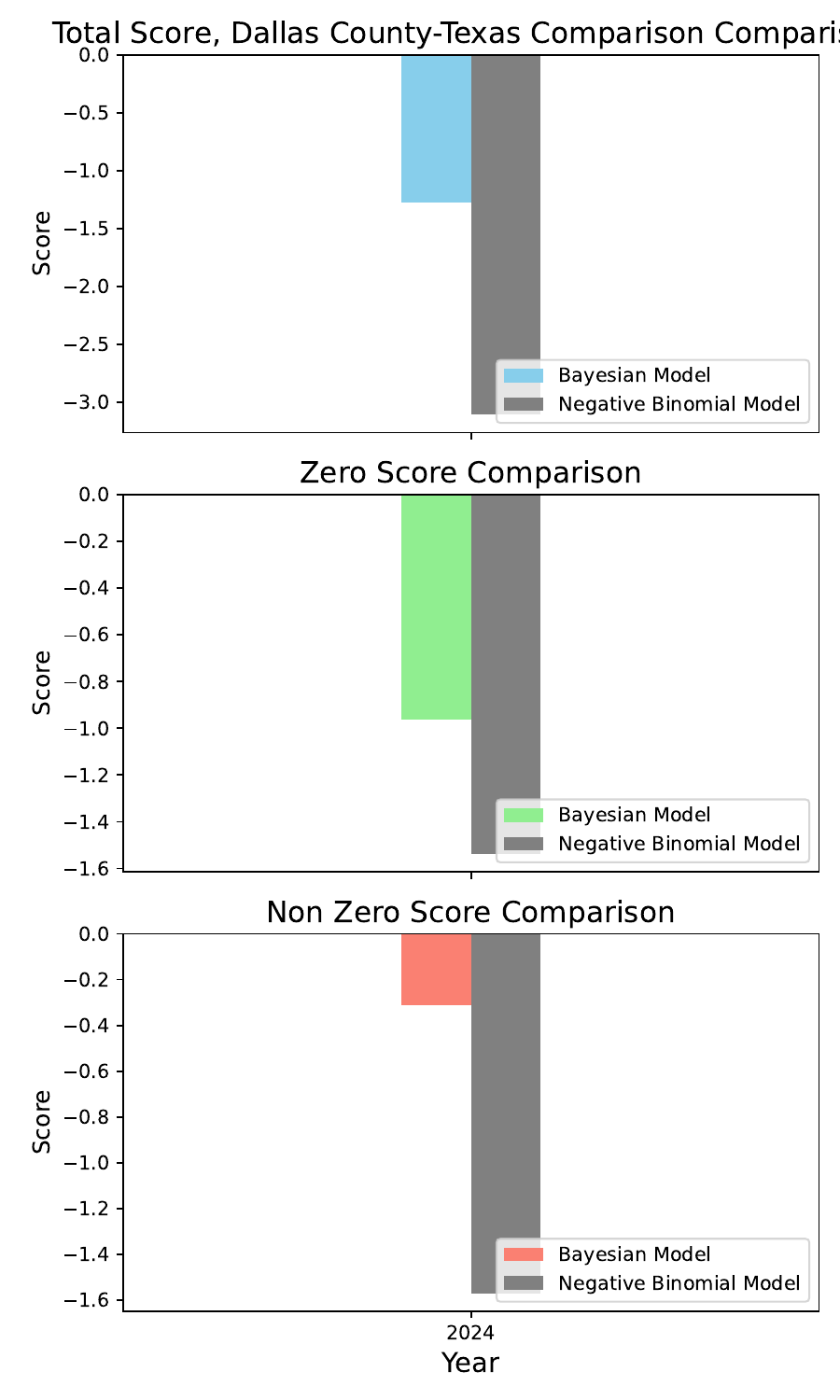}
        \caption{Dallas County Comparison}
        \label{fig:dallas-comparison}
    \end{subfigure}
    \caption{Model Score Comparisons Across Harris County and Dallas County, for 2024.}
    \label{fig:harris-dallas-comparison}
\end{figure}
Figures (\ref{fig:model_comparison-three_counties}) and (\ref{fig:harris-dallas-comparison}) summarize the results for the California and Texas counties. The performance of the Bayesian model in Los Angeles over all three years, 2021, 2022, and 2024, (there is no reported case for 2023) is seen to be consistently better than that of the Negative Binomial model. In Orange County, the Bayesian model outperforms the Negative Binomial model in 2022, whereas in 2023 and 2024, their performances are nearly identical. In Kern County, the Bayesian model demonstrates better performance in 2024. For Dallas and Harris County, the performance of our model is better. It means that, in nearly all counties and years, the Bayesian model shows better predictive performance (especially for non-zero cases) compared to the Negative Binomial model, indicating that it is a strong choice for counties with high infection rates. 
We must note that considering the shorter data period used for Dallas and Harris County, we can conclude that the model produces reliable results even when extensive historical data are unavailable.

\section{Global Warming Effect on WND Spillover}
\begin{figure*}[!t]
    \centering
    \begin{subfigure}[t]{1\textwidth}
        \centering
        \includegraphics[width=\textwidth]{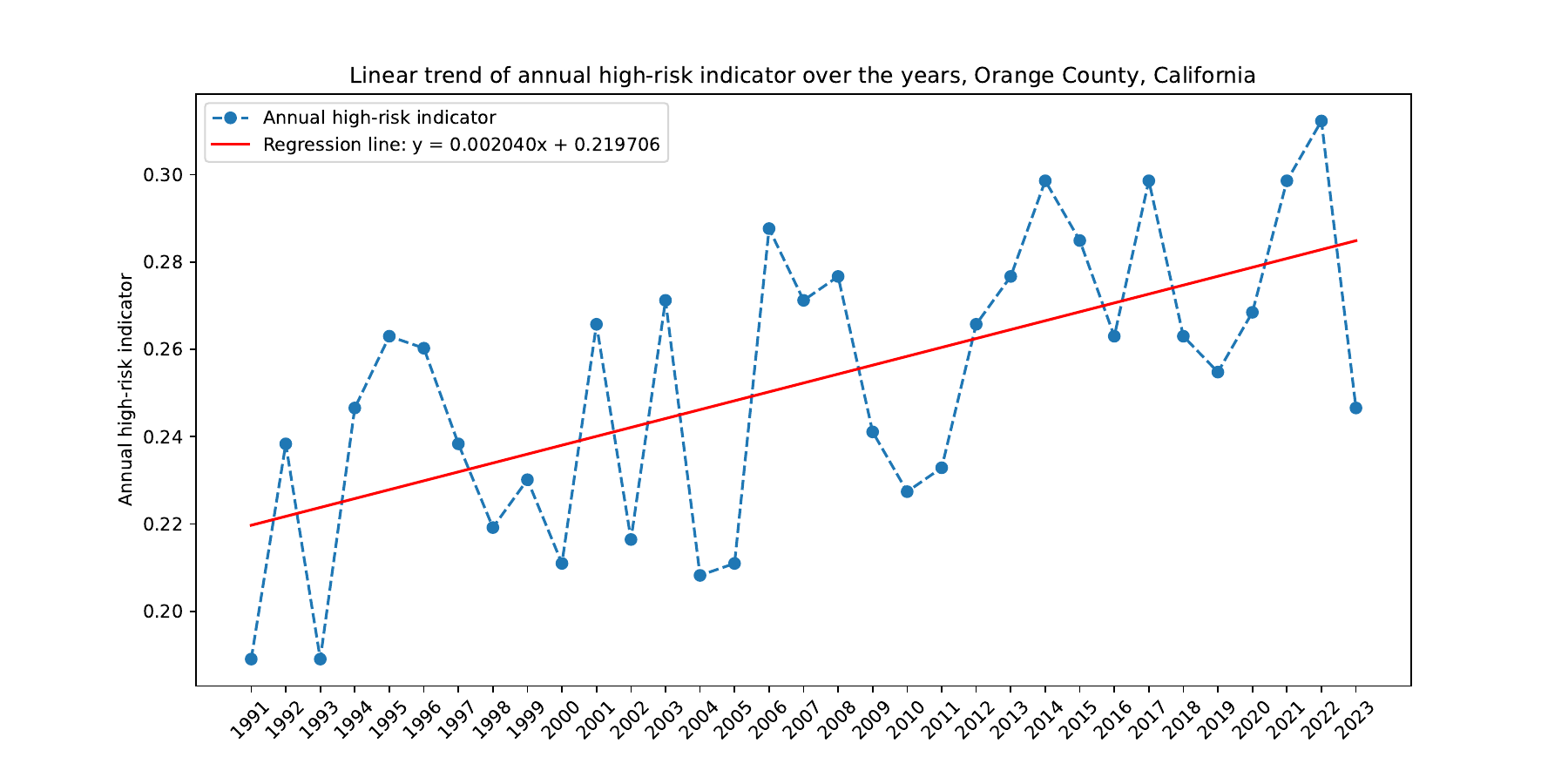}
        \caption{}
        \label{fig:Hum}
    \end{subfigure}
    \hfill
    \begin{subfigure}[t]{1\textwidth}
        \centering
        \includegraphics[width=\textwidth]{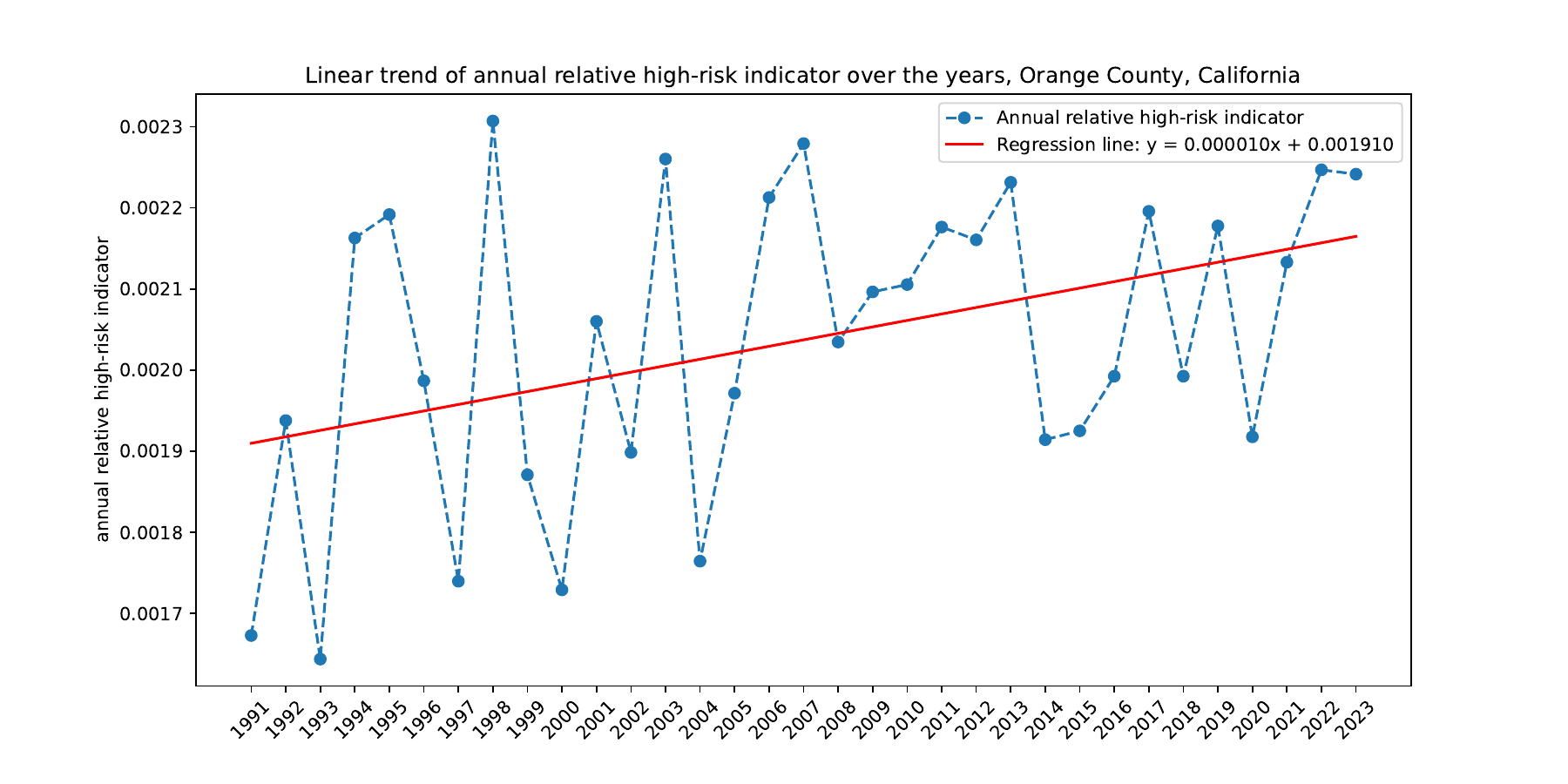}
        \caption{}
        \label{fig:Pre}
    \end{subfigure}
    \hfill
    \vspace{1em}
    \caption{The linear trends over three decades.}
    \label{fig: global annual}
\end{figure*}

Global warming has affected all aspects of human life, including health and hygiene problems. One of the most important aspects of global warming in human life is its effect on diseases and epidemics. This issue becomes more critical when epidemics are linked to vectors that are highly sensitive to weather factors. 
In recent years, several studies have investigated the relationship between climate change, global warming, and epidemic diseases. \cite{erazo2024contribution} offers an insightful analysis of the effects of climate change on West Nile disease in Europe, and \cite{kurane2010effect} explores the impact of global warming on dengue fever and Japanese encephalitis.
In this subsection, the objective is to determine whether there is evidence, based on our probabilistic risk assessment and model results, of the impact of global warming on West Nile disease. 
West Nile virus disease is a vector-borne disease whose vector's life cycle is highly sensitive to humidity, temperature, and precipitation. Consequently, this sensitivity prompts us to consider the relationship between global warming and these factors, leading to a higher risk of infection. Another reason for this interest is that if there is a correlation between the risk of spillover and global warming (or generally climate change), it is possible to implement mitigation strategies based on predictive models related to warming patterns.\\
In continuing, we have attempted to address two broad questions. First, whether there is an increasing trend in the number of days of high-risk correlated with global warming. Second, whether there is any trend in the combination of risky days and high-risk days. To address them first, we have defined the parameters of the annual high-risk indicator \textbf{$r_{year}$} and the annual relative high-risk indicator, \textbf{$r_{relative}$}. The annual high-risk indicator is defined as the fraction of high-risk days on all days of the year. The annual relative high-risk indicator is defined as the fraction of high-risk days between all risk days. To determine whether there is any trend in these parameters over the years, we used the predictive carrying capacity function (\ref{Prediction of Carrying Capacity}) and ran the model on the real climate parameters of each year (historical data). Consequently, we plotted the annual high-risk indicator and the annual relative high-risk indicator from 1991 to 2023 and fitted a regression line to the data. It is important to note that although West Nile disease began to spread in the United States after the 1990s, it is still possible to assess the risk of spillover to the human population in the absence of the pathogen, even before its introduction to the United States or any geographical location.\\
Figure \ref{fig: global annual} presents the results for Orange County, California. Panel (a) shows the trend of the annual high-risk indicator over three decades.
For the annual high-risk indicator, the fitted trend equation is $r_{year}=0.002040t+0.219706$. Based on the statistical test for this fitting p-value=0.000152, this indicates strong evidence against the null hypothesis that the slope is zero. In addition, the standard error of the slope is 0.000472, and the p-value of the Kolmogorov-Smirnov statistic is 0.28, indicating that it does not reject the null hypothesis that the residuals follow a normal distribution. Thus, the residuals of the regression model are consistent with the normal distribution.\\
The slope of the trend is seen to be significantly positive, which can be inferred as the effect of global warming. Now, based on the results, it is possible to suggest that there might be an increasing trend in the number of risky days over the years, potentially influenced by global warming. More significantly, it is not just that the number of high-risk days has increased over the years, but the balance has shifted towards having more high-risk days compared to low and medium-risk days (panel (b)). \\
Furthermore, for the annual relative high-risk indicator, the fitted trend equation is $r_{relative}=0.000008t+0.001910$. The statistical test p-value=0.017611 provides strong evidence against the null hypothesis that the slope is zero. The standard error of the slope is 0.000003, and the p-value for the Kolmogorov-Smirnov statistic is 0.85, indicating that we do not reject the null hypothesis that the residuals are normally distributed. Thus, the residuals of the regression model align with the assumption of normality.

\section{Conclusion}
In this context, the study presents a novel probabilistic framework for the assessment of the risk of the onset of WND spillover using the compartmental mechanistic model of mosquitoes in the Pipiens family. 
The model integrates environmental variables and uses Bayesian methods to set up the PDF to assess risk (onset and severity after the onset of the disease), allowing it to make effective predictions about long- and short-term risks. This approach provided a high degree of precision and was further validated using outbreak data from various cases. Furthermore, the analysis of global warming effects based on this probabilistic risk assessment method showed an increasing trend in the days of high risk; proactive public health interventions could be warranted in the face of climate change.

These results have major implications for public health policy and the management of vector-borne diseases. The possibility of probabilistically predicting the risk of WND spillover allows early interventions that could potentially dampen outbreaks before they expand. In addition, the link of climate change with increased risk emphasizes the implementation of adaptive strategies against the growing menace of vector-borne diseases in a warming world. These can also provide the basis for more directed, geographically oriented public health preparedness and response, and better prepare for the long-term risks of similar epidemics.

Although the Negative-Binomial model is considerably simpler than our Bayesian model, our approach may offer improved performance. Evaluating the models across a wider range of locations would help verify this potential and is a promising direction for future work.

Despite the accuracy of the model, there are certain limitations to consider.
First, the model relies heavily on historical climate data and the historical number of new cases in the human population, which may not fully capture unexpected variations or extreme events. 
Second, the compartmental model assumes specific mosquito and bird species as representatives, which could limit its generalizability to other regions with different species. 
Third, as is known, hundreds of species of mosquitoes and birds are involved in the transmission of West Nile Disease (WND). Focusing on a single family or species provides a generalization of what may occur in a specific location based on environmental parameters; however, it does not imply that our model (or any model) can fully and accurately explain what is happening in reality.

Four, the model’s predictions could be affected by incomplete or inaccurate data on environmental factors or disease transmission rates, leading to potential deviations in forecasting outcomes.
Finally, the carrying capacity estimated in this work from the number of new cases does not fully reflect the reality. For example, the same mosquito population can produce different outbreak sizes depending on the specific virus. However, practically using the number of new cases is the only way one can estimate the carrying capacity. Further research could address these limitations to enhance robustness.

Future research could explore expanding the model to include additional mosquito and bird species to improve generalizability across different regions. Incorporating real-time climate data and machine learning techniques could improve prediction accuracy, particularly for extreme weather events. Furthermore, the integration of human mobility patterns and changes in land use could provide a more comprehensive risk assessment framework, especially a network-based model. 


\end{document}